\documentclass[conference]{IEEEtran}
\IEEEoverridecommandlockouts

\usepackage{cite}
\usepackage{xspace}
\usepackage{url}
\usepackage{graphicx}
\usepackage{balance}  
\usepackage{xcolor}
\usepackage{amsmath}
\usepackage{enumitem}
\usepackage{bbm}
\usepackage{listings}
\usepackage{verbatim}
\usepackage{hyperref}
\usepackage{epstopdf}
\usepackage{amsthm}
\usepackage{tabularx}
\usepackage{subfigure}
\usepackage[linesnumbered,vlined,ruled,noend]{algorithm2e}

\DeclareMathOperator{\cost}{cost}
\DeclareMathOperator{\lcm}{lcm}

\newcommand{\ww}[1]{\textcolor{red}{Wentao: #1}}
\newcommand{\HL}[1]{\textcolor{black}{#1}}

\newcommand{\blue}[1]{\textcolor{black}{#1}}
\newcommand{\red}[1]{\textcolor{red}{#1}}

\newtheorem{definition}{Definition}
\newtheorem{theorem}{Theorem}

\newtheorem{example}{Example}
\newtheorem{observation}{Observation}

\def\BibTeX{{\rm B\kern-.05em{\sc i\kern-.025em b}\kern-.08em
    T\kern-.1667em\lower.7ex\hbox{E}\kern-.125emX}}

\begin{document}

\title{Factor Windows: Cost-based Query Rewriting for Optimizing Correlated Window~Aggregates}



\author{
	\IEEEauthorblockN{\dag Wentao Wu, \dag Philip A. Bernstein, \dag Alex Raizman, \ddag Christina Pavlopoulou$^*$\thanks{$^*$This work was done when Christina Pavlopoulou was at Microsoft.}}
	\IEEEauthorblockA{\dag \textit{Microsoft Corporation}, \ddag \textit{University of California, Riverside} \\
	\normalsize {
    \dag \{wentwu, philbe, alexr\}@microsoft.com, \ddag cpavl001@ucr.edu}}
}

\maketitle
\thispagestyle{plain}
\pagestyle{plain}

\begin{abstract}
Window aggregates are ubiquitous in stream processing.
In Azure Stream Analytics (ASA), a stream processing service hosted by Microsoft's Azure cloud, we see many customer queries that contain aggregate functions (such as \texttt{MIN} and \texttt{MAX}) over multiple correlated windows (e.g., tumbling windows of length five minutes and ten minutes) defined on the \emph{same} event stream.
In this paper, we present a cost-based optimization framework for optimizing such queries by sharing computation among multiple windows.
In particular, we introduce the notion of \emph{factor windows}, which are auxiliary windows that are not in the input query but may nevertheless help reduce the overall computation cost, and our cost-based optimizer can produce rewritten query plans that have lower costs than the original query plan by utilizing factor windows.
Since our optimization techniques are
at the level of query (plan) rewriting,
they can be implemented on any stream processing system that supports a declarative, SQL-like query language without changing the underlying query execution engine.
We formalize the shared computation problem, present the optimization techniques in detail, and report evaluation results over both synthetic and real datasets. 
Our results show that, compared to the original query plans, the rewritten plans output by our cost-based optimizer can yield significantly higher (up to 16.8$\times$) throughput.
\end{abstract}


\section{Introduction}

\begin{figure}
\centering
\subfigure[ASA query]{ \label{fig:multi-window-single-agg:query}
    \includegraphics[clip, trim=8.5cm 9cm 8.5cm 5.5cm, width=0.9\columnwidth]{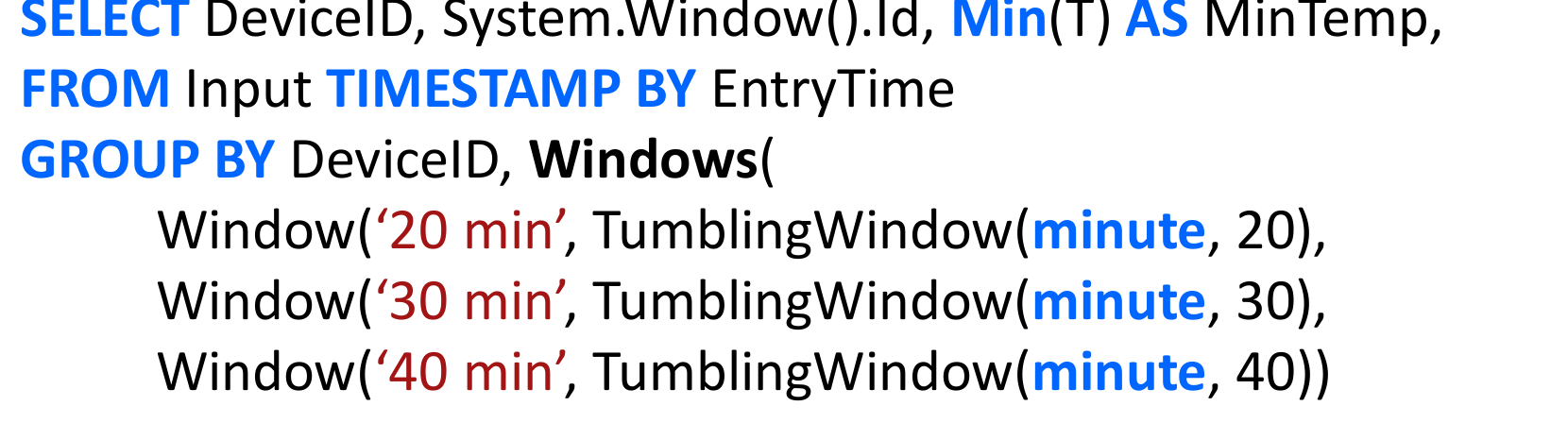}}
\subfigure[Translated Trill~\cite{ChandramouliGBDPTW14} expression]{ \label{fig:multi-window-single-agg:plan}
    \includegraphics[clip, trim=7cm 8.5cm 9.5cm 6.5cm, width=\columnwidth]{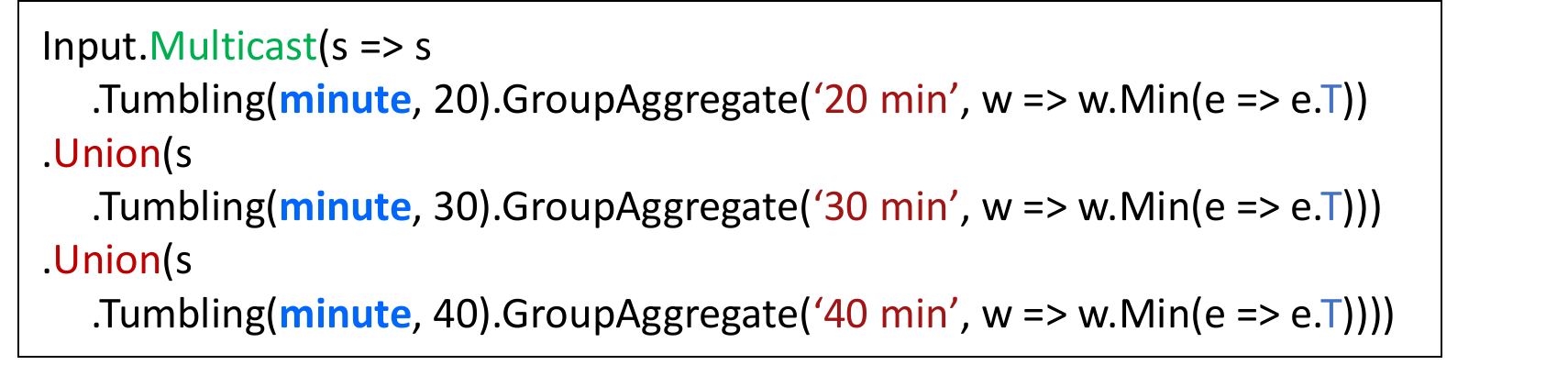}}
\vspace{-1em}
\caption{An ASA aggregation query over multiple windows.}
\label{fig:multi-window-single-agg}
\end{figure}

Near-real-time querying of data streams is required by many applications, such as algorithmic stock trading, fraud detection, process monitoring, and 
RFID event processing.
The importance of this technology has been growing due to the surge of demand for Internet of Things (IoT) and edge computing applications, leading to a variety of systems from both the open-source community (e.g., Apache Storm~\cite{ToshniwalTSRPKJGFDBMR14}, Apache Spark Streaming~\cite{ArmbrustDTYZX0S18,ZahariaDLHSS13}, Apache Flink~\cite{CarboneKEMHT15}) and the commercial world (e.g., Amazon Kinesis~\cite{kinesis}, Microsoft Azure Stream Analytics~\cite{asa}, Google Cloud Dataflow~\cite{dataflow}).
Although imperative programming/query interfaces, such as the functional expressions used in Trill~\cite{ChandramouliGBDPTW14} (see Figure~\ref{fig:multi-window-single-agg:plan} for an example), remain available in these stream processing systems, declarative SQL-like query interfaces are becoming increasingly popular. For example, Apache Spark recently introduced \emph{structured streaming}, a declarative streaming query API based on Spark SQL~\cite{ArmbrustDTYZX0S18}. Azure Stream Analysics (ASA), Microsoft's cloud-based stream processing service, also differentiates itself with a SQL interface.


Declarative query interfaces allow users of stream processing systems to focus on \emph{what} task is to be completed, rather than the details of \emph{how} to execute it.
When it comes to the question of efficient query execution, they rely on powerful query optimizers.
In the traditional world of database management systems, the success of declarative query languages heavily depends on \emph{cost-based query optimization}, which has been an active area for research since 1970's~\cite{SelingerACLP79}.
Unfortunately, in spite of the increasing popularity of declarative query interfaces in stream processing systems, cost-based query optimization of such systems remains underdeveloped --- most systems, if not all, rely on \emph{rule-based} query optimizers.

In this paper, we focus on cost-based optimization techniques for window aggregates, a ubiquitous category of streaming queries, in declarative stream processing systems.
In our experience with ASA,
users often want to perform the same aggregate function over the same data stream but with windows of different sizes.
They do this for a variety of reasons, such as learning about or debugging a stream by exploring its behavior over different time periods, reporting near real-time behavior of a stream over small windows as well as
much longer windows (e.g., an hour vs. a week), and simultaneously supporting different users whose dashboards display stream behavior over different window sizes.
\blue{For example, Microsoft's Azure IoT Central service~\cite{iot-central} hosts thousands of concurrently running dashboard queries that are window aggregates over event streams generated by various IoT devices from multiple users. It is very common in this scenario to see multiple (e.g., 5 to 10) queries over the same event stream but with varying window sizes, issued by downstream user applications that collect various telemetries from the same device reading.}

A straightforward implementation would evaluate the aggregate function over each window separately.
Although this implementation is relatively simple, it potentially wastes CPU cycles.
We start with an example to illustrate this inefficiency.

\begin{figure}
\centering
\subfigure[Query plan rewriting]{ \label{fig:window-set:plan-rewrite}
\includegraphics[clip, trim=3cm 3cm 3cm 3cm, width=\columnwidth]{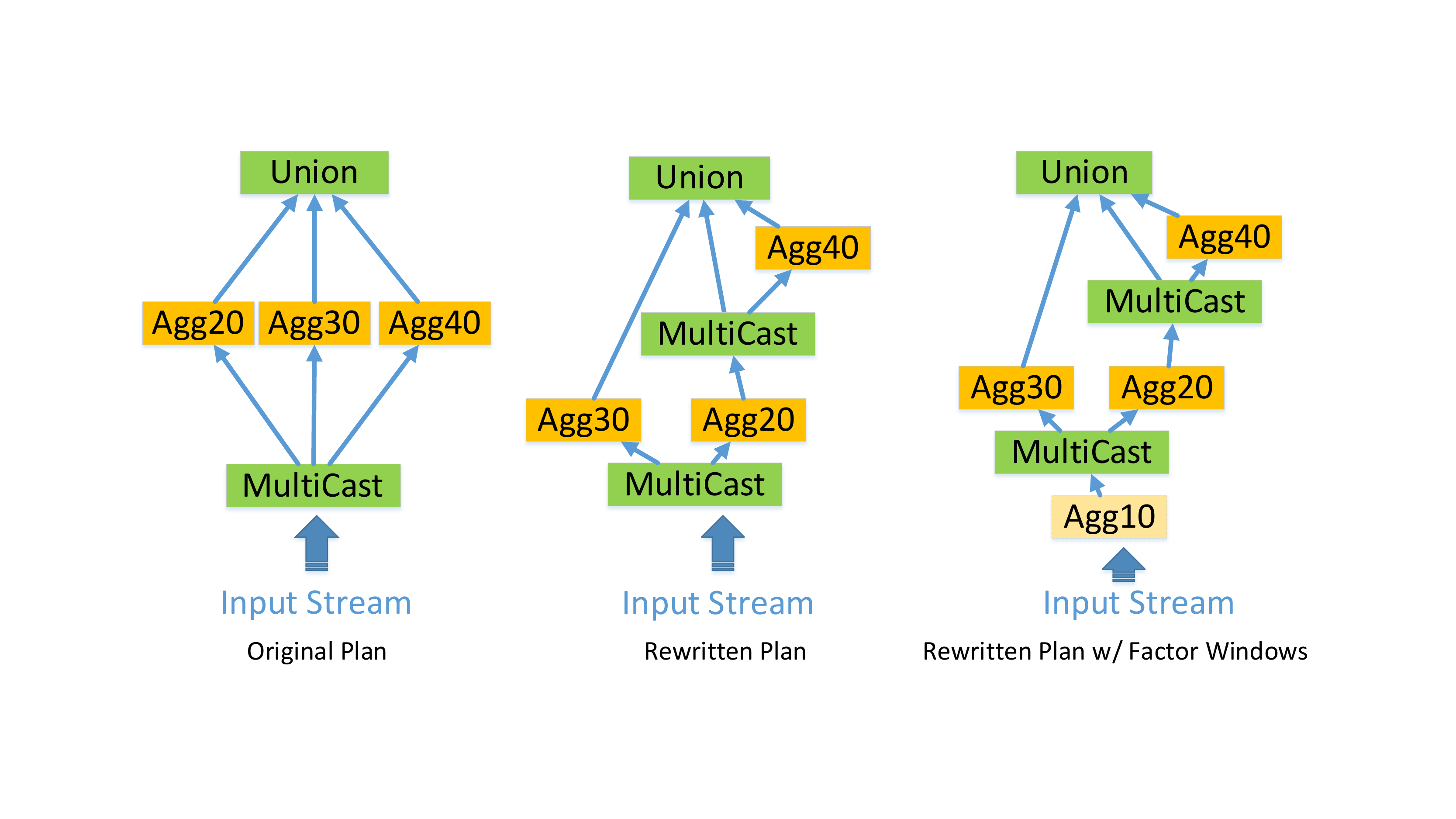}}
\subfigure[Translated Trill expression of the rewritten plan]{ \label{fig:window-set:trill-expr}
\includegraphics[clip, trim=6.8cm 8cm 8.5cm 7.3cm, width=\columnwidth]{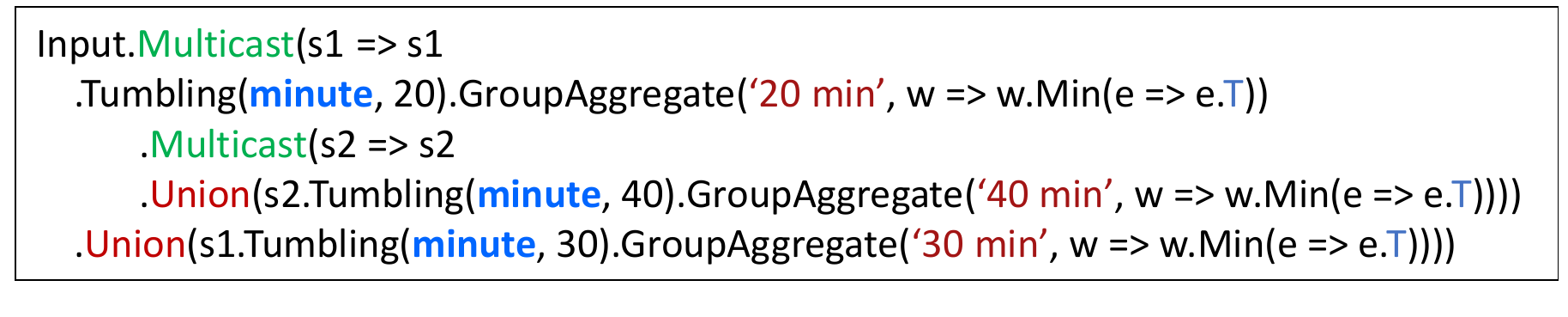}}
\subfigure[Translated Trill expression of the rewritten plan with factor windows]{ \label{fig:window-set:trill-expr:fw}
\includegraphics[clip, trim=6.8cm 8cm 8cm 6.5cm, width=\columnwidth]{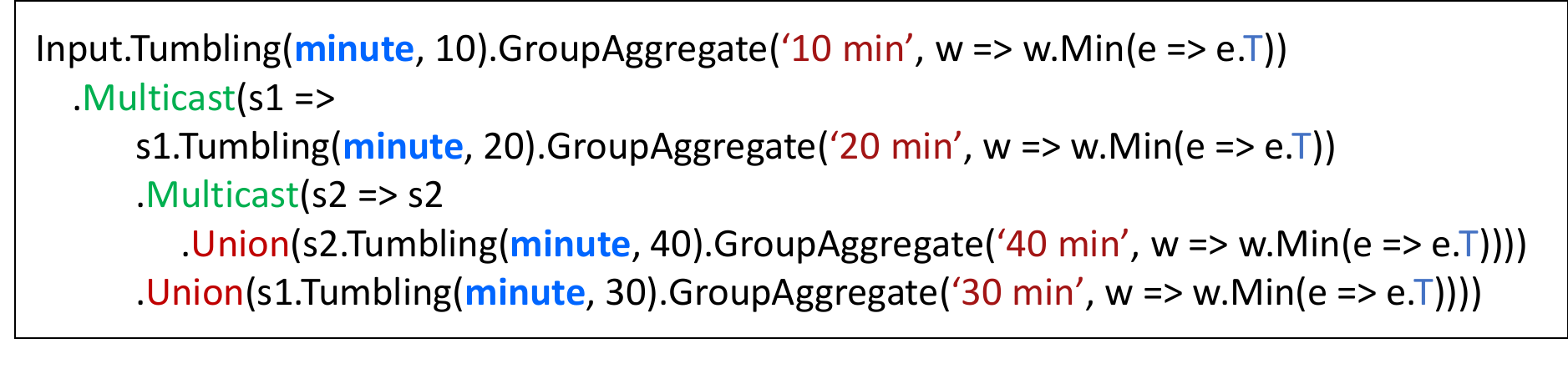}}
\vspace{-1em}
\caption{Rewritten query plans by cost-based optimization.}
\label{fig:window-set:query-rewrite}
\end{figure}

\begin{example}[Multi-window Aggregate Query]\label{example:multi-window-single-agg}
Figure~\ref{fig:multi-window-single-agg:query} presents a query with a single aggregate function, \verb|MIN|, over multiple windows.
It returns the minimum temperature reported by each device every 20, 30, and 40 minutes.
Figure~\ref{fig:multi-window-single-agg:plan} presents its execution plan in ASA, which is a Trill~\cite{ChandramouliGBDPTW14} expression that runs the aggregate over each window separately and then takes a union of the results.
The plan is shown graphically on the left side of Figure~\ref{fig:window-set:plan-rewrite}.
\end{example}

This execution plan is clearly inefficient.
For example, the \verb|MIN| function over the 40-minute tumbling window can be computed from two consecutive tuples output by the 20-minute tumbling window, instead of computing it directly from the input stream.
Such overlapping windows present an opportunity for optimization.

Our cost-based optimization technique exploits this opportunity by finding the cheapest way of computing the window aggregates in terms of the overall CPU overhead.
It produces the revised query plan shown graphically in the middle of Figure~\ref{fig:window-set:plan-rewrite}.
Instead of computing the aggregate function over the three windows separately,
the revised plan organizes the windows into a hierarchical structure.
As a result, downstream windows use sub-aggregates from their upstream windows as inputs.
For instance,
aggregates of the 40-minute window are computed from sub-aggregates that are outputs of the 20-minute window.
This revised plan's graph is translated into a Trill~\cite{ChandramouliGBDPTW14} expression shown in Figure~\ref{fig:window-set:trill-expr}.

In addition to exploiting shared computation among the \emph{existing} windows in the input query, we can further explore other \emph{auxiliary} windows that are not in the input query but may nevertheless help reduce the overall computation cost.
For this purpose, we introduce the notion of \emph{factor window}.
As an example, for the ASA query in Figure~\ref{fig:multi-window-single-agg:query}, we can insert a 10-minute tumbling window as a factor window, which leads to the revised query plan on the right side of Figure~\ref{fig:window-set:plan-rewrite}.
The corresponding Trill expression is given in Figure~\ref{fig:window-set:trill-expr:fw}.
Based on our experimental evaluation, query plans with factor windows can yield significantly higher throughput~\cite{streamBenchmark} than both the original plans and plans without using factor windows, on both synthetic and real datasets.

\paragraph*{\blue{Comparison to Window Slicing}}
\blue{
One prominent line of work on optimizing window aggregates is \emph{window slicing} (e.g.~\cite{KrishnamurthyWF06,LiMTPT05-no-pane,GuirguisSCL11,GuirguisSCL12,TraubGCBKRM19,TraubGCBKRM21}),
which chops the entire window into smaller chunks and then computes the aggregate over the whole window by aggregating sub-aggregates over the small chunks. Unlike window slicing, we do not proactively chop a window.
Instead, we exploit the internal overlapping relationships between correlated windows, which are ignored by window slicing techniques.
Recently, Traub et al. proposed Scotty, a ``general stream slicing'' framework that extends the scope of window types and aggregate functions where window slicing can be applied~\cite{TraubGCBKRM19,TraubGCBKRM21}.
Scotty offers a \emph{non-intrusive} implementation approach by writing ``connectors'' to existing stream processing engines such as Apache Flink.
Our approach shares the same non-intrusive aspiration, though it operates by \emph{query rewriting}.
One advantage of our approach is that it does not assume any extra support from the underlying stream processing engine, such as the ``\emph{user-defined operator}'' feature required by Scotty. For example, Scotty currently does not support Trill, and it is unclear how to write a Scotty ``connector'' for Trill.
Moreover, our approach does not require engine-specific implementation beyond the support of a SQL-like query interface.
For example, Scotty needs to handle checkpoints and state backends for Apache Flink~\cite{Scotty-GitHub}.
In our experimental evaluation, we compared our cost-based optimization approach with Scotty (Section~\ref{sec:evaluation:compare:scotty}).
For the types of windows and aggregate functions that are supported by both our approach and Scotty, we observe that our approach achieved similar and often much better throughput. On the other hand, Scotty supports more types of windows and aggregate functions. We leave the problem of extending our approach to these cases as future work.}



\paragraph*{Summary of contributions and Paper Organization}
\HL{To summarize, this paper makes the following contributions:
\begin{itemize}
    \item We introduce the \emph{window coverage graph} (WCG), a formal model and data structure that captures the overlapping relationships between windows (Section~\ref{section:window-set}).
    \item We propose a cost-based optimization framework using the WCG model, to minimize the computation cost of multi-window aggregate queries, as well as related query rewritings on the optimal, min-cost WCG (Section~\ref{section:cost-based-eval}).
    \item We extend the cost-based optimization framework by considering \emph{factor windows}, which are auxiliary windows that are not present in the query but can further reduce the overall computation cost (Section~\ref{section:factor-window}).
    \item We evaluate our proposed optimizations using both synthetic and real streaming datasets, with a focus on comparing the throughput of the original query plans and the optimized plans, without and with factor windows. Our results demonstrate that the optimized plans, especially the ones with factor windows, can outperform the original plans by having up to 16.8$\times$ throughput (Section~\ref{section:evaluation}).
\end{itemize}
}

\section{Overlaps Between Windows}\label{section:window-set}

We start with a formal study of the overlapping relationships between windows.
We then propose \emph{window coverage graph}, a formal model and data structure that captures overlapping relationships for a given set of windows.


\subsection{Preliminaries}

We follow the convention in the literature to represent a \emph{window} $W$ using two parameters~\cite{streamBenchmark}:
\begin{itemize}
    \item $r$ -- the \emph{range} of $W$ that represents its duration;
    \item $s$ -- the \emph{slide} of $W$ that represents the gap between its two consecutive firings.
\end{itemize}
Throughout this paper, we assume that $s$ and $r$ are integers and use the same time unit (e.g., second, minute, hour).
We assume $0<s\leq r$ and write $W\langle r, s\rangle$.
We call $W$ a \emph{hopping window} if $s < r$, or a \emph{tumbling window} if $s = r$.

A \emph{window set} $\mathcal{W}=\{W_1, ..., W_n\}$ represents a \emph{set} of windows with no duplicates.
An aggregate function $f$ defined over a window set $\mathcal{W}$ computes a result for each $W\in\mathcal{W}$ and takes a union of the results, i.e., $f(\mathcal{W})=\cup_{W\in\mathcal{W}}f(W)$.

\subsubsection{The Interval Representation of a Window}

As an alternative to the ``range-slide'' based representation, we can use a sequence of \emph{intervals} to represent the lifetime of a window~\cite{BargaGAH07}.
Without loss of generality, we assume the intervals are \emph{left-closed} and \emph{right-open}
and define the \emph{interval representation} of a window $W\langle r,s\rangle$ as
$W=\{[m\cdot s, m\cdot s + r)\},$
where $m\geq 0$ is an integer.
For example, the interval representation of window $W(10,2)$ is 
$\{[0, 10)$, $[2, 12)$, ... $\}$.

\subsection{Window Coverage and Partitioning}

Now consider two windows $W_1\langle r_1, s_1\rangle$ and $W_2\langle r_2, s_2\rangle$.
Using their interval representations, we also have
$W_1=\{[m_1\cdot s_1, m_1\cdot s_1 + r_1)\}$
and
$W_2=\{[m_2\cdot s_2, m_2\cdot s_2 + r_2)\},$
where $m_1\geq 0$ and $m_2\geq 0$ are integers.

\begin{definition}[Window Coverage]
\label{definition:window-coverage}
We say that $W_1$ is \emph{covered by} $W_2$, denoted $W_1 \leq W_2$, if $r_1 > r_2$ and for any interval $I=[a, b)$ in $W_1$ there exist intervals $I_a=[a, x)$ \textbf{\underline{and}} $I_b=[y, b)$ in $W_2$ such that \blue{$a < y$} and $x< b$.
(Note that, if $W_1$ is covered by $W_2$, then these two intervals are unique.)
As a special case, a window is covered by itself.
\end{definition}

\begin{figure}
\centering
    \includegraphics[width=0.6\columnwidth]{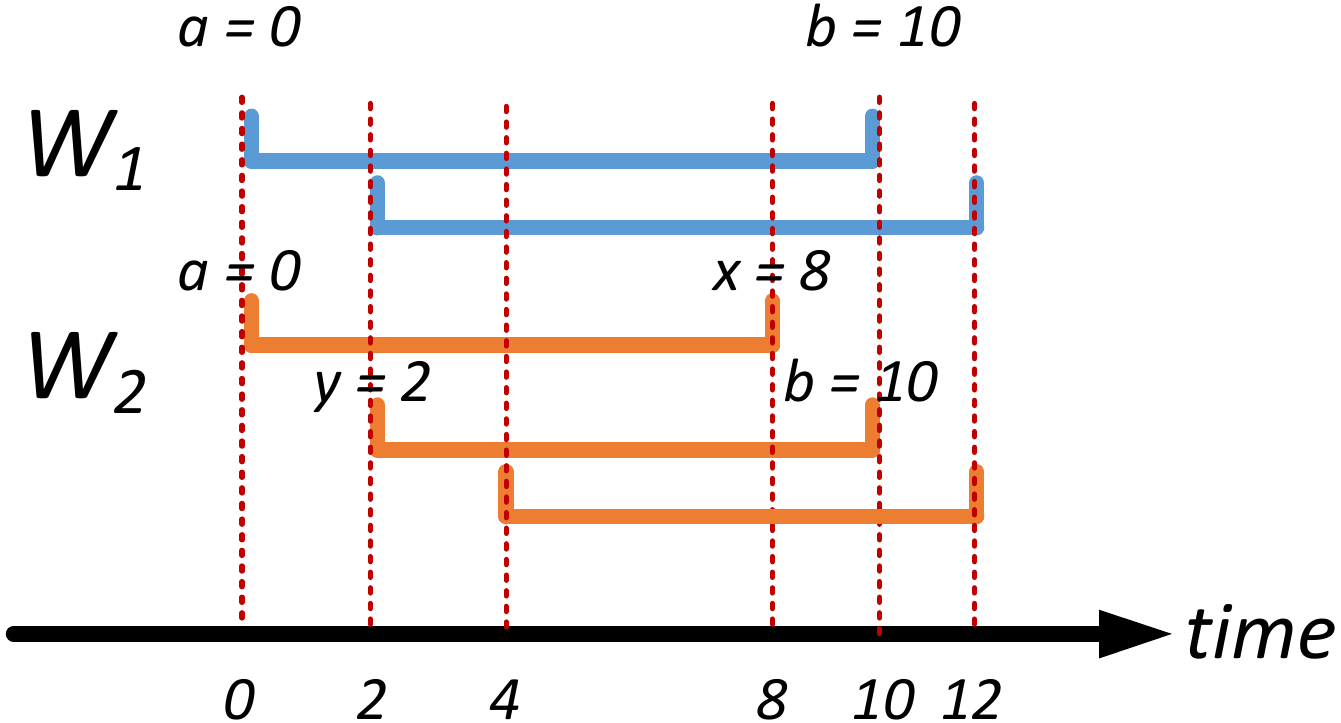}
\vspace{-0.5em}
\caption{An example of window coverage.}
\label{fig:window-coverage-example}
\end{figure}

\begin{example}[Window Coverage]\label{example:window-coverage}
Consider $W_1\langle s_1=2, r_1=10\rangle$ and $W_2\langle s_2=2, r_2=8\rangle$.
Figure~\ref{fig:window-coverage-example} plots the first two intervals of $W_1$ ($\{[0, 10)$, $[2, 12)\}$) and the first three intervals of $W_2$ ($\{[0, 8)$, $[2, 10)$, $[4, 12)\}$).
The first interval of $W_1$ is covered by the 1st and 2nd intervals of $W_2$, and the second interval of $W_1$ is covered by the 2nd and 3nd intervals of $W_2$.
\end{example}

The following theorem provides sufficient and necessary conditions for the window coverage relation (\blue{proofs are available in the appendix of this paper}): 
\begin{theorem}\label{theorem:window-coverage}
$W_1$ is covered by $W_2$ if and only if (1) $s_1$ is a multiple of $s_2$ and (2) $\delta_r =r_1-r_2$ is a multiple of $s_2$.
\end{theorem}

\begin{example}[Window Coverage Theorem]\label{example:window-coverage-thm}
Consider again the windows of Example~\ref{example:window-coverage}: $W_1\langle s_1=2, r_1=10\rangle$ and $W_2\langle s_2=2, r_2=8\rangle$.
We have $s_1/s_2=1$, so $s_1$ is a multiple of $s_2$, and $(r_1-r_2)/s_2=1$, so $r_1 - r_2$ is a multiple of $s_2$.
By Theorem~\ref{theorem:window-coverage}, $W_1$ is covered by $W_2$.

\end{example}

\subsubsection{A Partial Order}

The window coverage relation defines a \emph{partial order} over windows, as characterized by:
\begin{theorem}\label{theorem:transitivity}
The window coverage relation is \emph{reflexive}, \emph{antisymmetric}, and \emph{transitive}.
\end{theorem}

\subsubsection{Interval Coverage}

Suppose that $W_1\leq W_2$. For any interval $I=[a, b)$ in $W_1$, let $I_a=[a, x)$ and $I_b=[y, b)$ be the two intervals in $W_2$ specified by Definition~\ref{definition:window-coverage}.

\begin{definition}[Covering Interval Set]\label{definition:covering-set}
Let the set of intervals ``between'' $I_a$ and $I_b$ in $W_2$ be
$\mathcal{I}_{a,b} = \{[u, v): a\leq u \textbf{ and } v\leq b\}.$
We call $\mathcal{I}_{a,b}$ the \emph{covering (interval) set} of $I$.
\end{definition}
Clearly, $I_a,I_b\in\mathcal{I}_{a,b}$.
The cardinality $|\mathcal{I}_{a,b}|$ is independent of the choice of $a$ and $b$.
We call it the \emph{covering multiplier} of $W_2$ with respect to $W_1$, denoted $M(W_1, W_2)$.
An analytic form for the covering multiplier is given by:
\begin{theorem}\label{theorem:covering-multiplier}
If the window $W_1\langle r_1, s_1\rangle$ is covered by the window $W_2\langle r_2, s_2\rangle$, then
$M(W_1, W_2)=1+(r_1-r_2)/s_2.$
\end{theorem}

We now introduce the more general notion of ``interval coverage'' based on the above discussion.
\begin{definition}[Interval Coverage]
We say that an interval $I$ is \emph{covered by} a set of intervals $\mathcal{I}$ if $I=\cup_{J\in\mathcal{I}}J$.
\end{definition}

\begin{example}[Interval Coverage]
In Figure~\ref{fig:window-coverage-example}, for the first interval in $W_1$ the covering set consists of the first and second intervals in $W_2$, and for the second interval in $W_1$ consists of the second and third intervals in $W_2$.
\end{example}


\subsubsection{Interval/Window Partitioning}

A special case of interval coverage is when the intervals in the covering set are disjoint.
\begin{definition}[Interval Partitioning]
If an interval $I$ is covered by a set of intervals $\mathcal{I}$ such that the intervals in $\mathcal{I}$ are mutually exclusive, then $I$ is \emph{partitioned by} $\mathcal{I}$.
\end{definition}
We can further define ``window partitioning'' accordingly, which is a special case of window coverage:
\begin{definition}[Window Partitioning]
We say that $W_1$ is \emph{partitioned by} $W_2$, if $W_1$ is covered by $W_2$ \underline{\textbf{and}} each interval in $W_1$ is \emph{partitioned by} its covering set in $W_2$.
\end{definition}
 Figure~\ref{fig:window:coverage-partition} illustrates the difference between window partitioning and general window coverage.
Here each interval of $W_1$ is covered by two intervals of $W_2$, i.e., $M(W_1, W_2)=2$.
We now provide rigorous conditions for window partitioning: 

\begin{figure}
\centering
\subfigure[Window partitioning]{ \label{fig:window:partition}
\includegraphics[width=0.35\columnwidth]{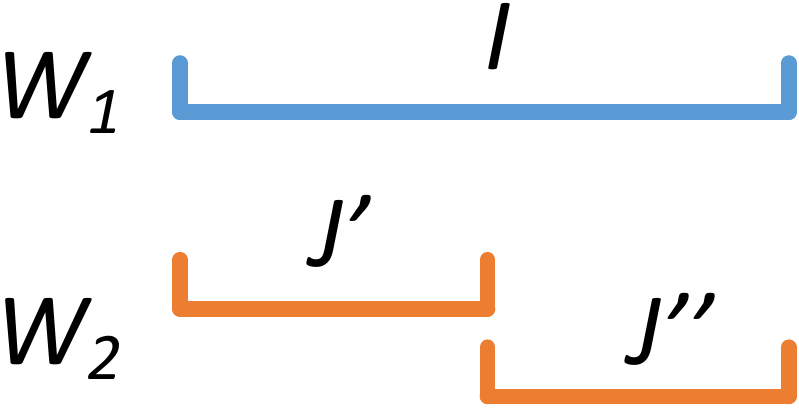}}
\hspace{0.1\columnwidth}
\subfigure[Window coverage]{ \label{fig:window:coverage}
\includegraphics[width=0.35\columnwidth]{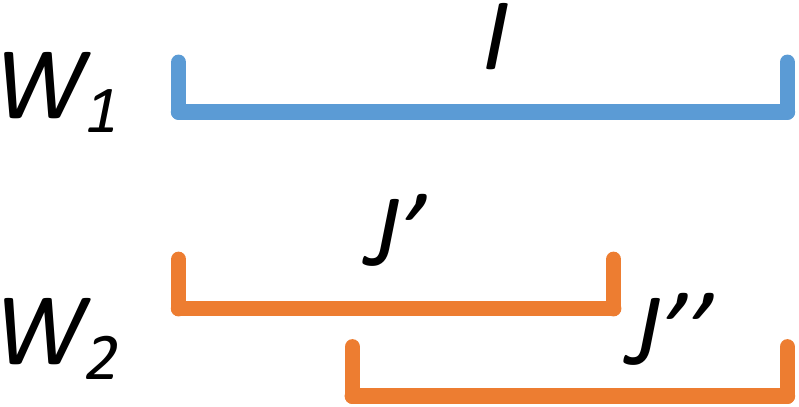}}
\vspace{-0.5em}
\caption{A comparison of window partitioning with general window coverage.}
\label{fig:window:coverage-partition}
\end{figure}


\begin{theorem}\label{theorem:window-partitioning}
$W_1$ is partitioned by $W_2$ if and only if (1) $s_1$ is a multiple of $s_2$, (2) $r_1$ is a multiple of $s_2$, and (3) $r_2=s_2$ (i.e., $W_2$ is a tumbling window).
\end{theorem}

\begin{example}[Window Partitioning]
In Example~\ref{example:window-coverage} $s_1/s_2=1$ and $r_1/s_2=5$.
So conditions (1) and (2) in Theorem~\ref{theorem:window-partitioning} hold. However, condition (3) is violated since $r_2\neq s_2$ (i.e., $W_2$ is not tumbling). As a result, $W_1$ cannot be partitioned by $W_2$.
\end{example}

\subsection{Window Coverage Graph (WCG)}

We define the \emph{window coverage graph} $\mathcal{G}=(\mathcal{W}, \mathcal{E})$ for a given window set $\mathcal{W}$ based on the partial order introduced by the window coverage relation.
For every $W_1, W_2\in\mathcal{W}$ such that $W_1\leq W_2$, we add an edge $e=(W_2, W_1)$ to the edge set $\mathcal{E}$.
The time complexity of constructing the WCG is $O(|\mathcal{W}|^2)$, given that checking the window coverage relationship takes only constant time (Theorems~\ref{theorem:window-coverage} and~\ref{theorem:window-partitioning}).

\section{Aggregates over WCG}\label{section:cost-based-eval}

We now study the problem of evaluating aggregate functions over a window set that is modeled by its WCG.
We first revisit a classic taxonomy of aggregate functions in the new context of window set and WCG.
We then present a cost-based framework for the WCG, with the goal of minimizing the overall computation cost.
We further present query rewriting techniques with respect to an optimal WCG.

\subsection{A Taxonomy of Aggregate Functions}\label{sec:shared-window:agg}

Let $f$ be a given aggregate function, e.g., \texttt{MIN}, \texttt{MAX}, \texttt{AVG}, and so on.
Gray et al. classified $f$ into three categories~\cite{GrayCBLRVPP97}:
\begin{itemize}
    \item \emph{Distributive} -- $f$ is distributive if there is some function $g$ s.t., for a table $T$, $f(T)=g(\{f(T_1), ..., f(T_n)\})$, where $\mathcal{T}=\{T_1, ..., T_n\}$ is a \emph{disjoint} partition of $T$. Typical examples include \texttt{MIN}, \texttt{MAX}, \texttt{COUNT}, and \texttt{SUM}. In fact, $f=g$ for \texttt{MIN}, \texttt{MAX}, and \texttt{SUM} but for \texttt{COUNT} $g$ should be \texttt{SUM}.
    \item \emph{Algebraic} -- $f$ is algebraic if there are functions $g$ and $h$ s.t. $f(T)=h(\{g(T_1), g(T_2),..., g(T_n)\})$. Typical examples are \texttt{AVG} and \texttt{STDEV}. For \texttt{AVG}, $g$ records the sum and count for each subset $T_i$ ($1\leq i\leq n$) and $h$ computes the average for $T_i$ by dividing the sum by the count.
    \item \emph{Holistic} -- $f$ is holistic if there is no constant bound on the size of storage needed to describe a sub-aggregate. Typical examples include \texttt{MEDIAN} and \texttt{RANK}.
\end{itemize}

Only \emph{distributive} or \emph{algebraic} aggregate functions can be computed by aggregating sub-aggregates~\cite{BensonGZMR20,TraubGCBKRM19}.
\blue{Although recent work~\cite{BensonGZMR20,TraubGCBKRM19} on window slicing ``supports'' holistic aggregate functions, the corresponding window slices contain all input events rather than sub-aggregates. Therefore, for holistic aggregate functions, we currently fall back to the default execution plan where each window is processed independently. We leave the exploration of better support for holistic aggregate functions as interesting future work.}

One important prerequisite in this taxonomy is that $\mathcal{T}=\{T_1, ..., T_n\}$ is a \emph{partition} of $T$. In our scenario, it means that if we want to evaluate $f$ over a window $W_1$ by aggregating sub-aggregates that have been computed over another window $W_2$, then $W_1$ has to be \emph{partitioned by} $W_2$.

\begin{theorem}\label{theorem:sub-aggregates}
Given that window $W_1$ is partitioned by window $W_2$, if the aggregate function $f$ is either distributive or algebraic, then $f$ over $W_1$ can be computed by aggregating sub-aggregates over $W_2$.
\end{theorem}

If $W_1$ is only covered (but not partitioned) by $W_2$, then the type of aggregate function $f$ that can be computed using Theorem~\ref{theorem:sub-aggregates} must be further restricted, such that
$f$ remains distributive or algebraic even if the $T_i$'s in $\mathcal{T}$ can overlap.
The aggregate functions \texttt{MIN} and \texttt{MAX} retain such properties, as stated by the following theorem: 

\begin{theorem}\label{theorem:distributive:min-max}
The aggregate functions \texttt{MIN} and \texttt{MAX} are distributive even if $\mathcal{T}$ is not disjoint.
\end{theorem}

\subsection{A Cost-based Optimization Framework}

Given a streaming query $Q$ that contains an aggregate function $f$ over a window set $\mathcal{W}$,
our goal is to \emph{minimize} the total computation overhead of evaluating $Q$.
A naive approach to evaluate $Q$ is to compute $f$ over each window of $\mathcal{W}$ one by one.
Clearly, this will do redundant computation if the windows in $\mathcal{W}$ ``overlap.''
To minimize computation one needs to maximize the amount of computation that is shared among overlapping windows.
We present a cost-based optimization framework that does this by exploiting the window coverage relationships
captured by the WCG of $\mathcal{W}$.

\subsubsection{Cost Modeling}
\label{sec:cost-modeling}


Let $\mathcal{W}=\{W_1, ..., W_n\}$ be a window set.
Given the WCG $\mathcal{G}=(\mathcal{W},\mathcal{E})$, we assign a \emph{weight} $c_i$ to each vertex (i.e., window) $W_i$ in $\mathcal{W}$ that represents its computation cost with respect to the (given) aggregate function $f$.
The total computation cost is simply the \emph{sum} of these weights, i.e.,
$C=\sum\nolimits_{i=1}^n c_i.$
Our goal is to minimize $C$.

We assume that the cost of computing $f$ is proportional to the number of events processed.
We further assume a steady input event rate $\eta\geq 1$.
Let
$R=\lcm(r_1, ..., r_n)$
be the \emph{least common multiple} of the ranges of the windows $W_1\langle r_1, s_1\rangle$, ..., $W_n\langle r_n, s_n\rangle$ in $\mathcal{W}$.
For each window $W_i$, \emph{the cost $c_i$ of computing $f$ over $W_i$ for events  in a period of length $R$} depends on two quantities:
\begin{itemize}[leftmargin=*]
    \item \emph{Recurrence count} $n_i$ -- the number of intervals (i.e., instances) of $W_i$ occurring during the period of $R$;
    \item \emph{Instance cost} $\mu_i$ -- the cost of evaluating an instance of $W_i$.
\end{itemize}
Clearly, $c_i=n_i\cdot\mu_i$. We next analyze the two quantities.

\begin{figure}
\centering
\includegraphics[width=0.5\columnwidth]{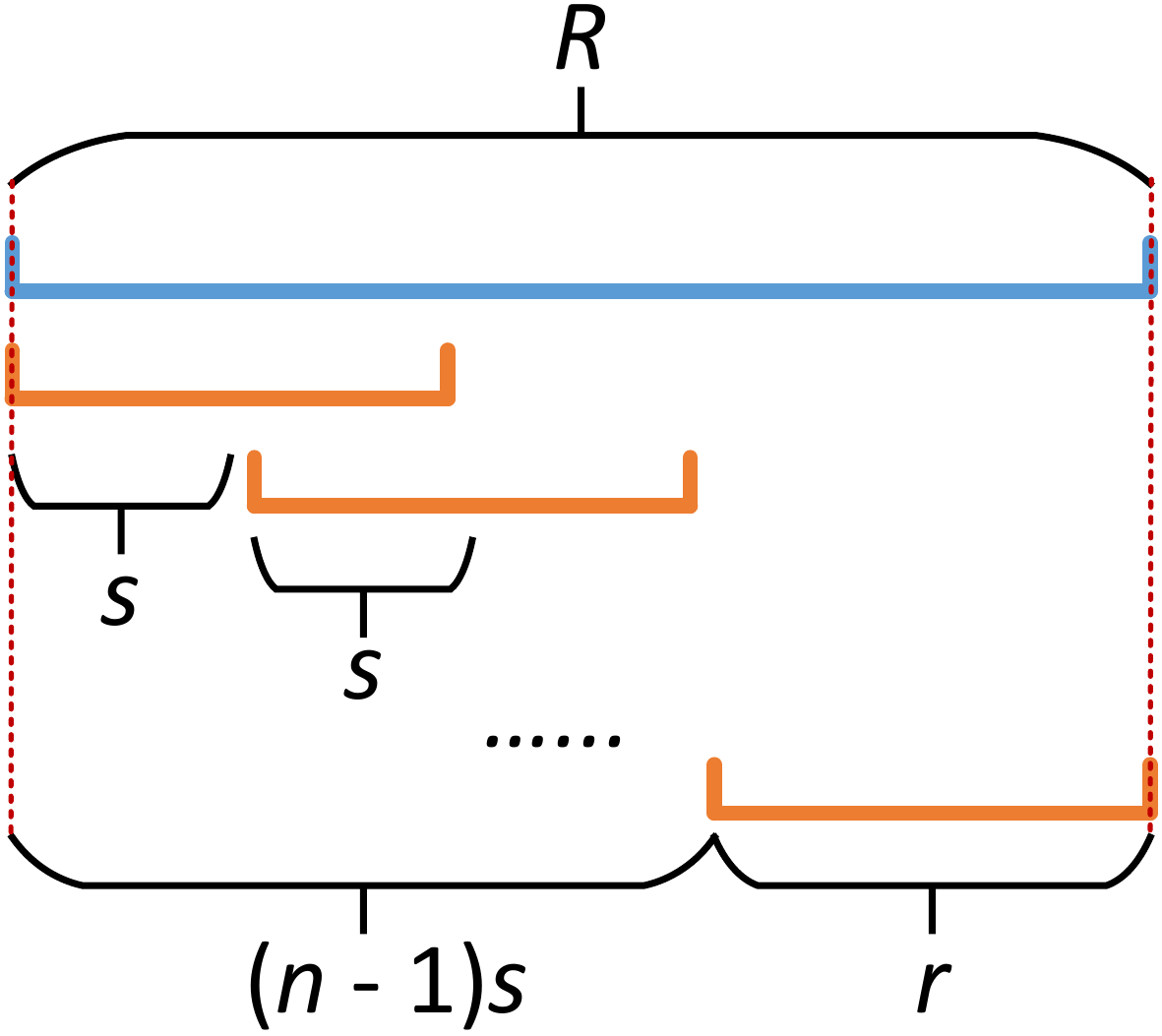}
\vspace{-1em}
\caption{Illustration of the recurrence count.}
\label{fig:recurrence-count}
\end{figure}

\paragraph*{Recurrence count}
For each window $W_i$, let $m_i=R/r_i$ be its \emph{multiplicity}.
The \emph{recurrence count} $n_i$ 
can be written as
\begin{equation}\label{eq:recurrence-count}
n_i = 1+(m_i-1)\frac{r_i}{s_i}.
\end{equation}
Figure~\ref{fig:recurrence-count} illustrates how we obtained the above formula for $n_i$.
Essentially, we have
$R=(n_i-1)\cdot s_i + r_i,$
which yields
$$n_i = 1 + \frac{R-r_i}{s_i}=1+\Big(\frac{R}{r_i} -1\Big)\frac{r_i}{s_i}=1+(m_i-1)\frac{r_i}{s_i}.$$
If $W_i$ is a tumbling window, then $n_i=m_i$.
In this paper we assume that $r_i$ is a multiple of $s_i$ so that $n_i$ is an integer.
\footnote{\blue{If we want $n_i$ to be an integer when $r_i$ is not a multiple of $s_i$, $m_i-1$ must be a multiple of $s_i$. Thus, $m_i-1= l_i\cdot s_i$ where $l_i$ is an integer, which yields $R=r_i(1+l_i\cdot s_i)$, for all $1\leq i\leq n$. Therefore, all $n_i$'s are integers only if there exist integers $l_1$, ..., $l_n$ such that $r_1(1+l_1\cdot s_1)=\cdots=r_n(1+l_n\cdot s_n)$. We leave the case when $n_i$'s may not be integers for future work.}}

\paragraph*{Instance cost}
Clearly, without any computation sharing, the instance cost of $W_i$ is
$\mu_i=\eta\cdot r_i.$
Sharing computation, however can reduce the computation cost. Consider $W_1\langle r_1, s_1\rangle$ and $W_2\langle r_2, s_2\rangle$. We have the following observation:
\begin{observation}\label{observation:1}
If $W_1$ is covered by (perhaps multiple) $W_2$'s, then the instance cost of $W_1$ can be reduced to
$$\blue{\mu_1=\min\limits_{W_2 \text{ s.t. } W_1\leq W_2}\{M(W_1, W_2)\}.}$$
\end{observation}

\subsubsection{Cost Minimization}

Algorithm~\ref{alg:cost-min} presents our procedure for finding the minimum overall cost based on the WCG, cost model, and Observation~\ref{observation:1}. It starts by constructing the WCG $\mathcal{G}$ with respect to the given window set $\mathcal{W}$ and aggregate function $f$ (line~\ref{alg:cost-min:wcg}) -- we need $f$ to know whether to use ``covered by'' or ``partitioned by'' when constructing WCG.\footnote{\HL{In our current implementation, we use ``covered by'' semantics when $f$ is \texttt{MIN} or \texttt{MAX}, and ``partitioned by'' when $f$ is \texttt{COUNT}, \texttt{SUM}, and \texttt{AVG}, which
are part of the SQL standard. Future work could expand these two lists with other aggregate functions.}}
We then process the windows one by one (lines~\ref{alg:cost-min:begin} to~\ref{alg:cost-min:cost-revise:end}).

For each window $W_i$, at line~\ref{alg:cost-min:initialization} we initialize its cost with $c_i=n_i\cdot (\eta\cdot r_i)$. (The initial cost is $c_i=m_i\cdot(\eta\cdot r_i)=\eta\cdot R$ if $W_i$ is a tumbling window.)
We then iterate over incoming edges $(W', W_i)$, revising the cost $c_i$ w.r.t. Observation~\ref{observation:1} (lines~\ref{alg:cost-min:cost-revise:begin} to~\ref{alg:cost-min:cost-revise:end}).
Finally, we remove all edges that do not correspond to the one that led to the minimum cost (lines~\ref{alg:cost-min:edge-remove:begin} to~\ref{alg:cost-min:end}).
The result is graph $\mathcal{G}_{\min}$, called the \emph{min-cost WCG} hereafter, which captures all minimum cost information.
It is the input to the query rewriting algorithm we will discuss in Section~\ref{sec:implementation:query-rewrite}.

\begin{algorithm}
\small
  \SetAlgoLined
  \KwIn{$\mathcal{W}=\{W_i\}_{i=1}^n$, a window set; $f$, aggregate function.}
  \KwOut{$\mathcal{G}_{\min}$, the min-cost WCG w.r.t. $\mathcal{W}$ and $f$.}
  \SetAlgoLined
  Construct the WCG $\mathcal{G}=(\mathcal{W}, \mathcal{E})$ \HL{w.r.t. ``covered by'' or ``partitioned by'' as determined by $f$}\;\label{alg:cost-min:wcg}
  \ForEach{$W_i\in\mathcal{W}$}{\label{alg:cost-min:begin}
    Initialize its cost $c_i\leftarrow n_i\cdot (\eta\cdot r_i)$\;\label{alg:cost-min:initialization}
    \ForEach{$W'\in\mathcal{W}$ s.t. $(W', W_i)\in\mathcal{E}$}{\label{alg:cost-min:cost-revise:begin}
        Revise cost $c_i\leftarrow\min\{c_i, n_i\cdot M(W_i, W')\}$\;\label{alg:cost-min:cost-revise:end}
    }
  }
  \ForEach{$W_i\in\mathcal{W}$}{\label{alg:cost-min:edge-remove:begin}
    Remove all incoming edges that do not correspond to (the final value of) $c_i$\;\label{alg:cost-min:end}
  }
  \Return{the result graph $\mathcal{G}_{\min}$}\;
  \caption{Find the min-cost WCG.}
\label{alg:cost-min}
\end{algorithm}

\begin{figure}[t]
\centering
\subfigure[Initial WCG]{ \label{fig:running-example:wcg}
\includegraphics[width=0.45\columnwidth]{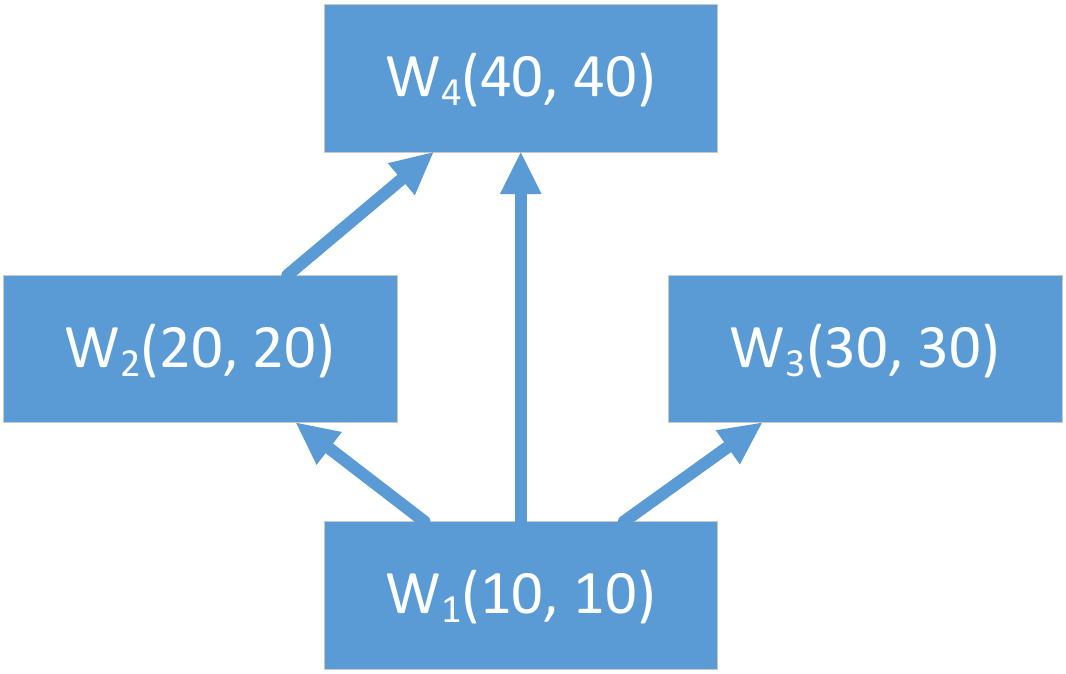}}
\subfigure[Min-cost WCG]{ \label{fig:running-example:min-cost-wcg}
\includegraphics[width=0.45\columnwidth]{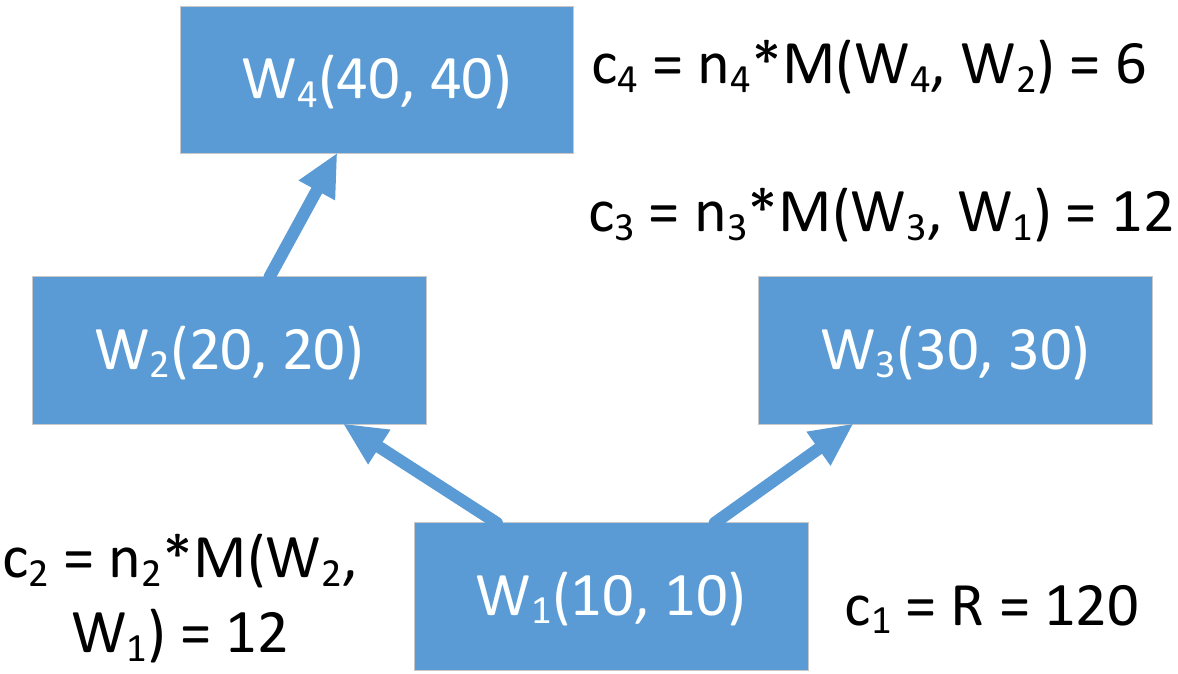}}
\vspace{-0.5em}
\caption{WCG and min-cost WCG for Example~\ref{example:running-query:min-cost-wcg}.}
\label{fig:running-example:wcgs}
\end{figure}

\vspace{-1.5em}
\begin{example}\label{example:running-query:min-cost-wcg}
Consider a query that contains four tumbling windows: $W_1\langle 10, 10\rangle$, $W_2\langle 20, 20\rangle$, $W_3\langle 30, 30\rangle$, and $W_4\langle 40, 40 \rangle$. \HL{It does not matter which aggregate function $f$ we choose here, since ``covered by'' and ``partitioned by'' semantics coincide when all windows in a window set are tumbling windows.}

Assuming an incoming event ingestion rate $\eta = 1$, the total cost of computing the four windows is $C=4\eta R=4R= 480$, where $R=\lcm\{10, 20, 30, 40\}=120$.

Figure~\ref{fig:running-example:wcgs} shows the initial WCG (Figure~\ref{fig:running-example:wcg}) and the final min-cost WCG (Figure~\ref{fig:running-example:min-cost-wcg}) by running Algorithm~\ref{alg:cost-min}, when exploiting the overlaps between the windows.
The total cost is therefore reduced to $C'=c'_1 + c'_2 + c'_3 + c'_4=120 + 12 + 12 + 6 = 150$,
a 62.5\% reduction from the initial cost $C=480$.
\end{example}

\vspace{-0.5em}
\paragraph*{\blue{Limitations}}
\blue{Since our cost-based optimization framework exploits the coverage relationships between windows, it cannot improve the execution plan if such opportunities are not present. For example, consider a set of tumbling windows where all ranges are ``mutually prime,'' e.g., $W_1(15, 15)$, $W_2(17, 17)$, and $W_3(19,19)$. In such cases, our cost model cannot lead to plans that improve over the default plan where each window is evaluated independently.}

\subsection{Query Rewriting}
\label{sec:implementation:query-rewrite}

To leverage the benefits of shared window computation, we rewrite the original ASA query plan with respect to the min-cost WCG $\mathcal{G}_{\min}$ based on the following observation:
\begin{theorem}
$\mathcal{G}_{\min}$ is a forest, i.e., a collection of trees.
\end{theorem}
The proof follows directly from noticing that each window in $G_{\min}$ has at most one incoming edge (due to lines 6 to 7).

Figure~\ref{fig:window-set:query-rewrite} shows how we revise the query execution plan in Example~\ref{example:multi-window-single-agg}.
Figure~\ref{fig:window-set:plan-rewrite} presents the original plan and the revised plan based on the min-cost WCG.
Figure~\ref{fig:window-set:trill-expr} presents the translated Trill expression~\cite{ChandramouliGBDPTW14}.
\blue{The appendix includes a formal description of this query rewriting procedure. Translation to query plans expressed by other streaming API's, such as the Apache Flink DataStream API~\cite{flink-datastream-api}, is similar.}


\vspace{-0.5em}
\section{Factor Windows}\label{section:factor-window}

We have been confining our discussion to sharing computation over windows in the given window set.
One can add auxiliary windows that are not in the window set but may nevertheless help reduce the overall computation cost.
We call them \emph{factor windows}.

\begin{definition}
Given a window set $\mathcal{W}$, a window $W$ is called a \emph{factor window} with respect to $\mathcal{W}$ if $W\not\in\mathcal{W}$ and there exists some window $W'\in\mathcal{W}$ such that $W'\leq W$.
\end{definition}
Note that we do not expose the results of factor windows to users, as they are not part of the user query.

\begin{figure}
\centering
\subfigure[Initial WCG]{ \label{fig:running-example:min-cost-wcg-no-fw}
\includegraphics[width=0.45\columnwidth]{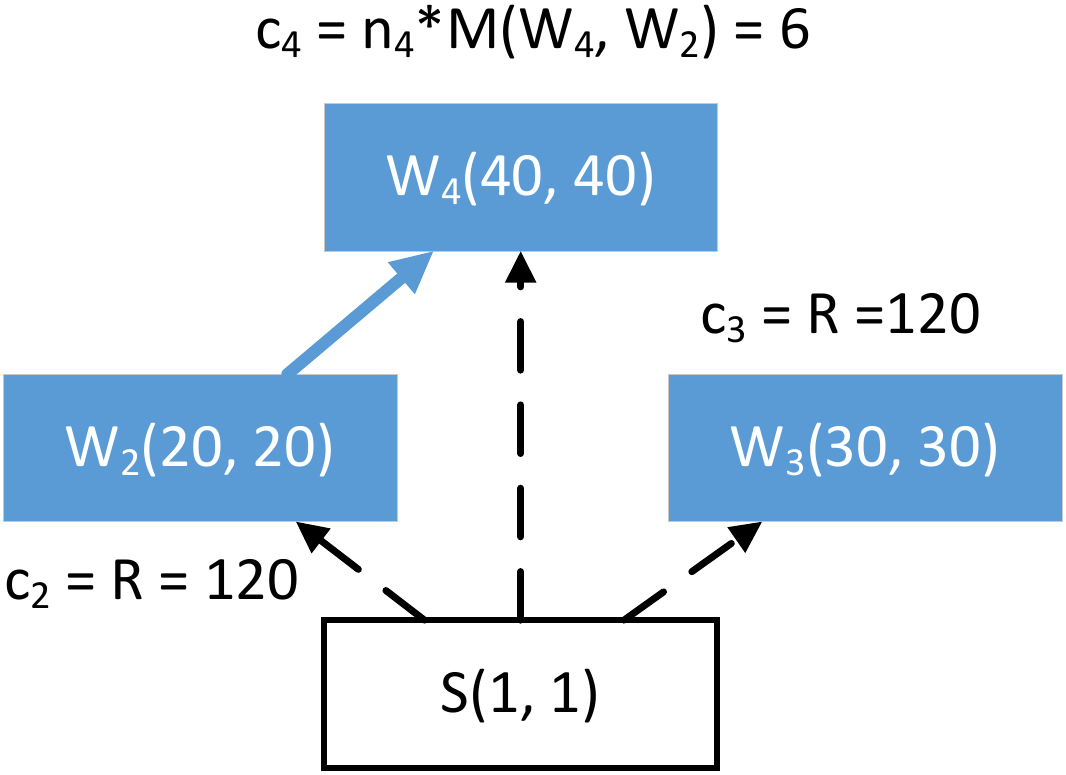}}
\subfigure[Min-cost WCG]{ \label{fig:running-example:min-cost-wcg-fw}
\includegraphics[width=0.45\columnwidth]{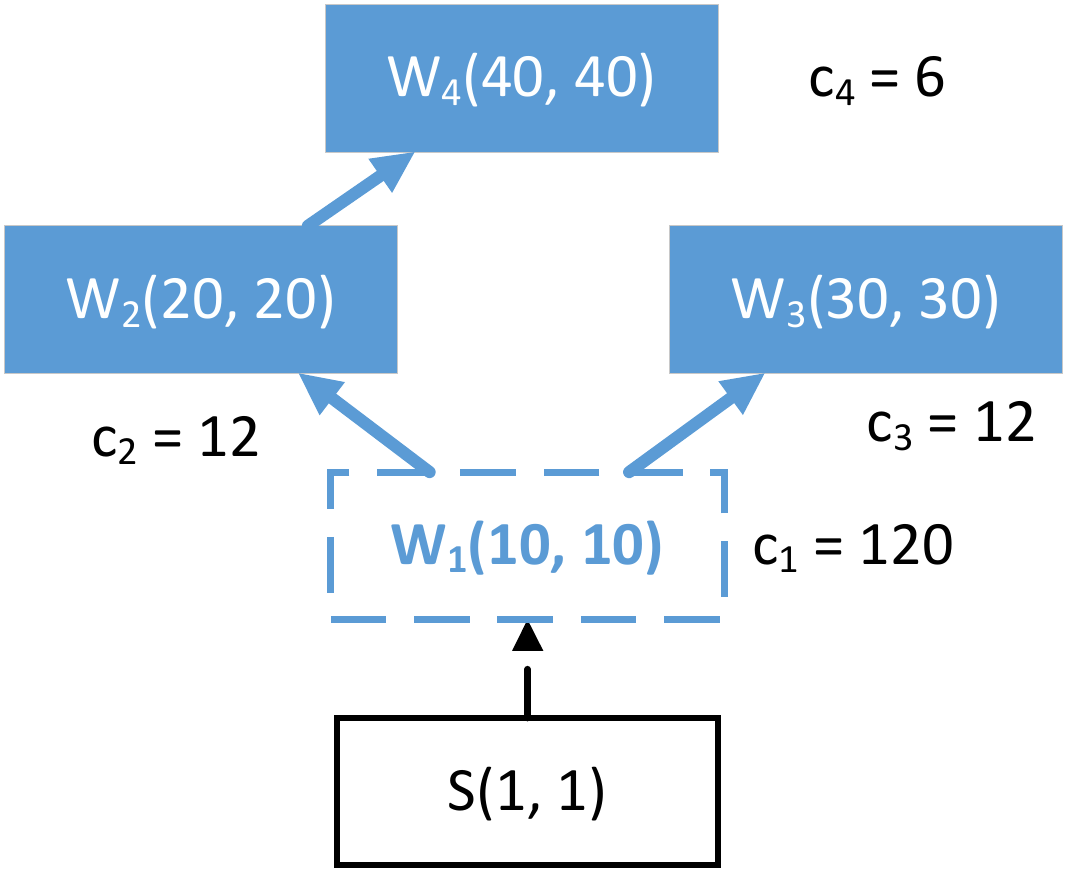}}
\vspace{-0.5em}
\caption{Min-cost WCGs for Example~\ref{example:multi-window-single-agg} with and without using factor windows.}
\label{fig:running-example:min-cost-wcgs}
\end{figure}

\begin{example}\label{example:factor-window}
Suppose we modify the query  in Example~\ref{example:running-query:min-cost-wcg} by removing the tumbling window $W_1(10, 10)$.
The resulting query $Q$ contains three tumbling windows $W_2(20,20)$, $W_3(30,30)$, and $W_4(40, 40)$.
The cost of directly computing them is
$C=3R=360,$
as here $R=\lcm\{20,30,40\}=120$ remains the same.

If we apply Algorithm~\ref{alg:cost-min} over $Q$, we get the min-cost WCG presented in Figure~\ref{fig:running-example:min-cost-wcg-no-fw}.
As a result, the overall cost is $C'=c'_2 + c'_3 + c'_4=120 + 120 + 6 = 246$,
a reduction of 31.7\% from the baseline cost $C=360$.

If we allow factor windows and apply Algorithm~\ref{alg:cost-min:factor-window:general} over $Q$, then we get the min-cost WCG  in Figure~\ref{fig:running-example:min-cost-wcg-fw}.
Window $W_1(10, 10)$ is ``added back'' as a factor window, which participates in evaluating $Q$ but does not expose its result to users.
As in Example~\ref{example:running-query:min-cost-wcg}, the overall cost now is
$C''=150$, which is 58.3\% less than the baseline cost $C=360$ and 39\% less than the cost $C'=246$ without using factor windows.
\end{example}

\vspace{-0.5em}
\subsection{Impact of Factor Window}
\label{section:factor-window:impact}

One natural question to ask is: When does a factor window help?
In the following, we provide a formal analysis.

\paragraph*{Augmented WCG}
For the WCG $\mathcal{G}=(\mathcal{W}, \mathcal{E})$ induced by the given window set $\mathcal{W}$ and aggregate function $f$, we add a virtual tumbling window $S\langle r=1, s=1\rangle$ into $\mathcal{W}$, and add an edge $(S, W)$ into $\mathcal{E}$ for each $W\in\mathcal{W}$ that has no incoming edges (i.e., $W$ is not covered by any other window).
However, if such an $S$ already exists in $\mathcal{W}$, we do not add another one.
Intuitively, $S$ represents a window consisting of \emph{atomic} intervals that emit an aggregate for each time unit; therefore $S$ covers all windows in $\mathcal{W}$.
The computation cost of $S$ is always $\eta\cdot R$, as it cannot be covered by any other window.
This augmented graph is  a directed acyclic graph (DAG) with a single ``root'' $S$.
From now on, when we refer to the WCG we mean its augmented version.

\begin{figure}
\centering
\subfigure[Interesting]{ \label{fig:wcg:patterns:interesting}
\includegraphics[width=0.4\columnwidth]{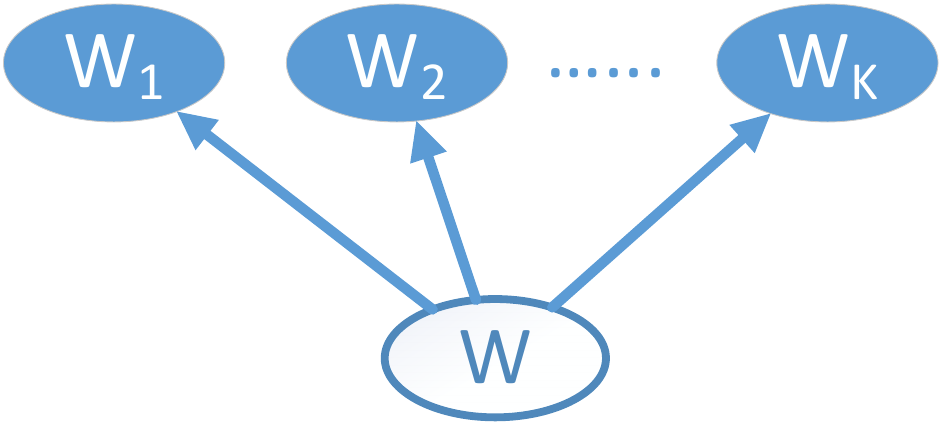}}
\hspace{0.1\columnwidth}
\subfigure[Uninteresting]{ \label{fig:wcg:patterns:uninteresting}
\includegraphics[width=0.4\columnwidth]{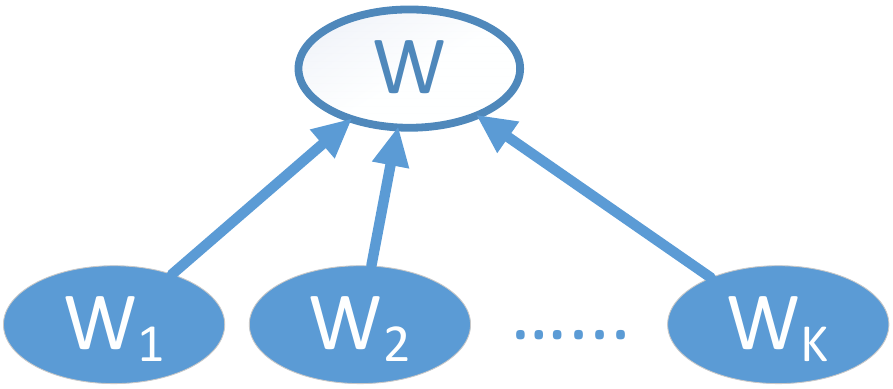}}
\vspace{-0.5em}
\caption{Two basic patterns in WCG ($K\geq 1$).}
\label{fig:wcg:patterns}
\end{figure}

\paragraph*{Two Basic Patterns}
Figure~\ref{fig:wcg:patterns} presents two basic patterns in (the augmented) WCG, for an arbitrary window $W\in\mathcal{W}$.
We are interested in the pattern in Figure~\ref{fig:wcg:patterns:interesting} but not the one in Figure~\ref{fig:wcg:patterns:uninteresting}, as $W$ can only affect the costs of its \emph{downstream} windows.
This eliminates windows in WCG without outgoing edges from consideration.

\begin{figure}
\centering
    \includegraphics[width=0.9\columnwidth]{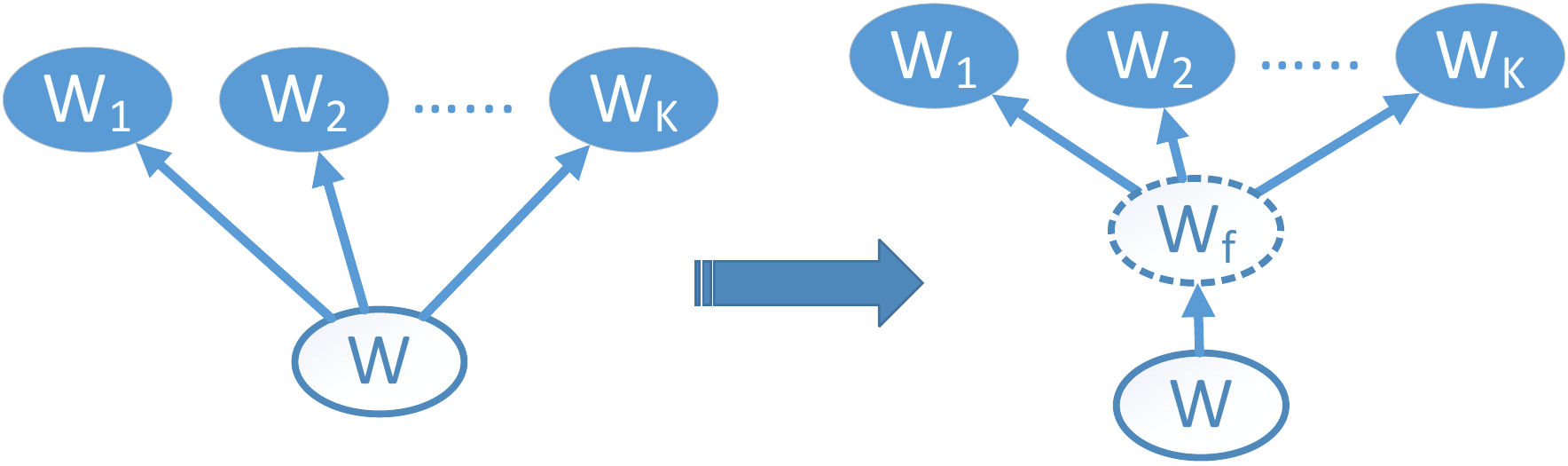}
\vspace{-1em}
\caption{Impact of factor window $W_f$.}
\label{fig:factor-window}
\vspace{-1em}
\end{figure}

\paragraph*{Analysis of Impact}
As shown in Figure~\ref{fig:factor-window}, let $W_f$ be a factor window inserted ``between'' $W$ and its downstream windows \blue{$W_1$, ..., $W_K$.}
We can do this for all ``intermediate'' vertices, i.e., windows with \emph{both} incoming and outgoing edges, in (the augmented) WCG, thanks to the virtual ``root'' $S$.
Clearly, $W_f\leq W$, and $W_j\leq W_f$ for $1\leq j\leq K$.
We now compare the overall computation costs with and without inserting $W_f$.
The cost with the factor window $W_f$ is
$c=\sum\nolimits_{j=1}^{K}\cost(W_j) + \cost(W_f) + \cost(W).$
On the other hand, the cost without $W_f$ is
$c'=\sum\nolimits_{j=1}^{K}\cost'(W_j) + \cost(W).$
\blue{We define the \emph{benefit} of $W_f$ as $\delta_f=c'-c$.}

Since $\cost(W_j)=n_j \cdot M(W_j, W_f)$, $\cost(W_f)=n_f \cdot M(W_f, W)$, and
$\cost'(W_j)=n_j\cdot M(W_j, W)$, it follows that
\begin{equation*}
\resizebox{\hsize}{!}{$
\blue{\delta_f = \sum\nolimits_{j=1}^{K} n_j \Big(M(W_j, W)-M(W_j, W_f)\Big) - n_f M(W_f, W).}
$}
\end{equation*}
By Theorem~\ref{theorem:covering-multiplier},
$M(W_j, W_f)=1+(r_j-r_f)/s_f$, $M(W_j, W)=1+(r_j - r_W)/s_W$, and
$M(W_f, W)=1+(r_f-r_W)/s_W$.
Substituting into the above equation, we obtain
\begin{equation*}
\blue{\delta_f = \sum\nolimits_{j=1}^{K} n_j\Big(\frac{r_j - r_W}{s_W}-\frac{r_j-r_f}{s_f}\Big) - n_f\Big(1+\frac{r_f-r_W}{s_W}\Big).}
\end{equation*}
We now define the following quantities to simplify notation:
(1) $\rho_j=r_j/r_f$ and $k_j=r_j/s_j$, for $1\leq j\leq K$;
(2) $k_f=r_f/s_f$; and (3) $k_W=r_W/s_W$.
With this notation, we have
\begin{equation}\label{eq:cost:reduction:general}
\resizebox{\hsize}{!}{$
\blue{\delta_f=n_f\Big(\sum\nolimits_{j=1}^{K}\frac{n_j}{n_f}\Big(k_f-\frac{r_j}{s_f}+\frac{r_j}{s_W}-k_W\Big)-(1+\frac{r_f}{s_W}-k_W)\Big).}
$}
\end{equation}
Inserting $W_f$ improves if and only if $\delta_f\geq 0$,  
i.e.,
\begin{equation}\label{eq:general-condition}
\blue{\sum\nolimits_{j=1}^{K}\frac{n_j}{n_f}\Big(k_f-\frac{r_j}{s_f}+\frac{r_j}{s_W}-k_W\Big) \geq 1+\frac{r_f}{s_W}-k_W.}
\end{equation}

\vspace{-0.5em}
\subsection{Candidate Generation and Selection}
\label{section:factor-window:candidate}

We can use Equation~\ref{eq:general-condition} to determine whether a factor window is beneficial.
The next problem is to find candidate factor windows that are beneficial, from which
we can select the best one.
\blue{Algorithm~\ref{alg:FindBestFactorWindow:covered-by} illustrates this candidate generation and selection procedure in detail.}

\begin{algorithm}[t]
\small
  \SetAlgoLined
  \KwIn{\blue{$W$, a window; $\{W_1, ..., W_K\}$, $W$'s downstream windows (ref. Figure~\ref{fig:factor-window}).}}
  \KwOut{\blue{The best factor window $W_f$ w.r.t. $W$ and $\{W_1, ..., W_K\}$.}}
  \SetAlgoLined

  \blue{// Construct the set $\mathcal{W}_f$ of candidate factor windows.}\label{alg:candidate-generation:begin}\\
  $\blue{\mathcal{W}_f\leftarrow\emptyset}$\;
  $\blue{s_d\leftarrow\gcd\{s_1, ..., s_K\}}$\;\label{alg:candidate-generation:s-gcd}
  $\blue{\mathcal{S}_f\leftarrow\{s_f: s_d \text{ mod } s_f = 0 \text{ and } s_f \text{ mod } s_W = 0\}}$\;\label{alg:candidate-generation:S_f}
  $\blue{r_{\text{min}}\leftarrow\min\{r_1, ..., r_K\}}$\;\label{alg:candidate-generation:r_min}
  \ForEach{$\blue{s_f\in\mathcal{S}_f}$}{
    $\blue{\mathcal{R}_f\leftarrow\{r_f: r_f \text{ mod } s_f = 0 \text{ and } r_f\leq r_{\text{min}}\}}$\;\label{alg:candidate-generation:R_f}
    \ForEach{$\blue{r_f\in \mathcal{R}_f}$}{
        \blue{Construct a candidate factor window $W_f\langle r_f, s_f\rangle$}\;
        \If{$\blue{W_f\leq W \text{ and } W_j\leq W_f \text{ for } 1\leq j\leq K}$}{\label{alg:candidate-generation:coverage}
            $\blue{\mathcal{W}_f\leftarrow \mathcal{W}_f\cup \{W_f\}}$\;\label{alg:candidate-generation:end}
        }
    }
  }
  \blue{// Find the best factor window from $\mathcal{W}_f$.}\label{alg:candidate-selection:begin}\\
  $\blue{\delta_f^{\max}\leftarrow 0}$, $\blue{W_f^{\max}\leftarrow\texttt{null}}$\;
  \ForEach{$\blue{W_f\in \mathcal{W}_f}$}{
    \blue{Compute the benefit $\delta_f$ of $W_f$ using Equation~\ref{eq:cost:reduction:general}}\;
    \If{$\blue{\delta_f \geq 0 \text{ and } \delta_f > \delta_f^{\max}}$}{
        $\blue{\delta_f^{\max}\leftarrow\delta_f}$, $\blue{W_f^{\max}\leftarrow W_f}$\; \label{alg:candidate-selection:end}
    }
  }
  \Return{$\blue{W_f^{\max}}$\;}
  \caption{\blue{Find the best factor window under ``covered by'' semantics.}}
\label{alg:FindBestFactorWindow:covered-by}
\end{algorithm}

\subsubsection{Candidate Generation}
\blue{It looks for \emph{eligible slides} $s_f$ and \emph{eligible ranges} $r_f$ as follows (lines~\ref{alg:candidate-generation:begin} to~\ref{alg:candidate-generation:end} of Algorithm~\ref{alg:FindBestFactorWindow:covered-by}):}
\begin{itemize}[leftmargin=*]
    \item \blue{\textit{Eligible slides}: Let $s_d=\gcd\{s_1, ..., s_K\}$. The set of eligible slides is
    $\mathcal{S}_f=\{s_f: s_d \text{ mod } s_f = 0 \text{ and } s_f \text{ mod } s_W = 0\}.$
    That is, $s_f$ must be a factor of $s_d$ and a multiple of $s_W$.}
    \item \blue{\textit{Eligible ranges}: Let $r_{\text{min}}=\min\{r_1, ..., r_K\}$. For each $s_f\in\mathcal{S}_f$, the set of eligible ranges is $\mathcal{R}_f=\{r_f: r_f \text{ mod } s_f = 0 \text{ and } r_f\leq r_{\text{min}}\}$, i.e., $r_f\leq r_{\text{min}}$ must be a multiple of $s_f$.}
\end{itemize}
For each eligible pair $(s_f, r_f)$, we construct a candidate factor window $W_f\langle r_f, s_f\rangle$ and further check the window coverage constraints in Figure~\ref{fig:factor-window}, i.e., $W_f\leq W$ and \blue{$W_j\leq W_f$ for $1\leq j\leq K$ (line~\ref{alg:candidate-generation:coverage}), to only keep valid candidates.}


\subsubsection{Candidate Selection}
Many candidate factor windows in $\mathcal{W}_f$ may be beneficial (i.e., Equation~\ref{eq:general-condition} holds).
Only the one that leads to the maximum cost reduction (i.e., benefit) should be added.
We thus compute the benefits of the candidates (by Equation~\ref{eq:cost:reduction:general}) and select the one with the maximum benefit \blue{(lines~\ref{alg:candidate-selection:begin} to~\ref{alg:candidate-selection:end} of Algorithm~\ref{alg:FindBestFactorWindow:covered-by})}.

\subsubsection{\blue{Time Complexity Analysis of Algorithm~\ref{alg:FindBestFactorWindow:covered-by}}}\label{sec:factor-window:find-best-fw:analysis}
\blue{Computing $s_d$ at line~\ref{alg:candidate-generation:s-gcd} takes $O(s_{\max}\log s_{\max})$ time using Euclid's algorithm~\cite{clrs}, where $s_{\max}=\max\{s_1, ..., s_K\}$.
Finding all eligible slides at line~\ref{alg:candidate-generation:S_f} takes $O(\lceil\frac{s_d}{s_W}\rceil)$ time.
Computing $r_{\min}$ at line~\ref{alg:candidate-generation:r_min} takes $O(K)$ time.
For each $s_f\in\mathcal{S}_f$, finding its eligible ranges at line~\ref{alg:candidate-generation:R_f} takes $O(\lceil\frac{r_{\min}}{s_f}\rceil)$ time.
For each $W_f\langle r_f, s_f\rangle$, it takes $O(K)$ time to check all related window coverage relationships at line~\ref{alg:candidate-generation:coverage}.
Hence, the candidate generation stage (lines~\ref{alg:candidate-generation:begin} to~\ref{alg:candidate-generation:end}) takes $O(s_{\max}\log s_{\max} + \lceil\frac{s_d}{s_W}\rceil + K + |\mathcal{S}_f|\cdot |\mathcal{R}_f|\cdot K)$ time.
To simplify our analysis, we assume it is dominated by $O(|\mathcal{S}_f|\cdot |\mathcal{R}_f|\cdot K)$.
Now consider the candidate selection stage (lines~\ref{alg:candidate-selection:begin} to~\ref{alg:candidate-selection:end}). Since we check Equation~\ref{eq:cost:reduction:general} once for each $W_f$, it takes $O(|\mathcal{S}_f|\cdot |\mathcal{R}_f|\cdot K)$ time in total.
Since $|\mathcal{S}_f|=O(\lceil\frac{s_d}{s_W}\rceil)$ and $|\mathcal{R}_f|=O(\lceil\frac{r_{\min}}{s_W}\rceil)$, it follows that the time complexity of Algorithm~\ref{alg:FindBestFactorWindow:covered-by} is $O(\lceil\frac{s_d}{s_W}\rceil\cdot\lceil\frac{r_{\min}}{s_W}\rceil\cdot K)$.}


\subsection{Putting Things Together}

Algorithm~\ref{alg:cost-min:factor-window:general} is the revised version of Algorithm~\ref{alg:cost-min} that returns the min-cost WCG when factor windows are allowed.
\blue{It first extends the original WCG by adding the best factor windows, found by Algorithm~\ref{alg:FindBestFactorWindow:covered-by}, for existing windows (lines~\ref{alg:factor-window:expand:begin} to~\ref{alg:factor-window:expand:end}).}
It then simply invokes Algorithm~\ref{alg:cost-min} on the extended WCG (rather than the original one) to find the new min-cost WCG that contains factor windows \blue{(line~\ref{alg:factor-window:min-wcg}).}

\begin{algorithm}[t]
\small
  \SetAlgoLined
  \KwIn{$\mathcal{W}=\{W_i\}_{i=1}^n$, a window set; $f$, aggregate function.}
  \KwOut{$\mathcal{G}_{\min}$, the min-cost WCG w.r.t. $\mathcal{W}$ and $f$, where factor windows are allowed.}
  \SetAlgoLined
  Construct the WCG $\mathcal{G}=(\mathcal{W}, \mathcal{E})$ \HL{w.r.t. ``covered by'' or ``partitioned by'' determined by $f$}\;\label{alg:factor-window:wcg}
  \ForEach{$W\in\mathcal{W}$}{\label{alg:factor-window:expand:begin}
    $W_f\leftarrow$ \underline{\emph{FindBestFactorWindow}}($W$, $W$'s downstream windows $\{W_1, ..., W_K\}$) \blue{using Algorithm~\ref{alg:FindBestFactorWindow:covered-by}}\;
    Expand $\mathcal{G}$ by adding $W_f$ and the corresponding edges (as shown in Figure~\ref{fig:factor-window})\;\label{alg:factor-window:expand:end}
  }
  $\mathcal{G}_{\min}\leftarrow$ Run lines~\ref{alg:cost-min:begin}-\ref{alg:cost-min:end} of Algorithm~\ref{alg:cost-min} over the expanded $\mathcal{G}$\;\label{alg:factor-window:min-wcg}
  \Return{the result graph $\mathcal{G}_{\min}$}\;
  \caption{Find the min-cost WCG when factor windows are allowed.}
\label{alg:cost-min:factor-window:general}
\end{algorithm}

Unlike Algorithm~\ref{alg:cost-min}, Algorithm~\ref{alg:cost-min:factor-window:general} is no longer optimal.
In fact, the cost minimization problem when factor windows are allowed is an instance of the Steiner tree problem~\cite{Karp72}, which is NP-hard.
Various approximate algorithms have been proposed for Steiner trees (e.g.,~\cite{ByrkaGRS10,RobinsZ00}), but we choose to stay with Algorithm~\ref{alg:cost-min:factor-window:general} because it is simple and easy to implement.
\blue{It would be interesting future work to characterize the gap between the factor windows found by Algorithm~\ref{alg:cost-min:factor-window:general} and the ones that could be found by an optimal solution.}\footnote{\blue{Note that we restricted ourselves to consider only a subset of all possible factor windows.
For example, for the WCG in Figure~\ref{fig:running-example:min-cost-wcg-no-fw}, our approach would not consider the factor window $W(15,15)$, as $\text{gcd}\{20,30,40\}=10 < 15$ (ref. line~\ref{alg:fw:tumbling:cand-gen:enum:begin} of Algorithm~\ref{alg:factor-window:partition-by:multiple}).
An ideal, optimal solution would have also considered such candidates. In fact, it needs to generate all valid candidate factor windows, instead of finding a ``locally optimal'' factor window for each input window (as Algorithm~\ref{alg:cost-min:factor-window:general} does), insert them into the WCG, and then solve the Steiner tree problem. Since the problem is NP-hard, the time complexity in the worst case would be exponential w.r.t. the size of the WCG.}}
\blue{However, even though the factor windows found by Algorithm~\ref{alg:cost-min:factor-window:general} may not be the optimal ones, the min-cost WCG with factor windows improves over the min-cost WCG without factor windows (returned by Algorithm~\ref{alg:cost-min}), since Algorithm~\ref{alg:cost-min:factor-window:general} only inserts a factor window if it is beneficial (lines~\ref{alg:factor-window:expand:begin} to~\ref{alg:factor-window:expand:end}).}

\paragraph*{\blue{Time Complexity Analysis of Algorithm~\ref{alg:cost-min:factor-window:general}}}
\blue{Construction of the WCG requires $O(|\mathcal{W}|^2)$ time as it needs to check each pair of windows to test their coverage relationship. For a given window $W\in\mathcal{W}$ and its downstream windows $W_1$, ..., $W_K$, 
it takes 
$O(\lceil\frac{s_d}{s_W}\rceil\cdot\lceil\frac{r_{\min}}{s_W}\rceil\cdot K)$ time to find its best factor window $W_f$ using Algorithm~\ref{alg:FindBestFactorWindow:covered-by}. Meanwhile, adding $W_f$ and the corresponding edges requires $O(K)$ time. Furthermore, running lines~\ref{alg:cost-min:begin} to~\ref{alg:cost-min:end} of Algorithm~\ref{alg:cost-min} over the expanded graph takes $O((2\cdot|\mathcal{W}|)^2)$ time.
Thus, the time complexity of Algorithm~\ref{alg:cost-min:factor-window:general} is $O(5|\mathcal{W}|^2 + |\mathcal{W}|\cdot M_{\mathcal{W}})$, where $M_{\mathcal{W}}=\max_{W\in\mathcal{W}}\{\lceil\frac{s_d}{s_W}\rceil\cdot\lceil\frac{r_{\min}}{s_W}\rceil\cdot K\}$.}
\subsection{The Case of ``Partitioned By''}
\label{section:factor-window:partition-by}

We can improve the procedure \textbf{FindBestFactorWindow} 
in Algorithm~\ref{alg:FindBestFactorWindow:covered-by}
if we restrict the window coverage relationships to ``partitioned by'' semantics, which works for more types of aggregate functions.
In this special case, the candidate factor windows are restricted to \emph{tumbling windows} (by Theorem~\ref{theorem:window-partitioning}).



\begin{algorithm}
\small
  \SetAlgoLined
  \KwIn{$W_f$, a factor window; $W$, a target window with downstream windows $W_1$, ..., $W_K$; $\lambda$, by Equation~\ref{eq:lambda}.}
  \KwOut{Return \texttt{true} if adding $W_f$ improves the overall cost, \texttt{false} otherwise.}
  \SetAlgoLined
  \If{$K\geq 2$}{\label{alg:factor-window-partition-by:K2:begin}
    \Return{\texttt{true}}\; \label{alg:factor-window-partition-by:K2:end}
  }
  // We have $K=1$ hereafter.\\
  \If{$k_1=1$}{\label{alg:factor-window-partition-by:K1:k_1_1:begin}
    \Return{\texttt{false}}\;\label{alg:factor-window-partition-by:K1:k_1_1:end}
  }\Else {\label{alg:factor-window-partition-by:K1:k_1_larger_than_1:begin}
    // We have $k_1>1$ hereafter.\\
    \If{$k_1\geq 3$ and $m_1\geq 3$}{
        \Return{\texttt{true}}\;
    }\Else{
        Compute $\frac{r_f}{r_W}$ and $\frac{\lambda}{\lambda - 1}=1+\frac{m_1}{(m_1-1)(k_1-1)}$\;
        \Return{\texttt{true} if $\frac{r_f}{r_W}\geq \frac{\lambda}{\lambda - 1}$, \texttt{false} otherwise}\;\label{alg:factor-window-partition-by:K1:k_1_larger_than_1:end}
    }
  }
  \caption{Determine whether a factor window would be beneficial under ``partitioned by'' semantics.}
\label{alg:factor-window-partition-by}
\end{algorithm}

\vspace{-0.5em}
\subsubsection{Revisit of Impact of Factor Windows}
We first revisit the problem of determining whether a factor window is beneficial, under ``partitioned by'' semantics.
Algorithm~\ref{alg:factor-window-partition-by} summarizes the procedure that determines whether a factor window $W_f$ would help in the case of ``partitioned by.''
Here, 
$\lambda$ is defined as
\begin{equation}\label{eq:lambda}
\lambda= \sum\nolimits_{j=1}^{K}\frac{n_j}{m_j}.
\end{equation}

The procedure in Algorithm~\ref{alg:factor-window-partition-by} looks complicated. We offer some intuition below to help understand it:

\vspace{0.5em}
\noindent{\blue{\bf (Case 1)}}
\blue{If $W_f$ has two or more downstream windows (i.e., when $K\geq 2$), then it improves the overall cost (lines~\ref{alg:factor-window-partition-by:K2:begin} to~\ref{alg:factor-window-partition-by:K2:end})}, since now at least one downstream window would benefit from reading sub-aggregates from $W_f$ (rather than from $W$). \HL{We provide more explanation using a special case (referring to Figure~\ref{fig:factor-window}) when $K=2$ and all windows are tumbling. We can simplify Equation~\ref{eq:cost:reduction:general} by noticing $k_f=k_W=1$, $r_f=s_f$, and $r_W=s_W$, since both $W_f$ and $W$ are tumbling windows:
\blue{$\delta_f=\sum\nolimits_{j=1}^{2}n_j\cdot\Big(\frac{r_j}{r_W}-\frac{r_j}{r_f}\Big)-n_f\cdot\frac{r_f}{r_W}.$}
Moreover, since all windows are tumbling, $n_j=m_j=R/r_j$ for $j\in\{1,2\}$, and $n_f=m_f=R/r_f$. As a result,
\blue{$\delta_f=R\cdot\Big(\frac{1}{r_W}-\frac{2}{r_f}\Big)\geq 0,$}
since $r_f\geq 2r_W$ by Theorem~\ref{theorem:window-partitioning}.}

\vspace{0.5em}
\noindent{\blue{\bf (Case 2)}}
\blue{If $W_f$ only has one downstream window $W_1$ that is tumbling (i.e., the case when $K=1$ and $k_1=1$), then it cannot reduce the overall cost (lines~\ref{alg:factor-window-partition-by:K1:k_1_1:begin} to~\ref{alg:factor-window-partition-by:K1:k_1_1:end})} because one now needs to use all sub-aggregates from $W$ to compute $W_f$ itself. Without $W_f$ one can use the same sub-aggregates to compute $W_1$ directly.
\blue{The case when $W_f$ has one unique downstream window $W_1$ that is not tumbling (i.e., when $K=1$ and $k_1 > 1$) can be understood in a similar way as ``\textbf{Case 1}'' above, since sub-aggregates from $W_f$ can reduce cost for \emph{intervals} in $W_1$ that overlap (lines~\ref{alg:factor-window-partition-by:K1:k_1_larger_than_1:begin} to~\ref{alg:factor-window-partition-by:K1:k_1_larger_than_1:end}).}

\vspace{0.5em}
\HL{We formally prove the correctness of Algorithm~\ref{alg:factor-window-partition-by} in the appendix, using Equations~\ref{eq:cost:reduction:general} and~\ref{eq:lambda}:}
\begin{theorem}\label{theorem:factor-window:partition-by}
Algorithm~\ref{alg:factor-window-partition-by} correctly determines whether $W_f$ would help when both $W_f$ and $W$ are tumbling windows.
\end{theorem}

\subsubsection{Revisit of Candidate Generation and Selection}
We now revisit the problems of candidate generation and selection under ``partitioned by'' semantics.

\vspace{0.5em}
\noindent\textbf{(Candidate Generation)} By restricting to tumbling windows under ``partitioned by'' semantics, we can significantly reduce the search space for potential candidates.
By Theorem~\ref{theorem:window-partitioning}, the range $r_f$ of a factor window $W_f$ must be a \emph{common factor} of the ranges $r_1$, ..., $r_K$ of all downstream windows $W_1$, ..., $W_K$ for a given target window $W$ (ref. Figure~\ref{fig:factor-window}).
Moreover, $r_f$ must also be a multiple of the range $r_W$ of the target window $W$.
As a result, one can enumerate all candidates by starting from the \emph{greatest common divisor} $r$ of $r_1$, ..., $r_K$ and look for all factors $r_f$ of $r$ that are also multiples of $r_W$.

\begin{figure}
\centering
\subfigure[Dependent]{ \label{fig:multi-factor-window:dependent}
\includegraphics[width=0.45\columnwidth]{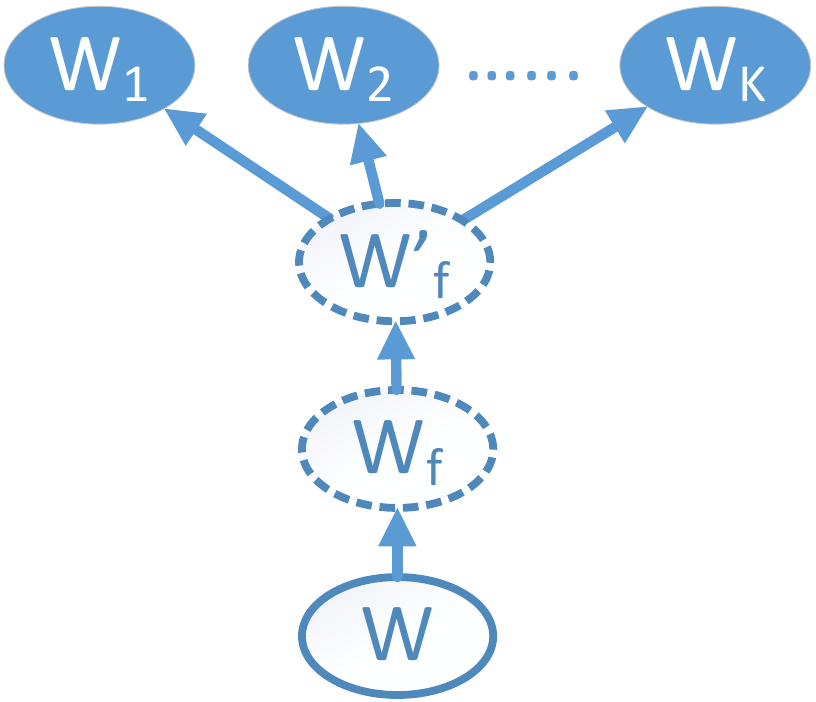}}
\subfigure[Independent]{ \label{fig:multi-factor-window:independent}
\includegraphics[width=0.45\columnwidth]{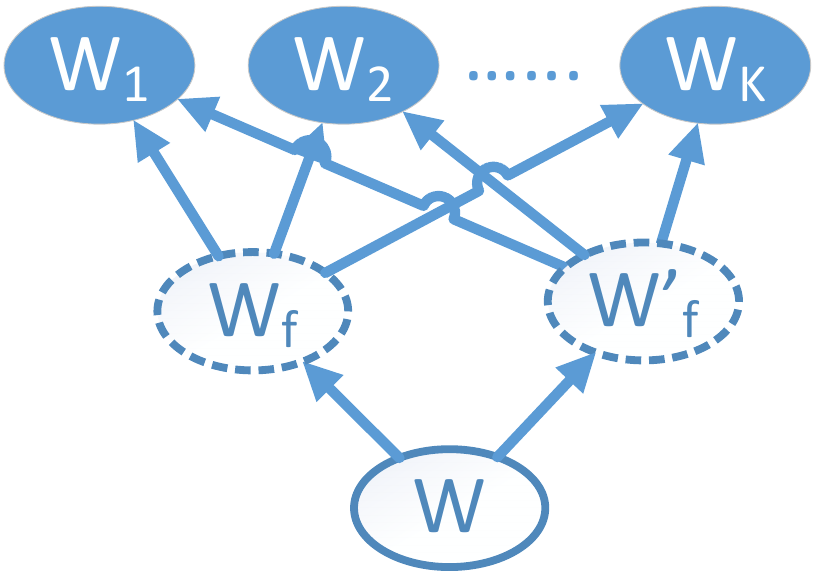}}
\vspace{-0.5em}
\caption{Dependent and independent factor windows with multiple candidates.}
\label{fig:multi-factor-window}
\end{figure}

\vspace{0.5em}
\noindent\textbf{(Candidate Selection)} To find the best factor window, we compare the benefits of two candidates $W_f$ and $W'_f$.
There are two cases as shown in Figure~\ref{fig:multi-factor-window}:
\begin{itemize}
    \item $W_f$ and $W'_f$ are \emph{dependent}, meaning either $W_f\leq W'_f$ or $W'_f\leq W_f$ -- see Figure~\ref{fig:multi-factor-window:dependent};
    \item $W_f$ and $W'_f$ are \emph{independent} -- see Figure~\ref{fig:multi-factor-window:independent}.
\end{itemize}

\paragraph*{Dependent Candidates}
Let $W_f$ and $W'_f$ be two eligible factor windows such that $W'_f\leq W_f$.
Then $W_f$ can be omitted as adding it cannot reduce the overall cost.
This can be understood by running Algorithm~\ref{alg:factor-window-partition-by} against $W_f$, by viewing $W'_f$ as $W_f$'s only (tumbling) downstream window.
Algorithm~\ref{alg:factor-window-partition-by} would return \verb|false| as this is the case when $K=1$ and $k_1=1$ \blue{(line~\ref{alg:factor-window-partition-by:K1:k_1_1:end})}.


\paragraph*{Independent Candidates} For the independent case, we have to compare the costs in more detail.
Specifically, let
$c_f =\sum\nolimits_{j=1}^{K}\cost(W_j) + \cost(W_f) + \cost(W) = \sum\nolimits_{j=1}^{K} n_j \cdot M(W_j, W_f) + n_f \cdot M(W_f, W) + \cost(W),$
and
$c'_f=\sum\nolimits_{j=1}^{K}\cost(W_j) + \cost(W'_f) + \cost(W)= \sum\nolimits_{j=1}^{K} n_j \cdot M(W_j, W'_f) + n'_f \cdot M(W'_f, W) + \cost(W).$


\begin{theorem}\label{theorem:factor-window:cost-compare}
Let $W_f$ and $W'_f$ be two \emph{independent} eligible factor windows under ``partitioned by'' semantics.
$c_f\leq c'_f$ iff
\begin{equation}\label{eq:factor-window-cost-compare}
\frac{r_f}{r'_f}\geq\frac{\lambda-\frac{r_f}{r_W}}{\lambda-\frac{r'_f}{r_W}}.
\end{equation}
Here $\lambda$ has been defined in Equation~\ref{eq:lambda}.
\end{theorem}

\begin{algorithm}[t]
\small
  \SetAlgoLined
  \KwIn{$W$, a window; $\{W_1, ..., W_K\}$, $W$'s downstream windows (ref. Figure~\ref{fig:factor-window}).}
  \KwOut{The best tumbling factor window $W_f$ that led to the minimum overall cost.}
  \SetAlgoLined
  \blue{Compute $\lambda$ using Equation~\ref{eq:lambda}\;}\label{alg:fw:tumbling:cand-gen:lambda}
  \blue{// Generate candidate tumbling factor windows.}\\
  $r_d\leftarrow\gcd(\{r_1, ..., r_K\})$\;\label{alg:fw:tumbling:cand-gen:enum:begin}
  \If{$r_d=r_W$}{
    \Return{$W$}\;\label{alg:fw:tumbling:cand-gen:enum:returnW}
  }
  \blue{$\mathcal{F}\leftarrow\{r_f: r_d \text{ mod } r_f = 0 \text{ and } r_f \text{ mod } r_W = 0\}$}\;\label{alg:fw:tumbling:cand-gen:enum:end}
  $\mathcal{W}_f\leftarrow\emptyset$\;\label{alg:fw:tumbling:cand-gen:begin}
  \ForEach{$r_f\in\mathcal{F}$}{
    Create a tumbling window $W_f\langle r_f, r_f \rangle$\;
    \blue{$b\leftarrow$\underline{\emph{Check}}($W_f$, $W$, $\{W_1, ..., W_K\}, \lambda$) by Algorithm~\ref{alg:factor-window-partition-by}}\;
    \If{$b=$\texttt{true}}{
        $\mathcal{W}_f\leftarrow\mathcal{W}_f\cup \{W_f$\}\;\label{alg:fw:tumbling:cand-gen:end}
    }
  }
  \blue{// Remove candidates that are not independent.}\\
  \ForEach{$W_f\in\mathcal{W}_f$}{\label{alg:fw:tumbling:prune:begin}
    \If{there exists $W'_f$ s.t. $W'_f\leq W_f$}{
        $\mathcal{W}_f\leftarrow\mathcal{W}_f- \{W_f\}$\;\label{alg:fw:tumbling:prune:end}
    }
  }
  \Return{the best $W_f\in\mathcal{W}_f$ by applying Theorem~\ref{theorem:factor-window:cost-compare}}\;
  \caption{Find the best factor window under ``partitioned by'' semantics.}
\label{alg:factor-window:partition-by:multiple}
\end{algorithm}

Algorithm~\ref{alg:factor-window:partition-by:multiple} presents the details of picking the best factor window for a target window $W$ and its downstream windows $W_1$, ..., $W_K$, under ``partitioned by'' semantics.
It starts by enumerating all candidates for $W_f$ based on the constraint that $r_f$ must be a common factor of $\{r_1,...,r_K\}$ \emph{and} a multiple of $r_W$ \blue{(lines~\ref{alg:fw:tumbling:cand-gen:enum:begin} to~\ref{alg:fw:tumbling:cand-gen:enum:end}).}
It simply returns $W$ if no candidate can be found \blue{(line~\ref{alg:fw:tumbling:cand-gen:enum:returnW}).}
It then looks for candidates of $W_f$ that are beneficial, using Algorithm~\ref{alg:factor-window-partition-by} \blue{(lines~\ref{alg:fw:tumbling:cand-gen:begin} to~\ref{alg:fw:tumbling:cand-gen:end}).}
It further prunes \emph{dependent} candidates that are dominated by others \blue{(lines~\ref{alg:fw:tumbling:prune:begin} to~\ref{alg:fw:tumbling:prune:end}).}
Finally, it finds the best $W_f$ by applying Theorem~\ref{theorem:factor-window:cost-compare} to compare the remaining candidates.


\begin{example}
Continuing with Example~\ref{example:factor-window},
Algorithm~\ref{alg:factor-window:partition-by:multiple} would generate three candidate factor windows $W(10, 10)$, $W(5, 5)$, and $W(2, 2)$, since all of them are beneficial according to Algorithm~\ref{alg:factor-window-partition-by} ($K=2$ indeed).
However, since both $W(5,5)$ and $W(2,2)$ cover $W(10, 10)$, these two candidates are removed and $W(10, 10)$ is the remaining, best candidate.
\end{example}

\vspace{-0.5em}
\paragraph*{\blue{Time Complexity Analysis of Algorithm~\ref{alg:factor-window:partition-by:multiple}}}
\blue{Computing $\lambda$ at line~\ref{alg:fw:tumbling:cand-gen:lambda} takes $O(K)$ time.
Computing $r_d$ at line~\ref{alg:fw:tumbling:cand-gen:enum:begin} takes $O(r_{\max}\log r_{\max})$ time using Euclid's algorithm~\cite{clrs}, where $r_{\max}=\max\{r_1, ..., r_K\}$. Computing $\mathcal{F}$ at line~\ref{alg:fw:tumbling:cand-gen:enum:end} takes $O(\lceil\frac{r_d}{r_W}\rceil)$ time.
Generating candidate tumbling factor windows (lines~\ref{alg:fw:tumbling:cand-gen:begin} to~\ref{alg:fw:tumbling:cand-gen:end}) takes $O(|\mathcal{F}|)$ time, as each run of Algorithm~\ref{alg:factor-window-partition-by} takes constant time. Pruning dependent candidates (lines~\ref{alg:fw:tumbling:prune:begin} to~\ref{alg:fw:tumbling:prune:end}) takes $O(|\mathcal{F}|^2)$ time due to pairwise comparison. Finally, finding the best candidate by applying Theorem~\ref{theorem:factor-window:cost-compare} takes $O(|\mathcal{F}|)$ time. Therefore, the time complexity of Algorithm~\ref{alg:factor-window:partition-by:multiple} is $O(K + r_{\max}\log r_{\max} + \lceil\frac{r_d}{r_W}\rceil +|\mathcal{F}|^2 + 2\cdot |\mathcal{F}|).$ To simplify our analysis, we assume it is dominated by $O(|\mathcal{F}|^2)$. Since $O(|\mathcal{F}|)=O(\lceil\frac{r_d}{r_W}\rceil)$, it follows that the time complexity of Algorithm~\ref{alg:factor-window:partition-by:multiple} is $O(\lceil\frac{r_d}{r_W}\rceil^2)$. This is in contrast to the  $O(\lceil\frac{s_d}{s_W}\rceil\cdot\lceil\frac{r_{\min}}{s_W}\rceil\cdot K)$ time complexity of Algorithm~\ref{alg:FindBestFactorWindow:covered-by}, which finds the best factor window following ``covered by'' semantics. In a real-world setting, we would expect $\lceil\frac{r_{\min}}{s_W}\rceil> \lceil\frac{r_d}{r_W}\rceil\approx \lceil\frac{s_d}{s_W}\rceil$, in which case Algorithm~\ref{alg:factor-window:partition-by:multiple} improves over Algorithm~\ref{alg:FindBestFactorWindow:covered-by} significantly. On the other hand, Algorithm~\ref{alg:factor-window:partition-by:multiple} may lose some optimization opportunities due to its reduced search space for candidate factor windows. We only use Algorithm~\ref{alg:factor-window:partition-by:multiple} when ``covered by'' semantics cannot be used to optimize the input aggregate function.}

\begin{figure*}[t]
\centering
\subfigure[\textbf{RandomGen}, ``partitioned by'']{ \label{fig:throughput:random:partitioned-by:W5}
    \includegraphics[width=0.45\textwidth]{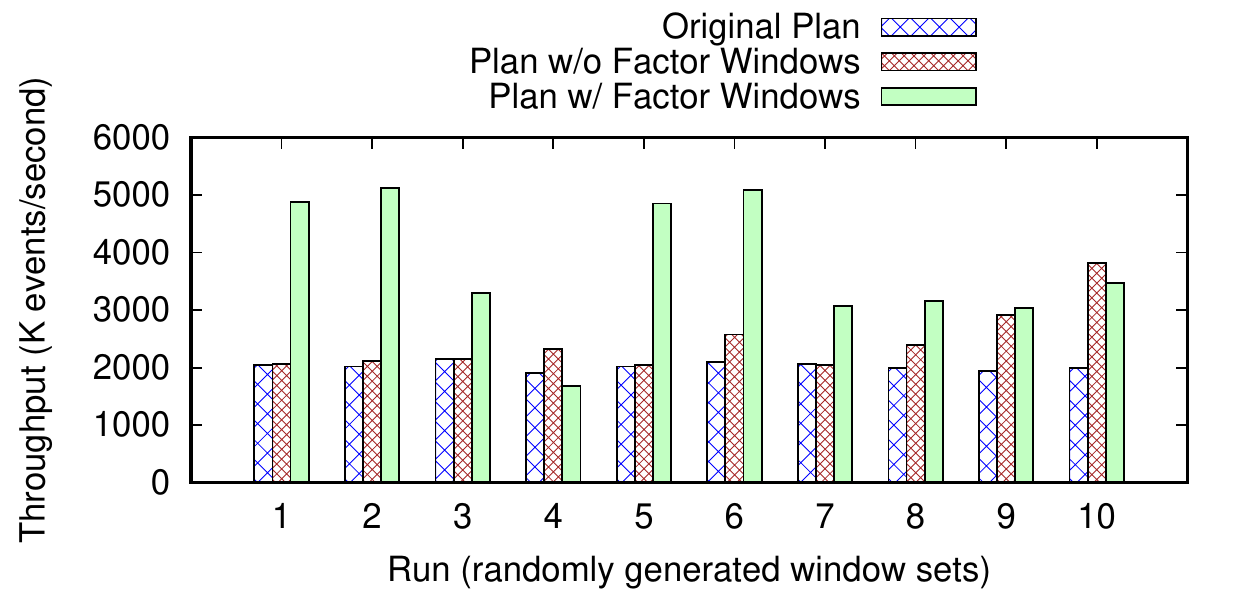}}
\subfigure[\textbf{RandomGen}, ``covered by'']{ \label{fig:throughput:random:covered-by:W5}
    \includegraphics[width=0.45\textwidth]{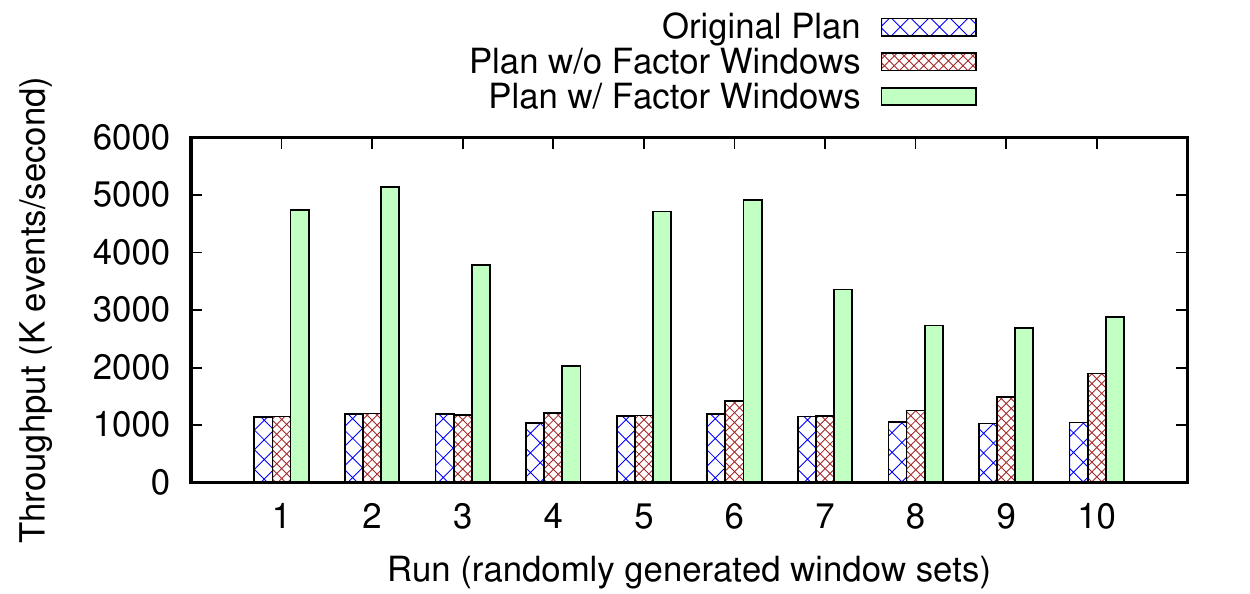}}
\subfigure[\textbf{SequentialGen}, ``partitioned by'']{ \label{fig:throughput:seq:partitioned-by:W5}
    \includegraphics[width=0.45\textwidth]{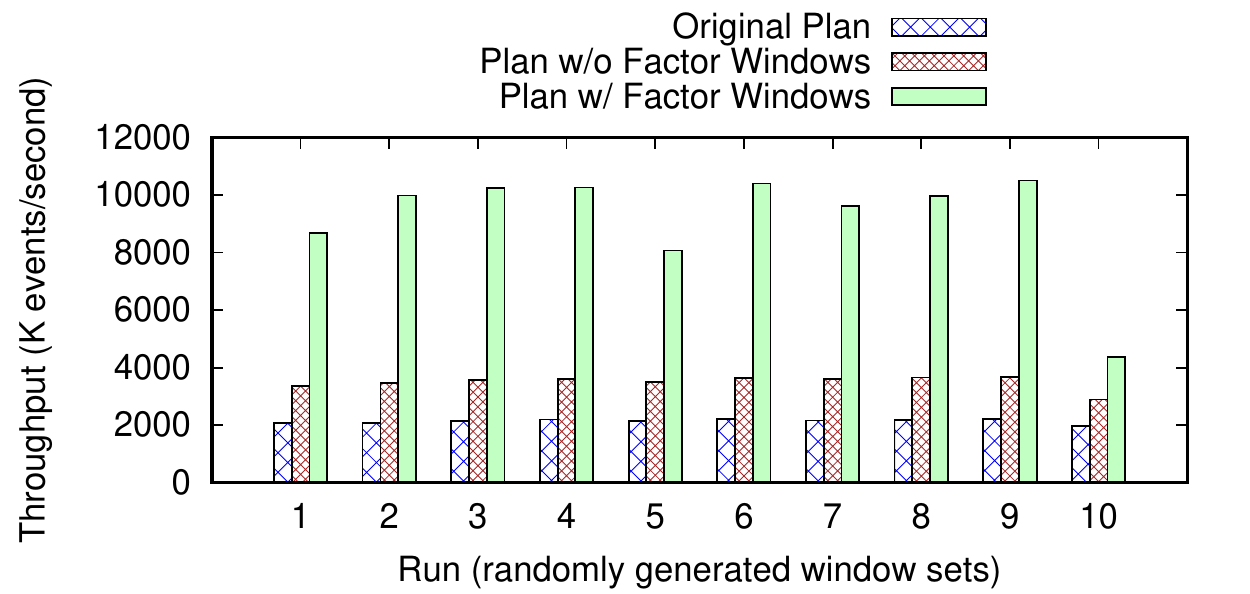}}
\subfigure[\textbf{SequentialGen}, ``covered by'']{ \label{fig:throughput:seq:covered-by:W5}
    \includegraphics[width=0.45\textwidth]{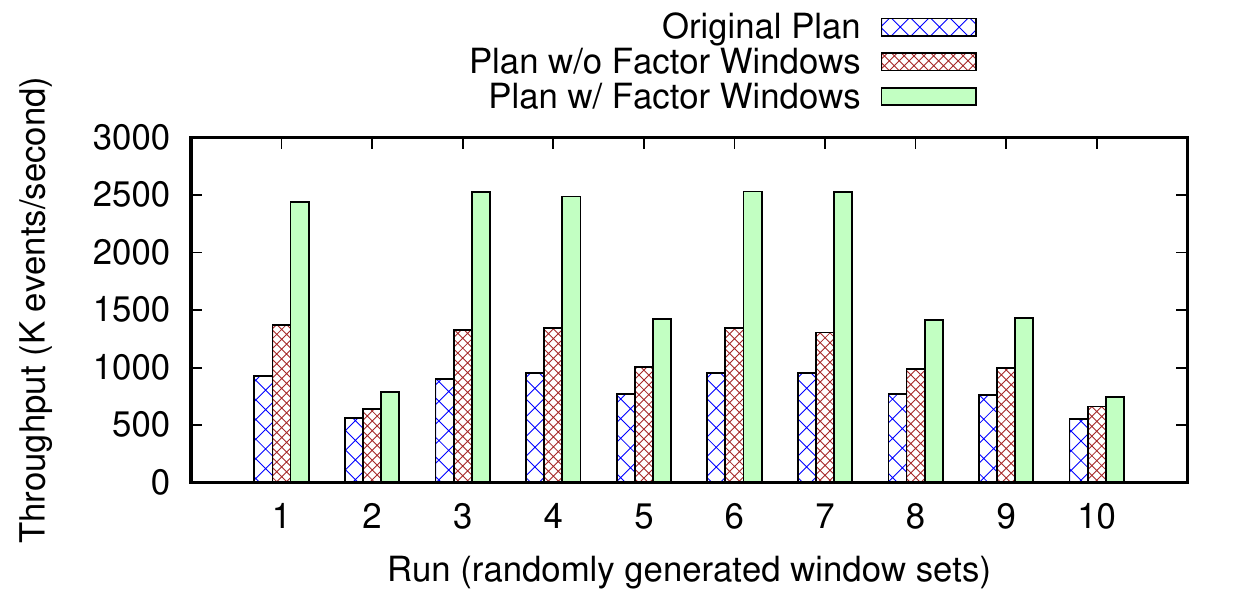}}
\vspace{-0.5em}
\caption{Throughput on window sets when processing 10 million input events from \textbf{Synthetic-10M} with $|\mathcal{W}|=5$.}
\label{fig:throughput:ws5}
\vspace{-2em}
\end{figure*}



\section{Evaluation}\label{section:evaluation}

We report experimental evaluation results in this section.
We observe that (1) the optimized query plan, even without factor windows, can significantly outperform the original query plan in terms of \emph{throughput}~\cite{streamBenchmark} (up to 2.5$\times$); (2) with factor windows, the throughput of the optimized query plan can be much higher (up to 16.8$\times$).
\blue{Moreover, our optimized plans can yield similar, and sometimes much higher, throughput compared to Scotty~\cite{TraubGCBKRM21}, one state-of-the-art window slicing technique.}
\blue{Meanwhile, our approach has negligible overhead and can scale up smoothly when increasing window-set size.}

\vspace{-0.5em}
\subsection{Experiment Settings}

\subsubsection{Setup}
We implemented our cost-based query optimizer in C\#. Given an input window-set aggregate query with its original query plan, it can produce the best query plans, with and without factor windows.
All query plans are represented as Trill expressions.
For each query plan, we measure its \emph{throughput}, which is defined as the number of events processed per unit time~\cite{streamBenchmark}.
We perform all experiments on a workstation equipped with 2.2 GHz Intel CPUs and 128 GB main memory.
\blue{All results are based on single-core executions.}

\subsubsection{Data Sets}

We used both synthetic and real data.
For synthetic data, we generated data streams with 1 million and 10 million events, denoted as \textbf{Synthetic-1M} and \textbf{Synthetic-10M} respectively, where the events arrive at a constant pace.
For real data, we used the same dataset as used in~\cite{CaiB0C21}, which was derived from the DEBS 2012 Grand Challenge~\cite{DEBS2012} dataset that consists of monitoring data from manufacturing equipment. Specifically, we pair the given timestamps with the values of the column \texttt{mf01}, i.e., the ``electrical power main-phase 1'' sensor reading. 
This dataset contains roughly 32 million events and is denoted as \textbf{Real-32M}.
\blue{We used ``\texttt{MIN}'' as the aggregate function, which can be supported by both ``covered by'' and ``partitioned by'' semantics.}

\begin{algorithm}[t]
\small
  \SetAlgoLined
  \KwIn{$S$, the ``seed'' slides; $R$, the ``seed'' ranges; $k_s$, $k_r$: the multipliers; $N$, the size of the window set;
  \emph{tumbling}: whether each window is tumbling or not.}
  \KwOut{$\mathcal{W}$, the window set generated.}
  \SetAlgoLined
  $\mathcal{W}\leftarrow\emptyset$\;
  \For{$1\leq i\leq N$} {
    \If{tumbling}{
        $r_0\leftarrow$Random($R$)\;
        $r\leftarrow$Random($\{2r_0, ..., k_r\cdot r_0\}$)\;
        $\mathcal{W}\leftarrow\mathcal{W}\cup\{W\langle r, r\rangle\}$\;
    }\Else{
        $s_0\leftarrow$Random($S$)\;
        $s\leftarrow$Random($\{2s_0, ..., k_s\cdot s_0\}$)\;
        $\mathcal{W}\leftarrow\mathcal{W}\cup\{W\langle 2s, s\rangle\}$\;
    }
  }
  \Return{the window set $\mathcal{W}$}\;
  \caption{The \textbf{RandomGen} window-set generator.}
\label{alg:random-window-gen}
\end{algorithm}

\subsubsection{Generation of Window Sets}
We generated window sets using the following approaches.
\begin{itemize}[leftmargin=*]
    \item \textbf{(RandomGen)} We generate each window $W\langle r, s\rangle$ randomly. Specifically, to generate a tumbling window where $s=r$, we first pick a ``seed'' range $r_0$ uniformly randomly from a given list and then choose $r$ uniformly randomly from $\{2r_0, ..., k_r\cdot r_0\}$. We purposely avoid choosing $r=r_0$ to test the effectiveness of our cost-based optimizer when exploring factor windows, as $W\langle r_0, r_0\rangle$ is a valid factor window in this case that should be considered by the optimizer. To generate a hopping window, we operate in a similar manner by first picking a ``seed'' slide $s_0$ uniformly randomly from a given list and then choosing $s$ uniformly randomly among $\{2s_0, ..., k_s\cdot s_0\}$; we finally set $r=2s$ and return $W\langle r, s\rangle$.
    Algorithm~\ref{alg:random-window-gen} summarizes this procedure. 
    \item \textbf{(SequentialGen)} In practice, the windows contained by a window set may be more correlated than those generated by \textbf{RandomGen}. Here, we focus on a common case that we observed in the real world, where the windows follow a ``sequential'' pattern in terms of either the range or the slide size. We presented such an example in Figure~\ref{fig:multi-window-single-agg}. This motivates us to implement the \textbf{SequentialGen} window-set generator that aims for capturing this sequential pattern.
    Specifically, unlike in \textbf{RandomGen} where $r$ is randomly selected from $\{2r_0, ..., k_r\cdot r_0\}$ when generating tumbling windows , we simply pick $r$ sequentially following the order $2r_0$, ..., $k_r\cdot r_0$. Similarly, we pick $s$ sequentially following the order $2s_0$, ..., $k_s\cdot s_0$ when generating hopping windows.
\end{itemize}

\vspace{-0.5em}
\subsection{Results on Synthetic Data}

For the parameters in \textbf{RandomGen} and \textbf{SequentialGen}, we set the window-set size $N\in\{5, 10\}$, the ``seed'' slides $S=\{5, 10, 20\}$ \blue{(only for generating hopping windows, where ranges are fixed as twice the slides)}, the ``seed'' ranges $R=\{2, 5, 10\}$ \blue{(only for generating tumbling windows)}, and $k_s=k_r=50$.
For each window-set size $N$, we generated 10 window sets for both tumbling and hopping windows.
We also set $\eta=1$ in our cost model.

\subsubsection{Throughput}

\blue{Figure~\ref{fig:throughput:ws5} reports the throughput results observed on \textbf{Synthetic-10M} for window sets of size 5 generated by both \textbf{RandomGen} and \textbf{SequentialGen}. The results on \textbf{Synthetic-10M} with window sets of size 10, as well as the results on \textbf{Synthetic-1M} are included in the appendix.}

\paragraph*{Observations on window sets by \textbf{RandomGen}}
(1) For the window sets containing \emph{tumbling} windows, the ``partitioned by'' semantics were leveraged when constructing the window coverage graph (WCG) and exploring factor windows.
As illustrated in Figure~\ref{fig:throughput:random:partitioned-by:W5}, compared to the original plan, the rewritten plan \emph{without} factor windows can boost throughput by up to 1.9$\times$, whereas the plan \emph{with} factor windows can boost the throughput by up to 2.5$\times$.
(2) For the window sets containing \emph{hopping} windows, the general ``covered by'' semantics were used to create WCG's and factor windows.
Figure~\ref{fig:throughput:random:covered-by:W5} 
presents the results. 
We observe similar patterns as we observed on tumbling windows, where factor windows yield significantly larger throughput (by up to 4.3$\times$).
\blue{(3) In a couple of cases, the optimized plans are slightly worse than the original plans.
This is possible, since our cost model does not use throughput as the cost metric.
However, such cases are rare based on our evaluation, and in the appendix we show that our cost metric is highly correlated with throughput.}

\paragraph*{Observations on window sets by \textbf{SequentialGen}}
The observations are similar to those on window sets generated by \textbf{RandomGen}.
Again, using factor windows significantly boosts the throughput (by up to 4.8$\times$ and 2.8$\times$ for ``partitioned by'' and ``covered by'' semantics, respectively).
We further notice that the rewritten query plans \emph{without} factor windows are more effective than they were in the case of \textbf{RandomGen}.
This is not surprising, though, as the improved correlation between windows generated by \textbf{SequentialGen} leads to more overlaps and thus more sharing opportunities.

\begin{table}
\small
\centering
\begin{tabularx}{\columnwidth}{|l|X|X|X|X|}
\hline
\textbf{Setup} & \textbf{w/o FW (Mean)} & \textbf{w/o FW (Max)} & \textbf{w/ FW (Mean)} & \textbf{w/ FW (Max)}\\
\hline
\hline
R-5-tumbling &	1.21$\times$ & 1.92$\times$ & 1.85$\times$ & 2.54$\times$\\
R-10-tumbling &	1.34$\times$ &	1.77$\times$ &	1.88$\times$ &	3.38$\times$\\
R-5-hopping &	1.18$\times$ &	1.82$\times$ &	3.26$\times$ &	4.29$\times$ \\
R-10-hopping &	1.34$\times$ &	1.71$\times$ &	3.20$\times$ &	6.15$\times$ \\
\hline
\hline
S-5-tumbling & 1.63$\times$ &	1.67$\times$ &	4.28$\times$ &	4.81$\times$\\
S-10-tumbling &	1.98$\times$ &	2.05$\times$ &	7.91$\times$ &	\textbf{9.38}$\times$\\
S-5-hopping &	1.34$\times$ &	1.48$\times$ &	2.17$\times$ &	2.81$\times$\\
S-10-hopping &	1.58$\times$	& 1.73$\times$ &	2.92$\times$ &	3.79$\times$\\
\hline
\end{tabularx}
\caption{Summary of throughput boosts on \textbf{Synthetic-10M}, where `R' stands for window sets generated by \textbf{RandomGen}, `S' stands for window sets generated by \textbf{SequentialGen}, and `5' and `10' are the sizes of the window sets generated.}
\vspace{-1em}
\label{tab:throughput-summary}
\end{table}

\paragraph*{Summary}
In Table~\ref{tab:throughput-summary}, we summarize the mean and max throughput boosts of the rewritten query plans (without and with factor windows) over the original query plans, observed when processing \textbf{Synthetic-10M} under different experimental setups for window-set generation.
With factor windows, we can achieve up to 9.4$\times$ throughput boost on \textbf{Synthetic-10M}.

\begin{table}
\small
\centering
\begin{tabularx}{\columnwidth}{|l|X|X|X|X|}
\hline
\textbf{Setup} & \textbf{w/o FW (Mean)} & \textbf{w/o FW (Max)} & \textbf{w/ FW (Mean)} & \textbf{w/ FW (Max)}\\
\hline
\hline
R-5-tumbling &	1.19$\times$ &	1.78$\times$ &	1.43$\times$ &	1.91$\times$\\
R-10-tumbling &	1.30$\times$ &	1.71$\times$ &	1.53$\times$ &	2.86$\times$\\
R-5-hopping	& 1.09$\times$ & 1.39$\times$ &	1.54$\times$ &	2.63$\times$\\
R-10-hopping &	1.18$\times$ &	1.39$\times$ &	1.46$\times$ &	3.53$\times$\\
\hline
\hline
S-5-tumbling &	1.63$\times$ &	1.67$\times$ &	4.12$\times$ &	4.85$\times$\\
S-10-tumbling &	1.90$\times$ &	1.97$\times$ &	7.53$\times$ &	\textbf{9.14}$\times$\\
S-5-hopping & 1.12$\times$ & 1.30$\times$ &	1.22$\times$ & 1.77$\times$ \\
S-10-hopping & 1.22$\times$ & 1.51$\times$ & 1.45$\times$ &	2.31$\times$\\
\hline
\end{tabularx}
\caption{Summary of throughput boosts on \textbf{Real-32M}, with the same notation as in Table~\ref{tab:throughput-summary}.}
\vspace{-1em}
\label{tab:throughput-summary:DEBS}
\end{table}

\vspace{-0.5em}
\subsection{Results on Real Data}
We further tested the throughput of window sets over the real dataset \textbf{Real-32M}.
Table~\ref{tab:throughput-summary:DEBS} summarizes the results on throughput boosts of the rewritten query plans, without and with factor windows, over the original plans, and the details are included in the appendix.
Overall, using factor windows can achieve throughput boost up to 9.1$\times$ over \textbf{Real-32M}.

\vspace{-0.5em}
\subsection{\blue{Scalability Tests}}

\blue{To understand the scalability of our cost-based optimization approach, we increased the window-set size $|\mathcal{W}|$ to 15 and 20.
Table~\ref{tab:throughput-summary:scalability} summarizes the throughput results on \textbf{Synthetic-10M}; the details are in the appendix.
Overall, the query plans generated by our approach scale up smoothly when increasing the window-set size, with throughput boost up to 16.8$\times$.}

\begin{table}
\small
\centering
\begin{tabularx}{\columnwidth}{|l|X|X|X|X|}
\hline
\textbf{Setup} & \textbf{w/o FW (Mean)} & \textbf{w/o FW (Max)} & \textbf{w/ FW (Mean)} & \textbf{w/ FW (Max)}\\
\hline
\hline
\blue{R-15-tumbling} &	1.55$\times$ &	1.96$\times$ &	2.97$\times$ &	4.34$\times$\\
\blue{R-20-tumbling} &	1.49$\times$ &	2.29$\times$ &	2.10$\times$ &	4.83$\times$\\
\blue{R-15-hopping} &	1.55$\times$ &	1.95$\times$ &	4.67$\times$ &	6.59$\times$\\
\blue{R-20-hopping} &	1.68$\times$ &	2.20$\times$ &	4.23$\times$ &	7.65$\times$\\
\hline
\hline
\blue{S-15-tumbling} &	2.43$\times$ &	2.49$\times$ &	11.29$\times$ &	13.83$\times$\\
\blue{S-20-tumbling} &	2.42$\times$ &	2.53$\times$ &	14.28$\times$ &	\textbf{16.82}$\times$\\
\blue{S-15-hopping} &	1.85$\times$ &	2.09$\times$ &	3.51$\times$ &	4.68$\times$\\
\blue{S-20-hopping} &	1.91$\times$ &	2.15$\times$ &	4.02$\times$ &	5.32$\times$\\
\hline
\end{tabularx}
\caption{\blue{Summary of results on scalability test with $\mathcal{W}\in\{15, 20\}$, in terms of throughput boosts on \textbf{Synthetic-10M}.
The notation here is the same as in Tables~\ref{tab:throughput-summary} and~\ref{tab:throughput-summary:DEBS}.}}
\vspace{-2em}
\label{tab:throughput-summary:scalability}
\end{table}

\vspace{-0.5em}
\subsection{\blue{Query Optimization Overhead}}

\begin{figure}[t]
\centering
    \includegraphics[width=0.9\columnwidth]{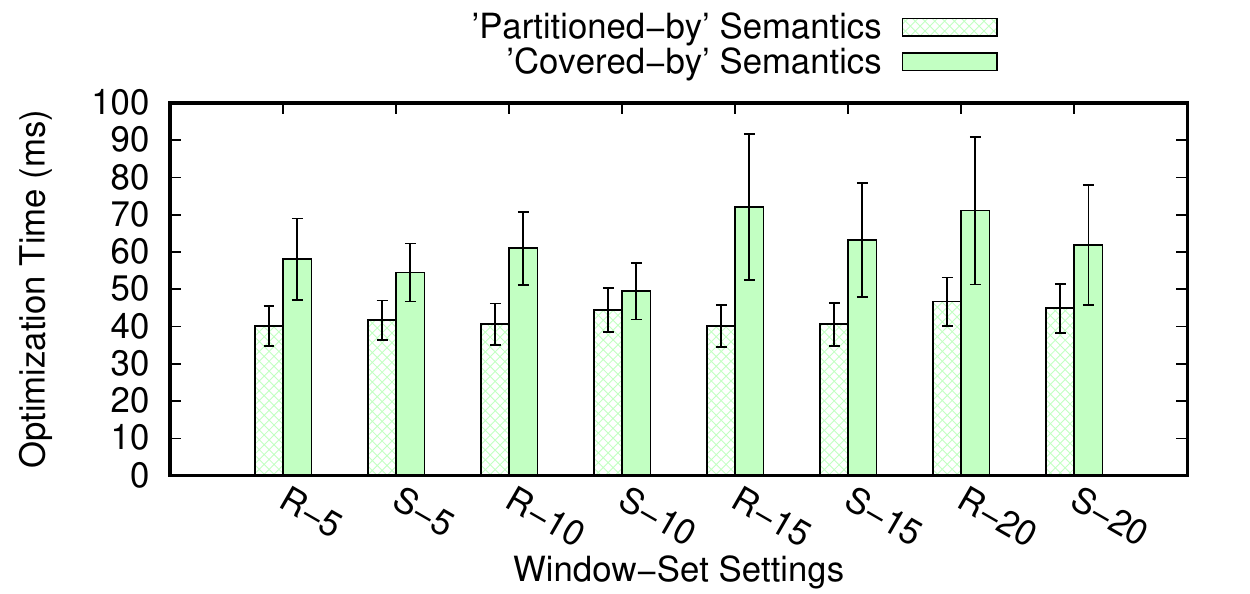}
\vspace{-0.5em}
\caption{\blue{Factor-window based optimization overhead (average time and standard deviation) with increasing window-set size from 5 to 20. `R' and `S' are shorthands for ``\textbf{RandomGen}'' and ``\textbf{SequentialGen}''.}}
\label{fig:optimization-overhead}
\end{figure}

\blue{Figure~\ref{fig:optimization-overhead} presents the average time spent on query optimization and its standard deviation (shown with error bars), when enabling factor windows and varying window-set size from 5 to 20. For each setting, the average and standard deviation were measured based on the 10 window sets generated by either \textbf{RandomGen} or \textbf{SequentialGen}. We observe that the optimization overhead is very small overall ($<$100 milliseconds for the settings that we tested). Moreover, the optimization overhead of ``covered by'' semantics is higher than that of ``partitioned by'' semantics. This makes sense considering the larger search space with ``covered by'' semantics.}

\begin{figure*}[t]
\centering
\subfigure[\textbf{RandomGen}, ``partitioned by'']{ \label{fig:scotty-compare:random:partitioned-by:W10}
    \includegraphics[width=0.45\textwidth]{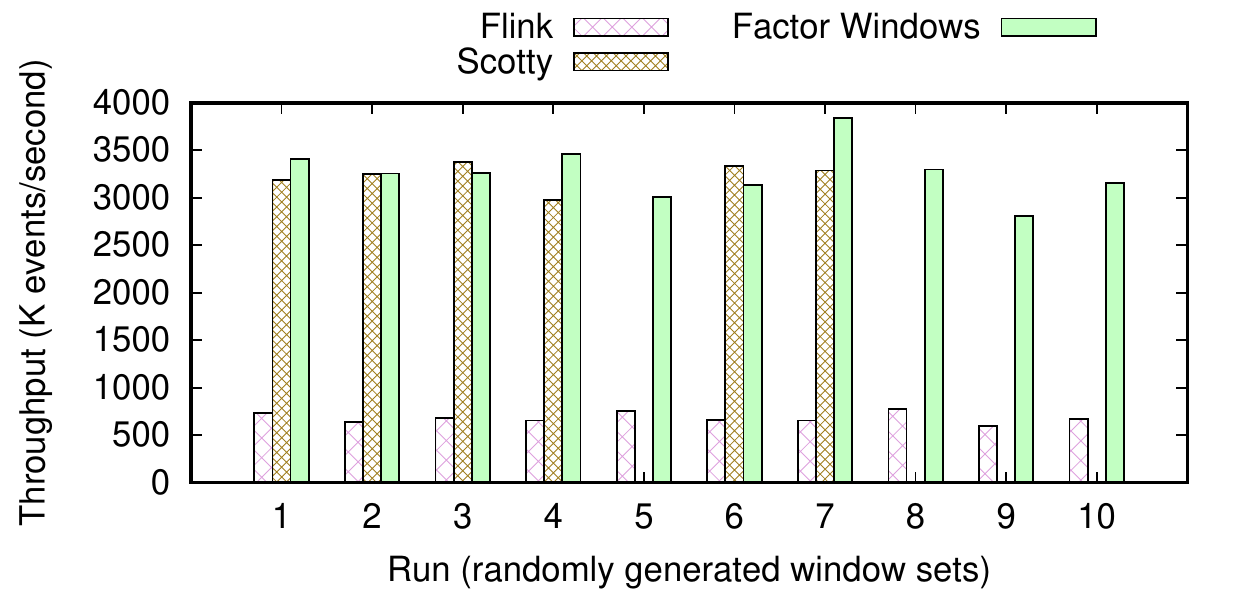}}
\subfigure[\textbf{RandomGen}, ``covered by'']{ \label{fig:scotty-compare:random:covered-by:W10}
    \includegraphics[width=0.45\textwidth]{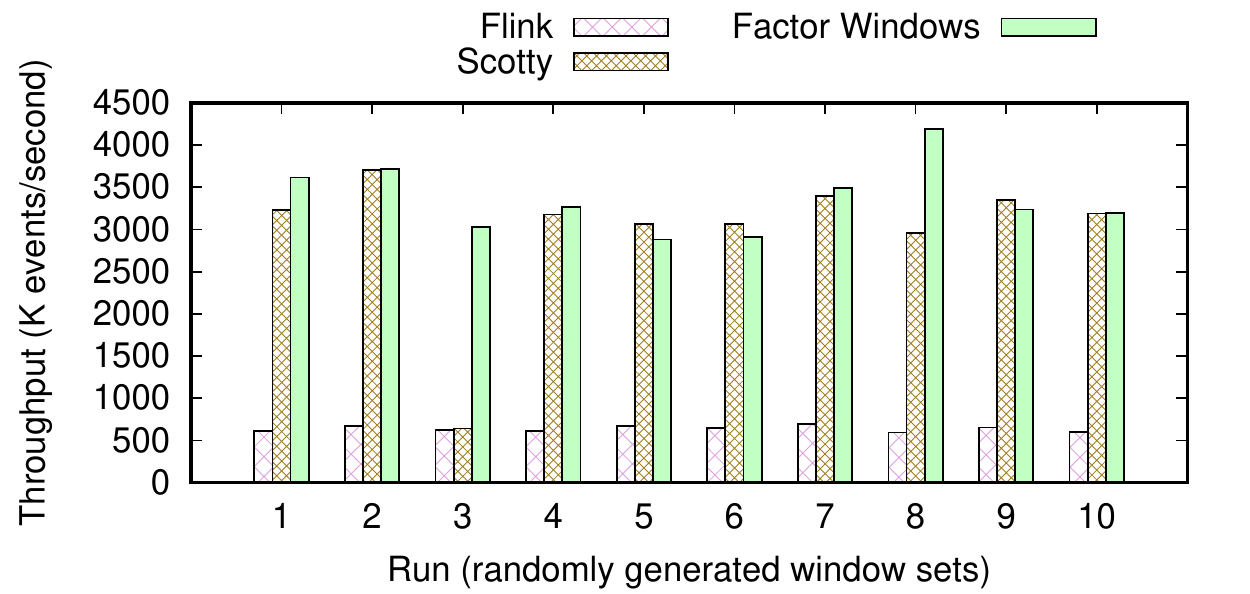}}
\subfigure[\textbf{SequentialGen}, ``partitioned by'']{ \label{fig:scotty-compare:seq:partitioned-by:W10}
    \includegraphics[width=0.45\textwidth]{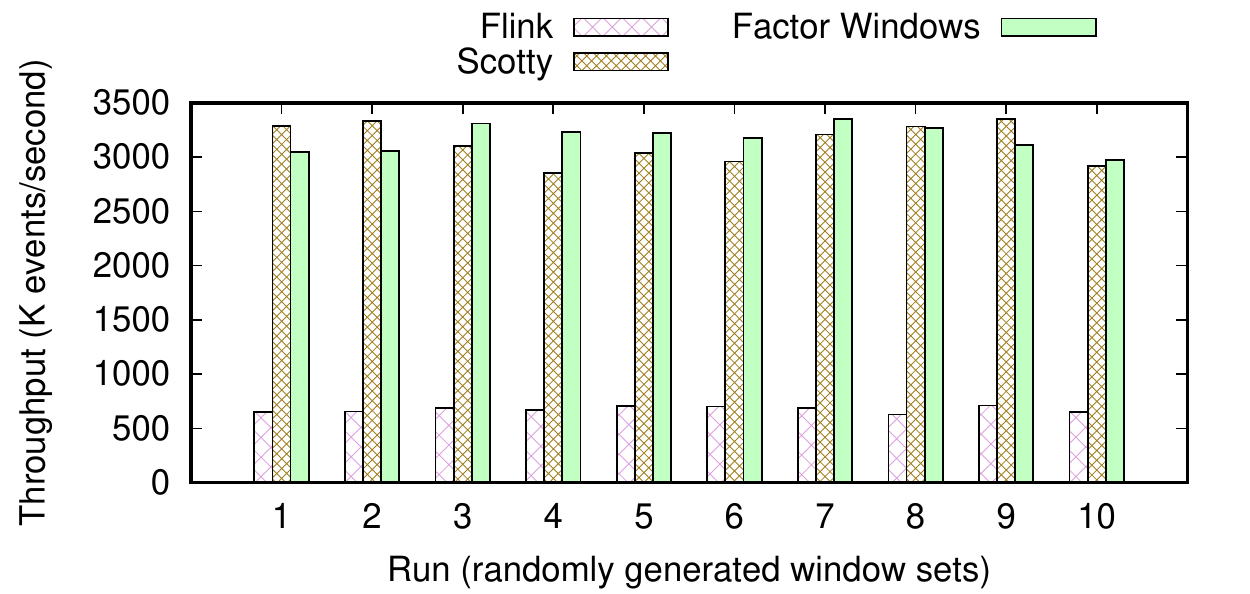}}
\subfigure[\textbf{SequentialGen}, ``covered by'']{ \label{fig:scotty-compare:seq:covered-by:W10}
    \includegraphics[width=0.45\textwidth]{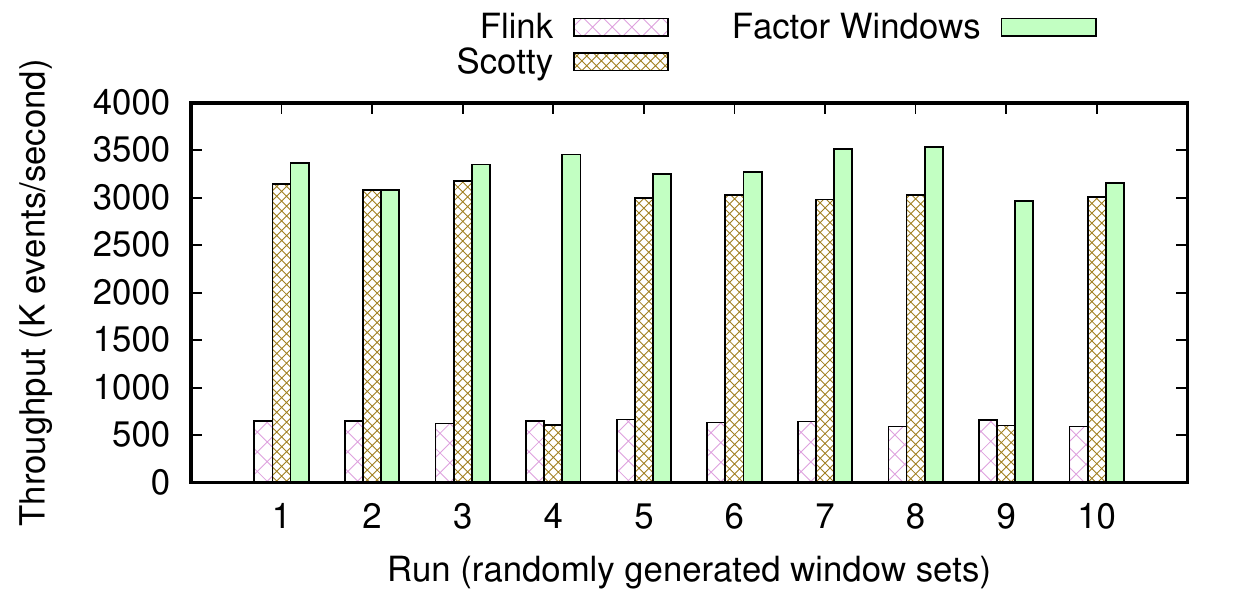}}
\vspace{-0.5em}
\caption{\blue{Comparison with Scotty~\cite{TraubGCBKRM21} in terms of throughput on window sets with $|\mathcal{W}|=10$.}}
\label{fig:scotty-compare:ws10}
\vspace{-1.5em}
\end{figure*}

\vspace{-0.5em}
\subsection{\blue{Comparison with Window Slicing}}\label{sec:evaluation:compare:scotty}

\blue{We compare our cost-based optimization approach with Scotty~\cite{TraubGCBKRM21}, one state-of-the-art window slicing technology.
Since Scotty does not support Trill, we translate our optimized query plans into Apache Flink queries expressed by its DataStream API~\cite{flink-datastream-api}, following a similar query rewriting procedure described in Section~\ref{sec:implementation:query-rewrite}.
We compare the throughput of Flink, Scotty, and our optimized plans with factor windows, using the same data generator developed by Scotty for benchmarking its own performance~\cite{TraubGCBKRM21,Scotty-GitHub}.
In our experiments we set the window-set size $|\mathcal{W}|\in\{5, 10\}$. We did not further increase $|\mathcal{W}|$ since Scotty cannot process some window sets with $|\mathcal{W}|=10$ (see Figure~\ref{fig:scotty-compare:random:partitioned-by:W10}).
Figure~\ref{fig:scotty-compare:ws10} shows the results with $|\mathcal{W}|=10$. The results with $|\mathcal{W}|=5$ are in the appendix.}

\blue{We have two observations. First, both Scotty and our factor-window based optimization significantly outperform the default Flink query execution plan, where each window aggregate is evaluated independently. Second, our approach can yield similar, and sometimes much higher, throughput compared to Scotty. This holds for both window sets generated by \textbf{RandomGen} (Figures~\ref{fig:scotty-compare:random:partitioned-by:W10} and~\ref{fig:scotty-compare:random:covered-by:W10}) and \textbf{SequentialGen} (Figures~\ref{fig:scotty-compare:seq:partitioned-by:W10} and~\ref{fig:scotty-compare:seq:covered-by:W10}), where we observe up to 5.7$\times$ throughput boost (excluding cases where the throughput of Scotty is unavailable).}

\vspace{-0.5em}
\section{Related Work}\label{section:related-work}
\blue{
The related work on stream query processing and optimization is overwhelming (see~\cite{HirzelSSGG13} for a survey).
We focus our discussion on optimization techniques dedicated to window aggregates~\cite{CarboneKH19,LiMTPT05}.
In addition to the \emph{window slicing} techniques discussed in the introduction (e.g.~\cite{KrishnamurthyWF06,LiMTPT05-no-pane,GuirguisSCL11,GuirguisSCL12,CarboneTKHM16,TangwongsanHSW15,TraubGCBKRM19,TraubGCBKRM21}), there has been a flurry of recent work that accelerates window aggregation via better utilization of modern hardware, such as Grizzly~\cite{GrulichBZTBCRM20} and LightSaber~\cite{TheodorakisKPP20}.}
This line of work is orthogonal to ours. However, it may be worthwhile to consider combining it with our cost-based optimization framework, which we leave for future work.

Cost-based query optimization is the standard practice in batch processing systems~\cite{SelingerACLP79},
but is not popular in stream processing systems.
There is little work on cost modeling in the streaming world~\cite{ViglasN02}.
One reason might be the difficulty of defining a single cost criterion, as streaming systems may need to honor various performance metrics simultaneously, such as latency, throughput, and resource utilization~\cite{ChandramouliGBRS11}.
Although the application of static cost-based query optimization is limited~\cite{AyadN04}, dynamic query optimization (a.k.a., adaptive query processing) at runtime has been extensively studied in the context of streaming (e.g.,~\cite{HellersteinA00,Bernstein19,DeshpandeH04,FloratouAGRR17,MaiZPXSVCKMKDR18,NehmeRB09,NehmeWLRB13,RamanDH03,VenkataramanPOA17}).
Our current cost model is static and it is interesting future work to investigate how to dynamically adjust cost estimates at runtime by keeping track of the input event rates.

In recent years, a number of distributed streaming systems have been built as open-source or proprietary software (e.g., Storm~\cite{ToshniwalTSRPKJGFDBMR14}, Spark Streaming~\cite{ArmbrustDTYZX0S18}, Flink~\cite{CarboneKEMHT15}, MillWheel~\cite{AkidauBBCHLMMNW13}, Dataflow~\cite{AkidauBCCFLMMPS15}, Quill~\cite{ChandramouliFGE16}, etc.).
While most of these systems provide users with \emph{imperative} programming interfaces, the adoption of \emph{declarative}, SQL-like query interfaces~\cite{ArasuBW06}, similar to the one that ASA exposes, has been increasingly popular.
For example, both Spark Streaming and Flink now support SQL queries on top of data streams.
Moving to the declarative interface raises the level of abstraction and enables compile-time query optimization.
The optimization techniques proposed in this paper can be implemented in either imperative or declarative systems.
We demonstrated the latter for the ASA SQL query compiler (Section~\ref{sec:implementation:query-rewrite}),
but our algorithms are not tied to the ASA SQL language and can be applied in other streaming systems that support declarative query languages.

\section{Conclusion}\label{section:conclusion}

We proposed a cost-based optimization framework to optimize the evaluation of aggregate functions over multiple correlated windows.
It leverages the window coverage graph (WCG) that we introduced to capture the inherent overlapping relationships between windows.
We introduced factor windows into the WCG to help reduce the overall computation overhead.
Evaluation results show that the optimized query plans can significantly outperform the original plans in terms of throughput, especially when factor windows are enabled, without the need for runtime support from stream processing engines.


\clearpage
{
\bibliographystyle{abbrv}
\bibliography{windowSetOpt}
}

\clearpage
\appendices

\begin{figure*}[t]
\centering
\subfigure[\textbf{RandomGen}, ``partitioned by'']{ \label{fig:throughput:random:partitioned-by:W10}
    \includegraphics[width=0.45\textwidth]{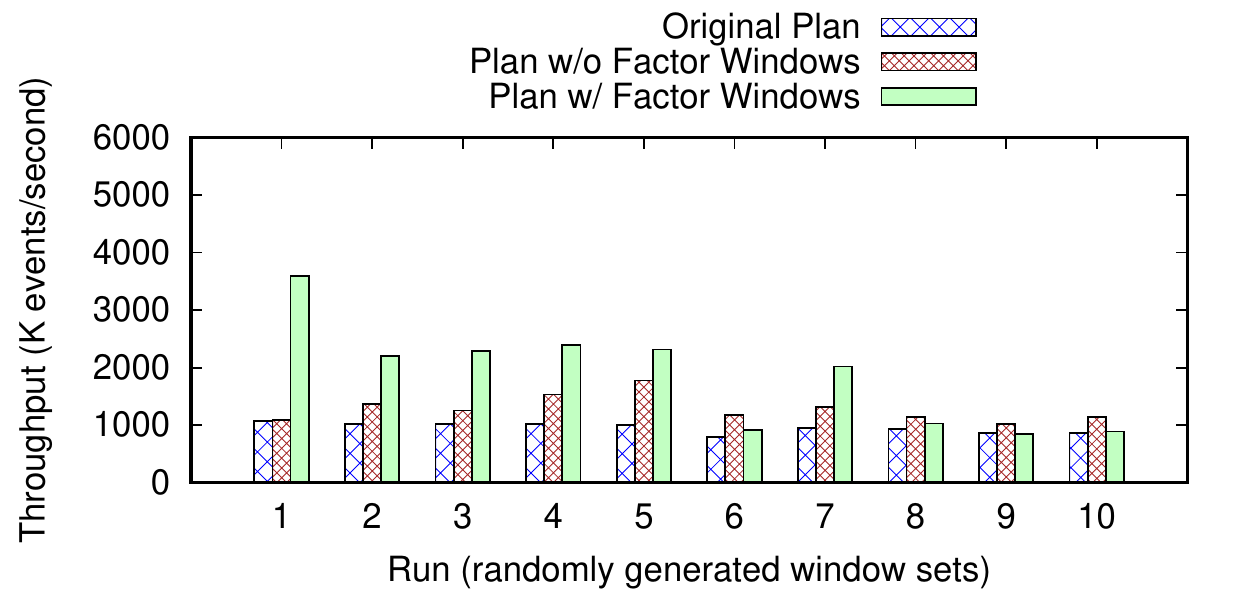}}
\subfigure[\textbf{RandomGen}, ``covered by'']{ \label{fig:throughput:random:covered-by:W10}
    \includegraphics[width=0.45\textwidth]{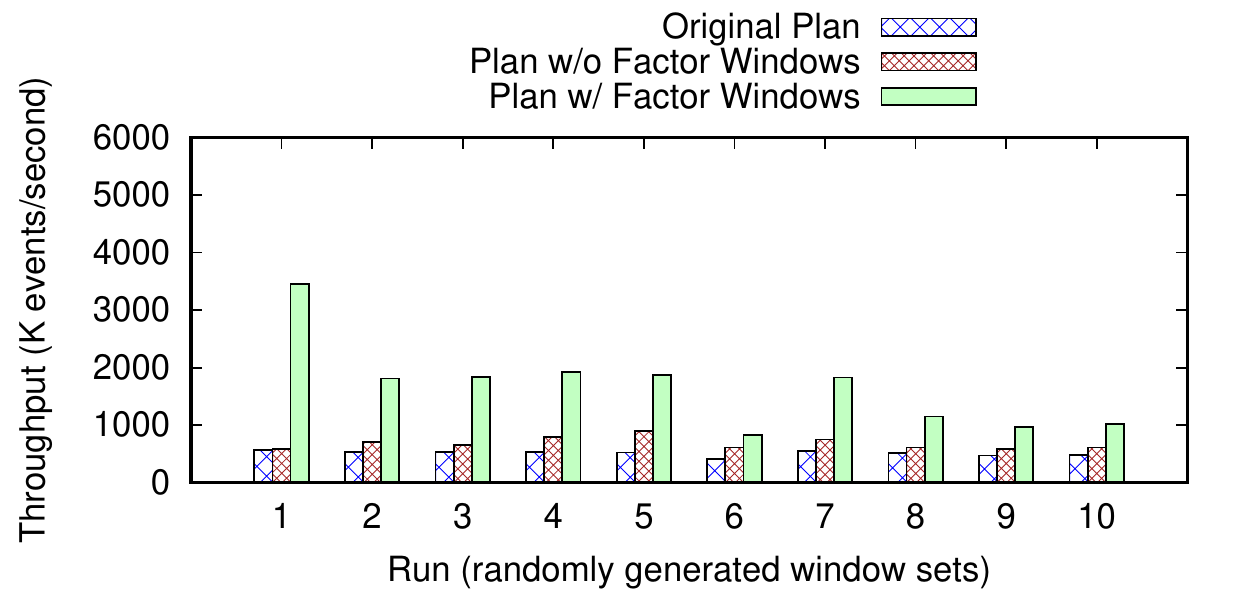}}
\subfigure[\textbf{SequentialGen}, ``partitioned by'']{ \label{fig:throughput:seq:partitioned-by:W10}
    \includegraphics[width=0.45\textwidth]{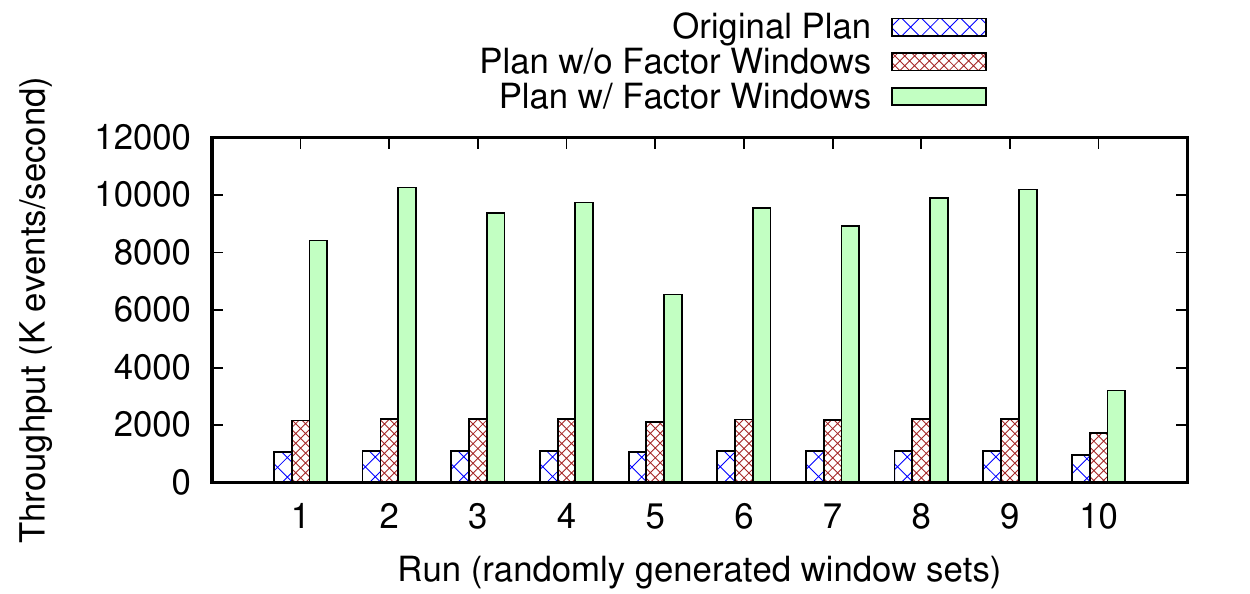}}
\subfigure[\textbf{SequentialGen}, ``covered by'']{ \label{fig:throughput:seq:covered-by:W10}
    \includegraphics[width=0.45\textwidth]{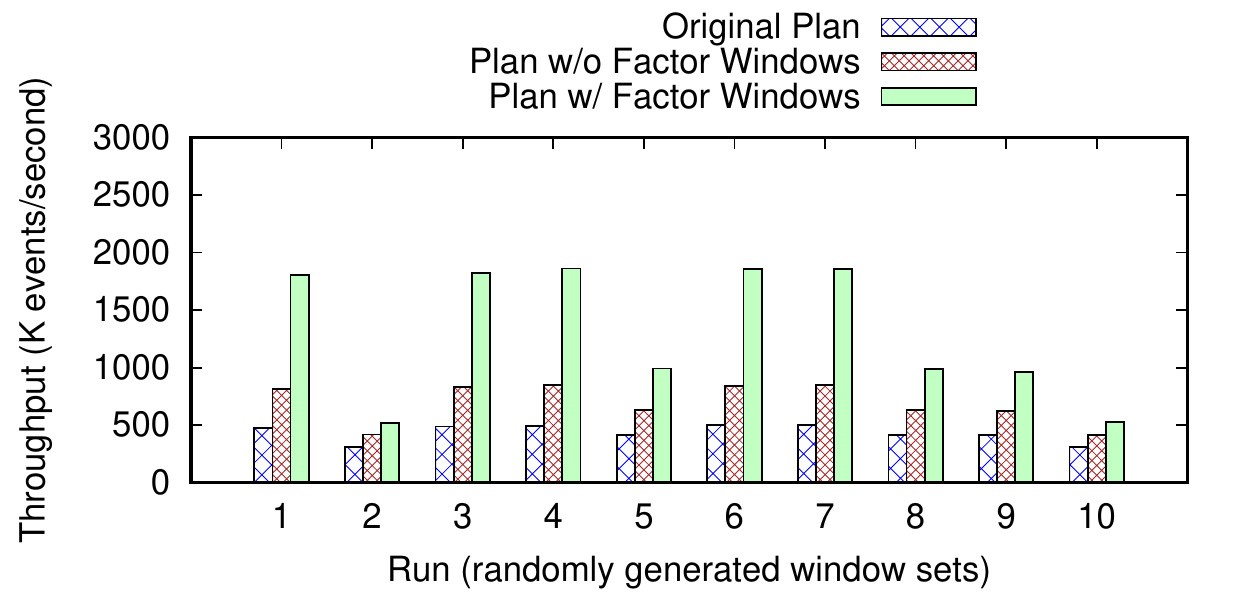}}
\vspace{-0.5em}
\caption{\blue{Throughput on window sets when processing 10 million input events from \textbf{Synthetic-10M} with $|\mathcal{W}|=10$.}} 
\label{fig:throughput:ws10}
\vspace{-1.5em}
\end{figure*}

\begin{figure*}
\centering
\subfigure[\textbf{RandomGen}, ``partitioned by'']{ \label{fig:throughput:random:partitioned-by:W5:1m}
    \includegraphics[width=0.45\textwidth]{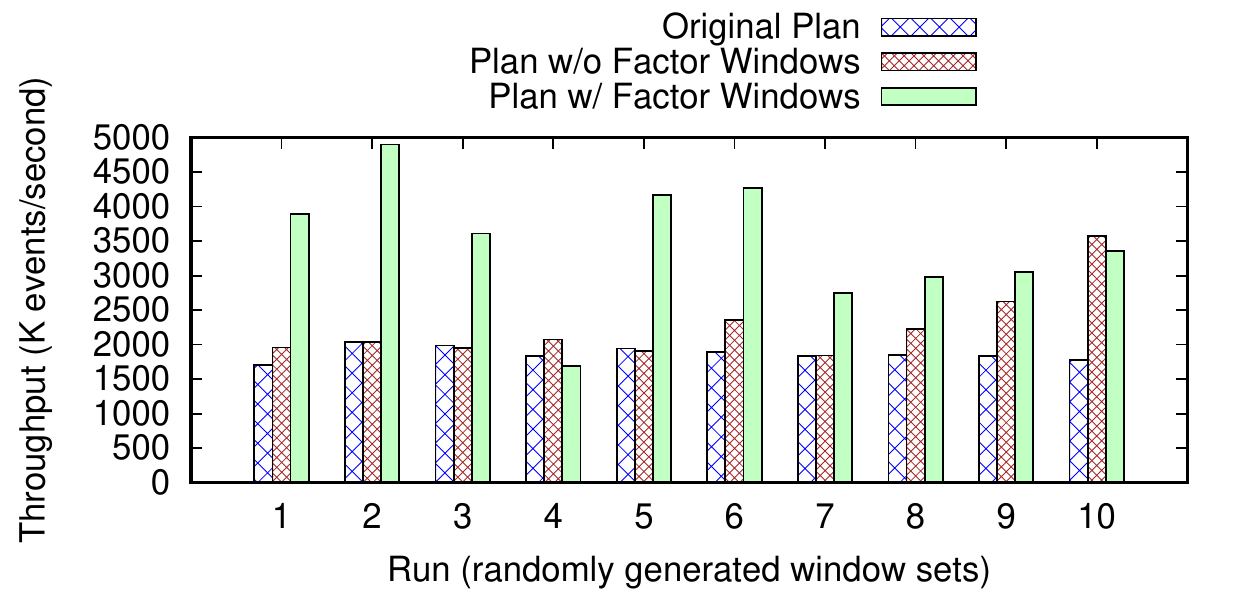}}
\subfigure[\textbf{RandomGen}, ``covered by'']{ \label{fig:throughput:random:covered-by:W5:1m}
    \includegraphics[width=0.45\textwidth]{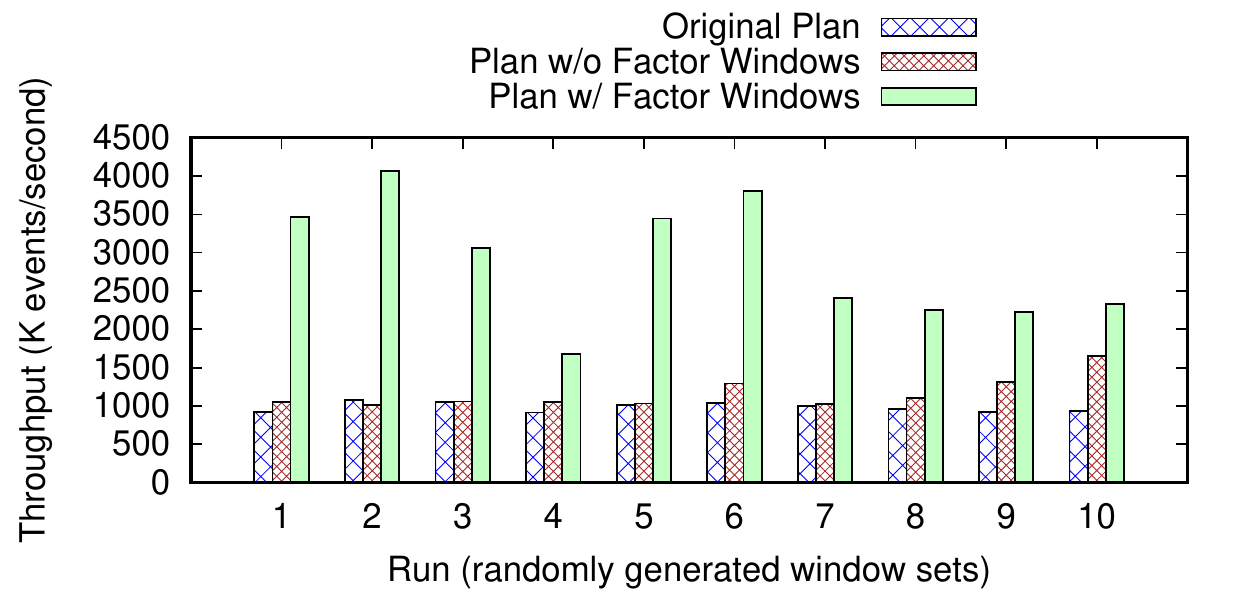}}
\subfigure[\textbf{SequentialGen}, ``partitioned by'']{ \label{fig:throughput:seq:partitioned-by:W5:1m}
    \includegraphics[width=0.45\textwidth]{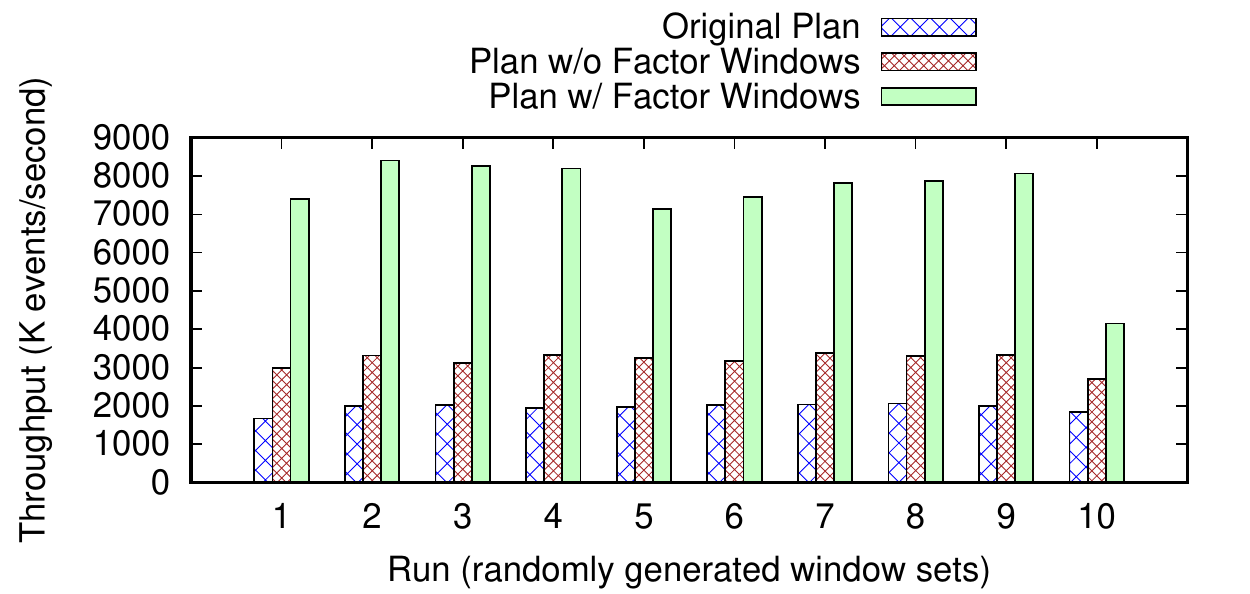}}
\subfigure[\textbf{SequentialGen}, ``covered by'']{ \label{fig:throughput:seq:covered-by:W5:1m}
    \includegraphics[width=0.45\textwidth]{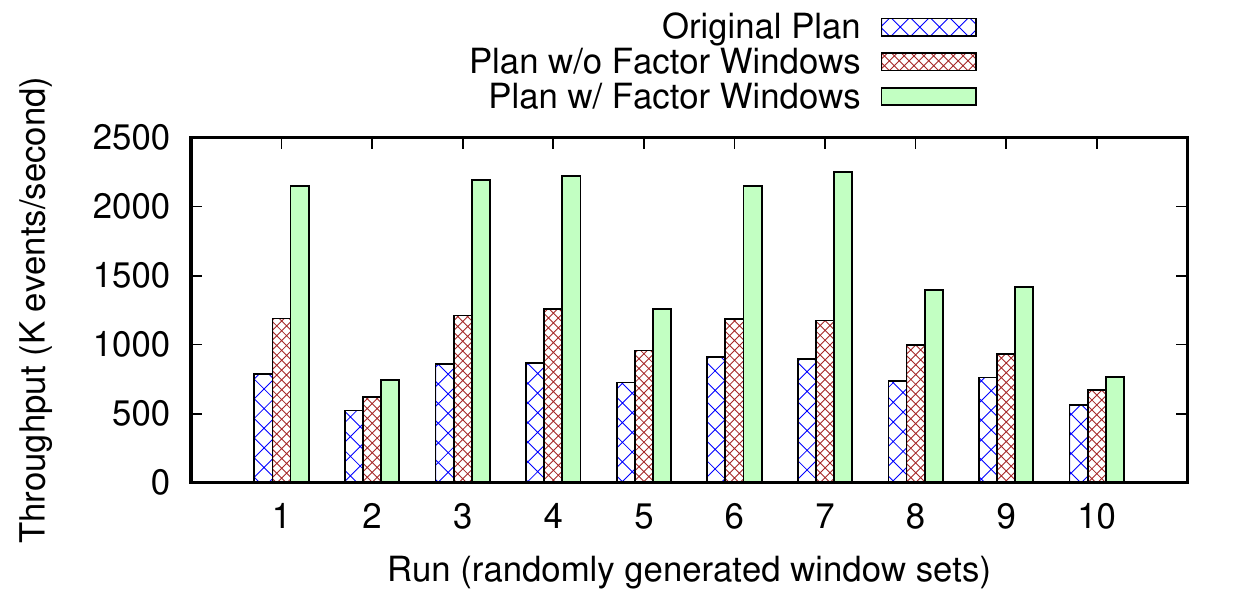}}
\vspace{-0.5em}
\caption{\blue{Throughput on window sets when processing 1 million input events from \textbf{Synthetic-1M} with $|\mathcal{W}|=5$.}}
\label{fig:throughput:ws5:1m}
\vspace{-1.5em}
\end{figure*}

\begin{figure*}[!ht]
\centering
\subfigure[\textbf{RandomGen}, ``partitioned by'']{ \label{fig:throughput:random:partitioned-by:W10:1m}
    \includegraphics[width=0.45\textwidth]{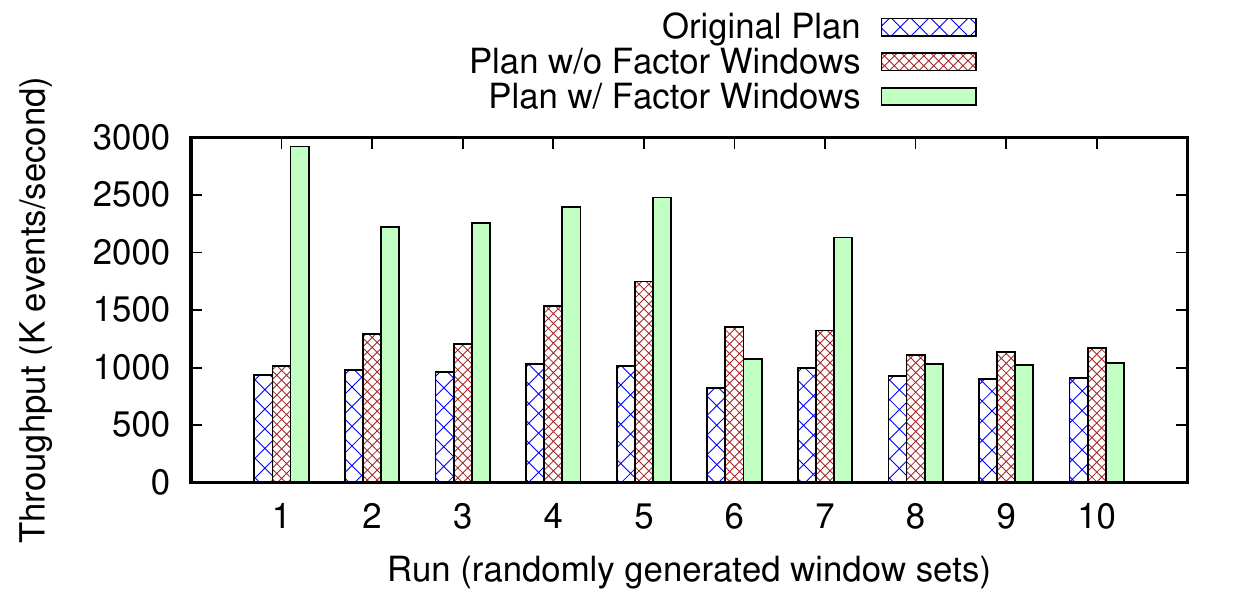}}
\subfigure[\textbf{RandomGen}, ``covered by'']{ \label{fig:throughput:random:covered-by:W10:1m}
    \includegraphics[width=0.45\textwidth]{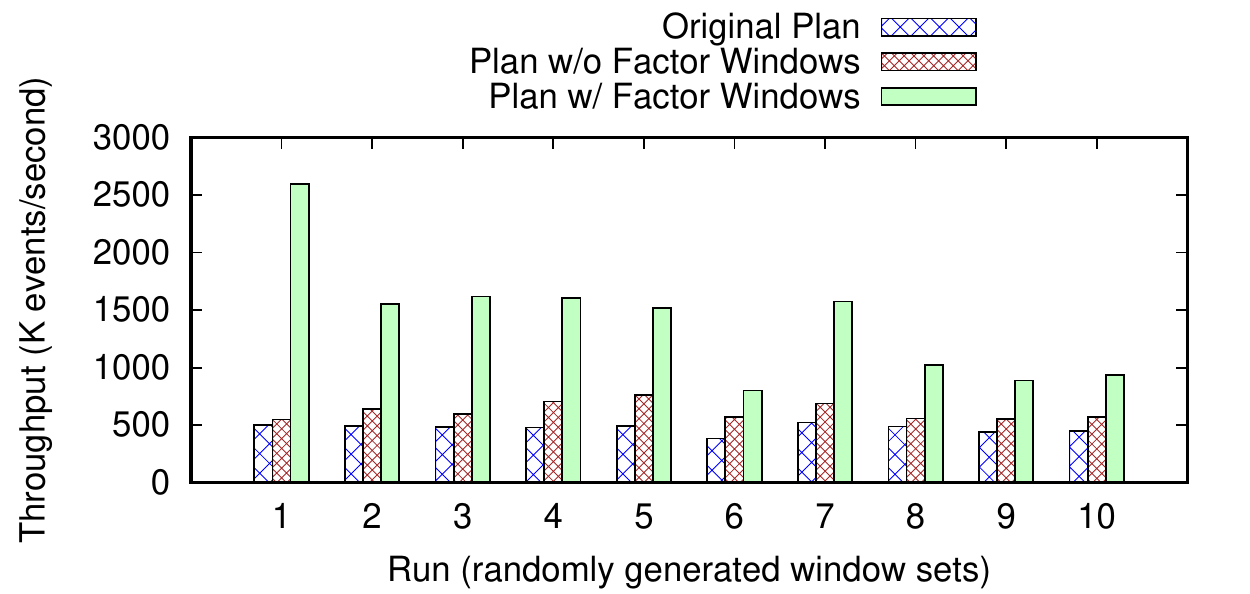}}
\subfigure[\textbf{SequentialGen}, ``partitioned by'']{ \label{fig:throughput:seq:partitioned-by:W10:1m}
    \includegraphics[width=0.45\textwidth]{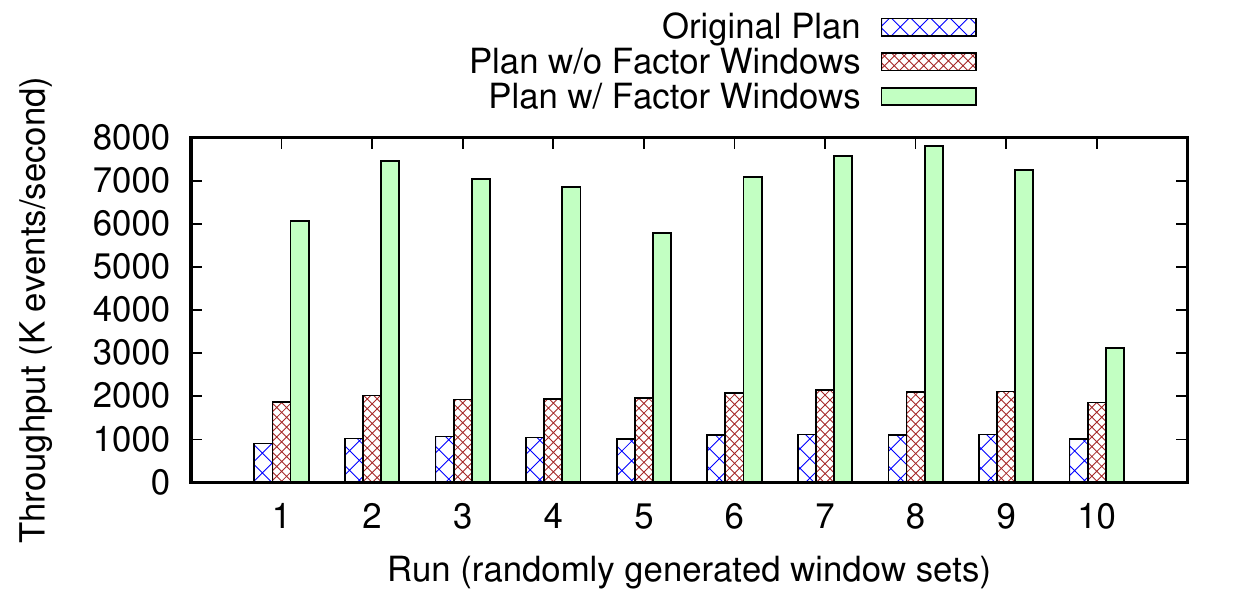}}
\subfigure[\textbf{SequentialGen}, ``covered by'']{ \label{fig:throughput:seq:covered-by:W10:1m}
    \includegraphics[width=0.45\textwidth]{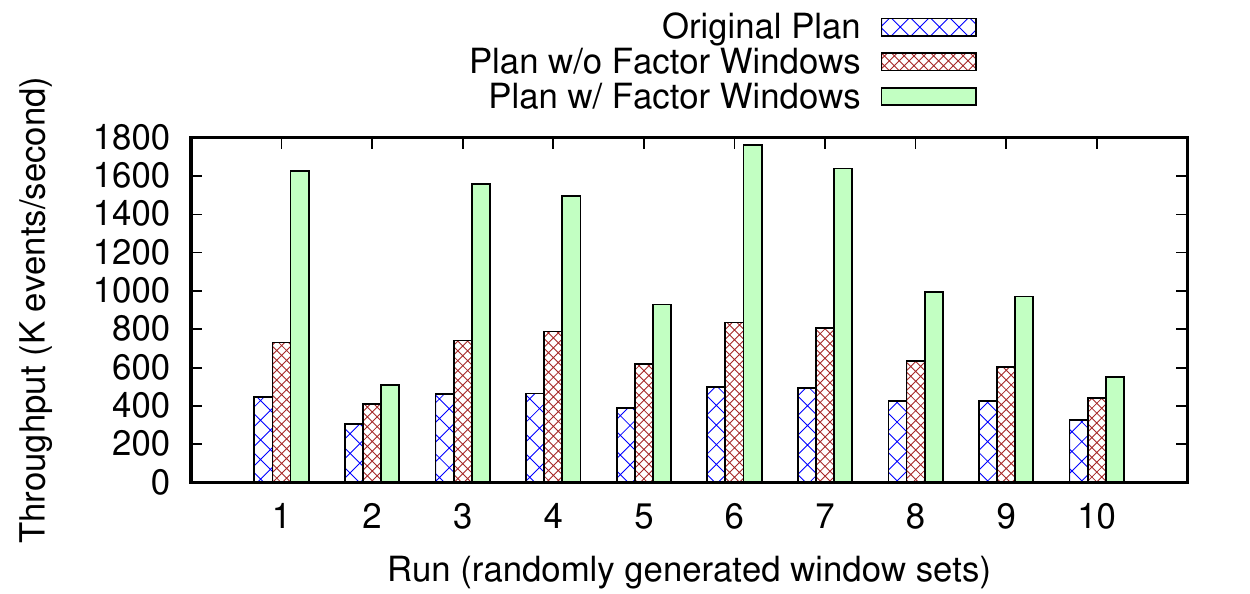}}
\vspace{-0.5em}
\caption{\blue{Throughput on window sets when processing 1 million input events from \textbf{Synthetic-1M} with $|\mathcal{W}|=10$.}} 
\label{fig:throughput:ws10:1m}
\vspace{-1.5em}
\end{figure*}

\begin{figure*}
\centering
\subfigure[\textbf{RandomGen}, ``partitioned by'']{ \label{fig:DEBS:throughput:random:partitioned-by:W5}
    \includegraphics[width=0.45\textwidth]{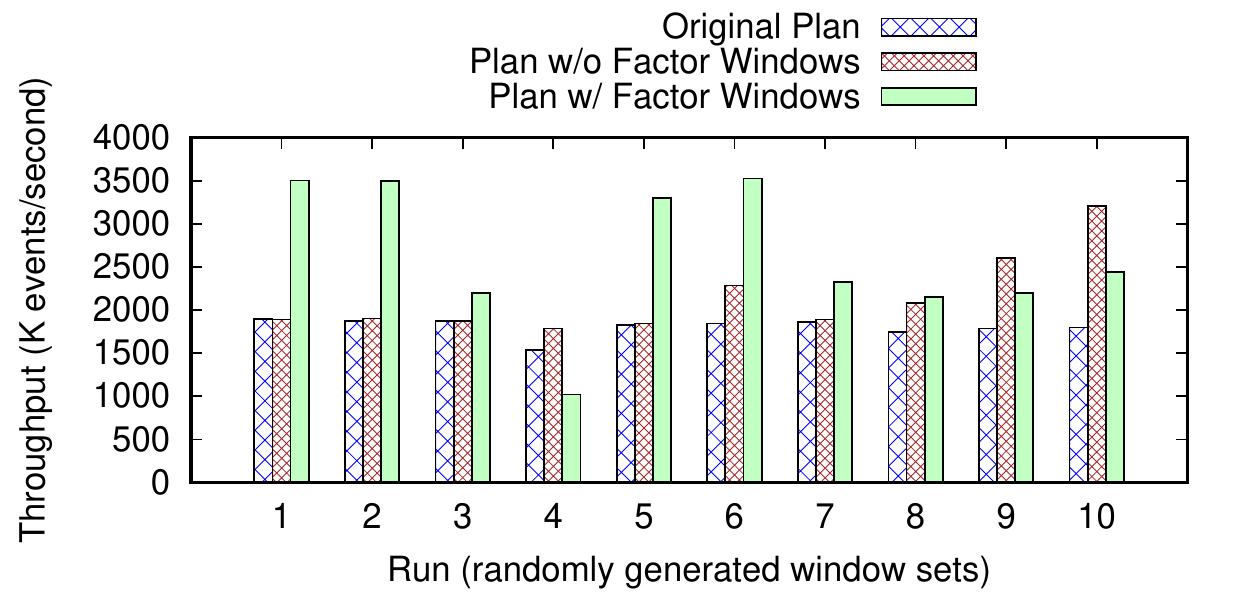}}
\subfigure[\textbf{RandomGen}, ``covered by'']{ \label{fig:DEBS:throughput:random:covered-by:W5}
    \includegraphics[width=0.45\textwidth]{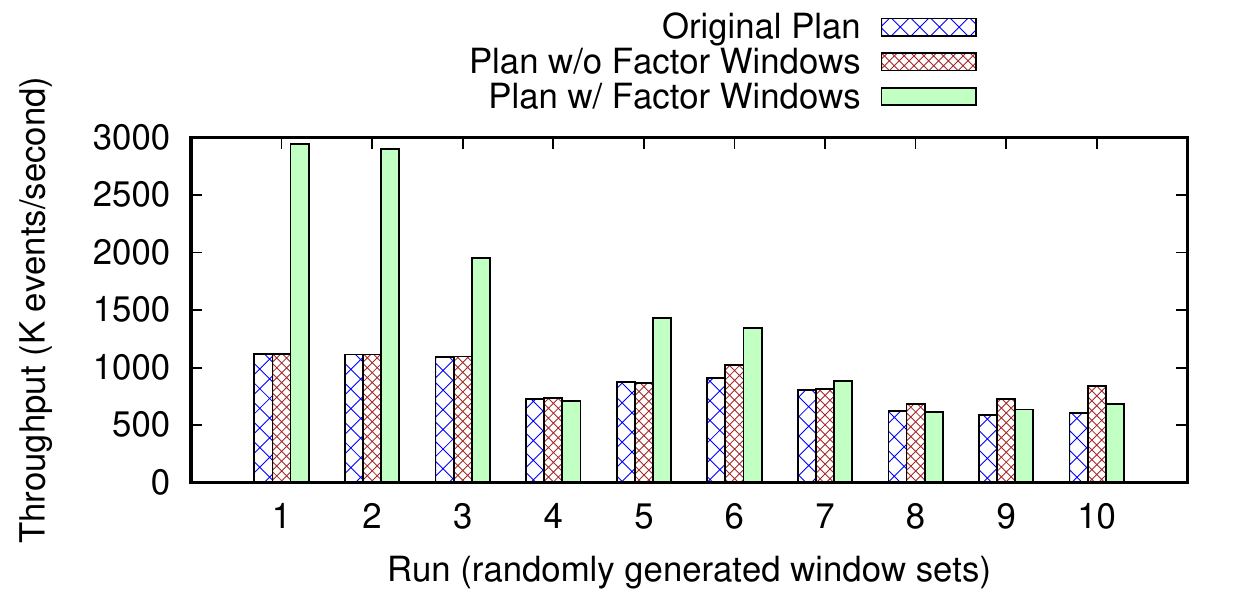}}
\subfigure[\textbf{SequentialGen}, ``partitioned by'']{ \label{fig:DEBS:throughput:seq:partitioned-by:W5}
    \includegraphics[width=0.45\textwidth]{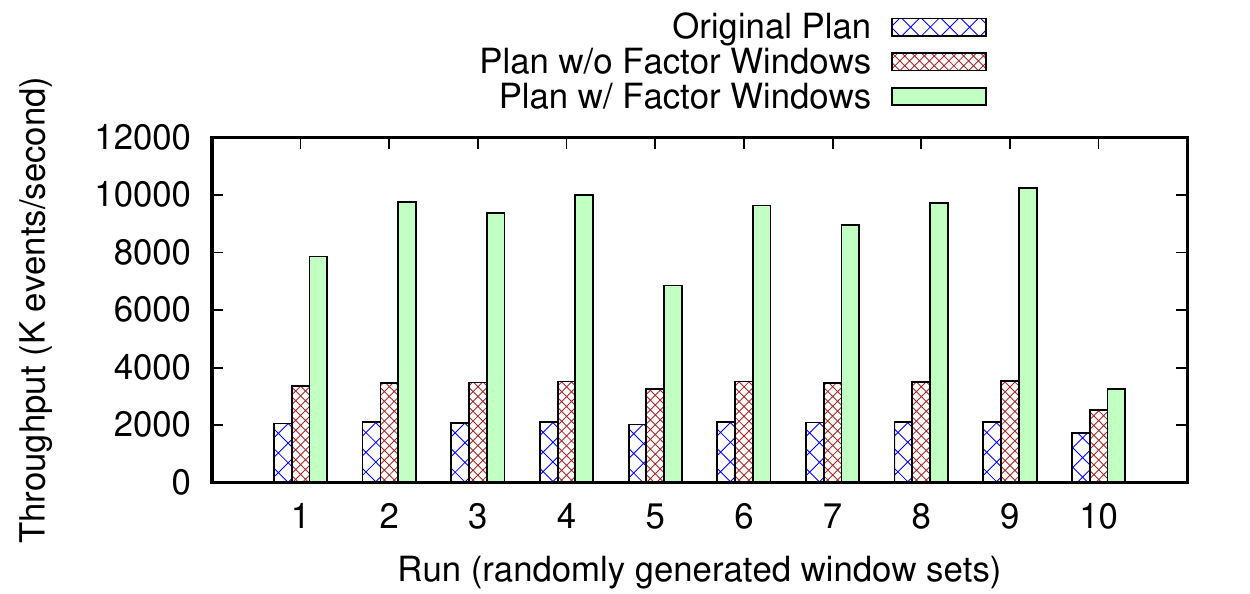}}
\subfigure[\textbf{SequentialGen}, ``covered by'']{ \label{fig:DEBS:throughput:seq:covered-by:W5}
    \includegraphics[width=0.45\textwidth]{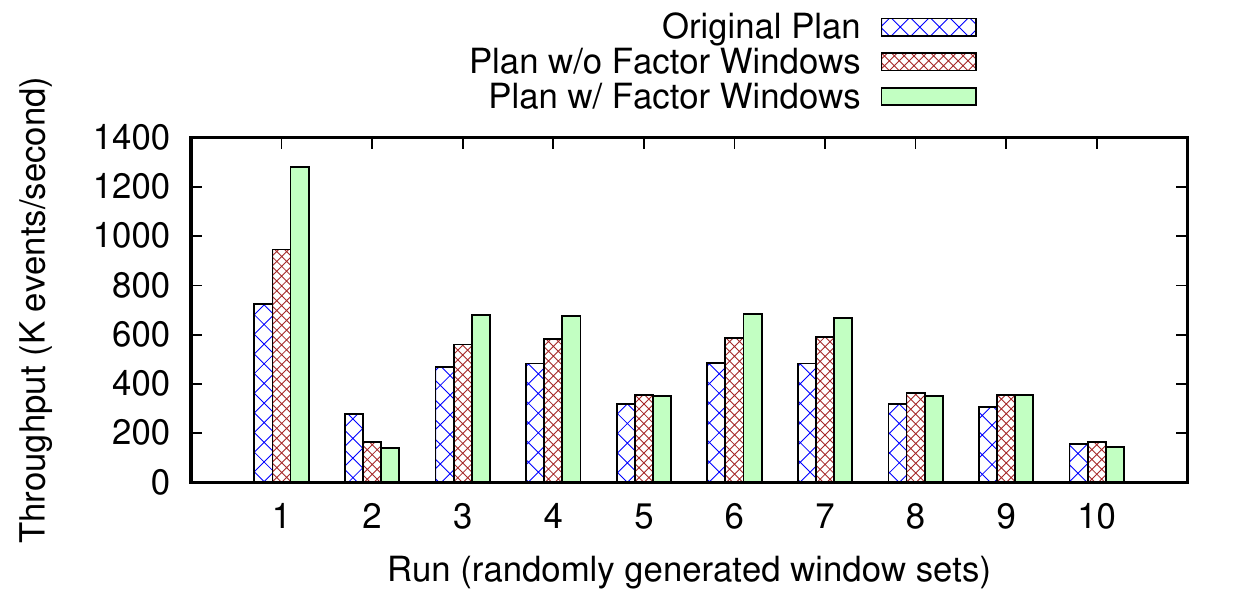}}
\vspace{-0.5em}
\caption{\blue{Throughput when processing 32 million input events from \textbf{Real-32M} with $|\mathcal{W}|=5$.}} 
\label{fig:DEBS:ws5}
\vspace{-1.5em}
\end{figure*}

\begin{figure*}[!ht]
\centering
\subfigure[\textbf{RandomGen}, ``partitioned by'']{ \label{fig:DEBS:throughput:random:partitioned-by:W10}
    \includegraphics[width=0.45\textwidth]{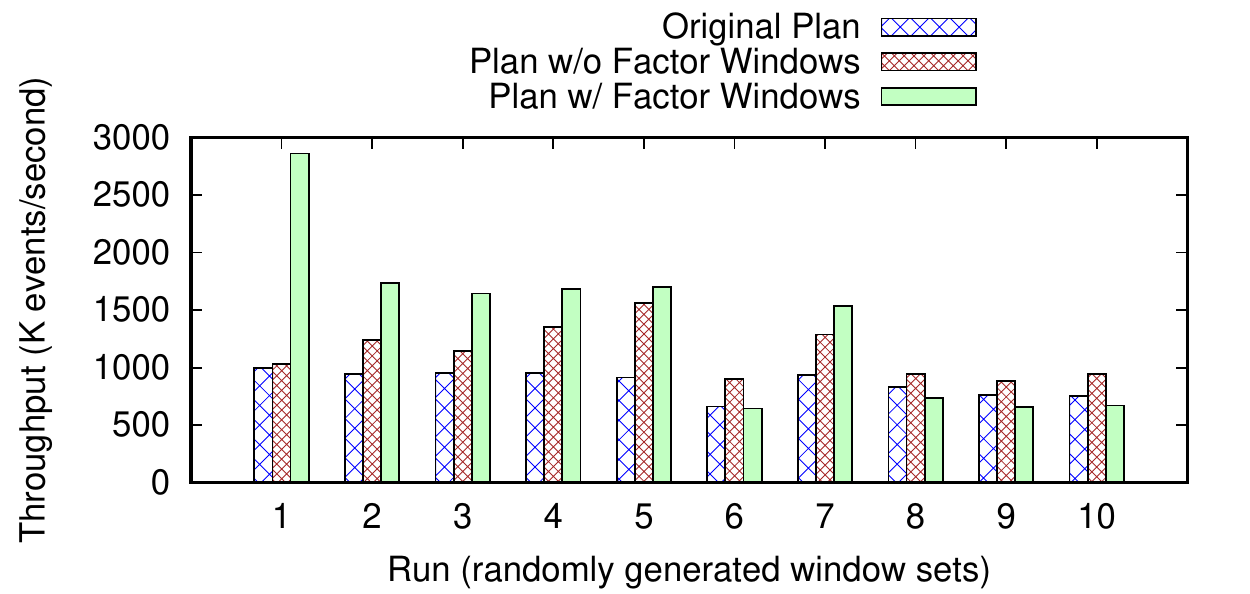}}
\subfigure[\textbf{RandomGen}, ``covered by'']{ \label{fig:DEBS:throughput:random:covered-by:W10}
    \includegraphics[width=0.45\textwidth]{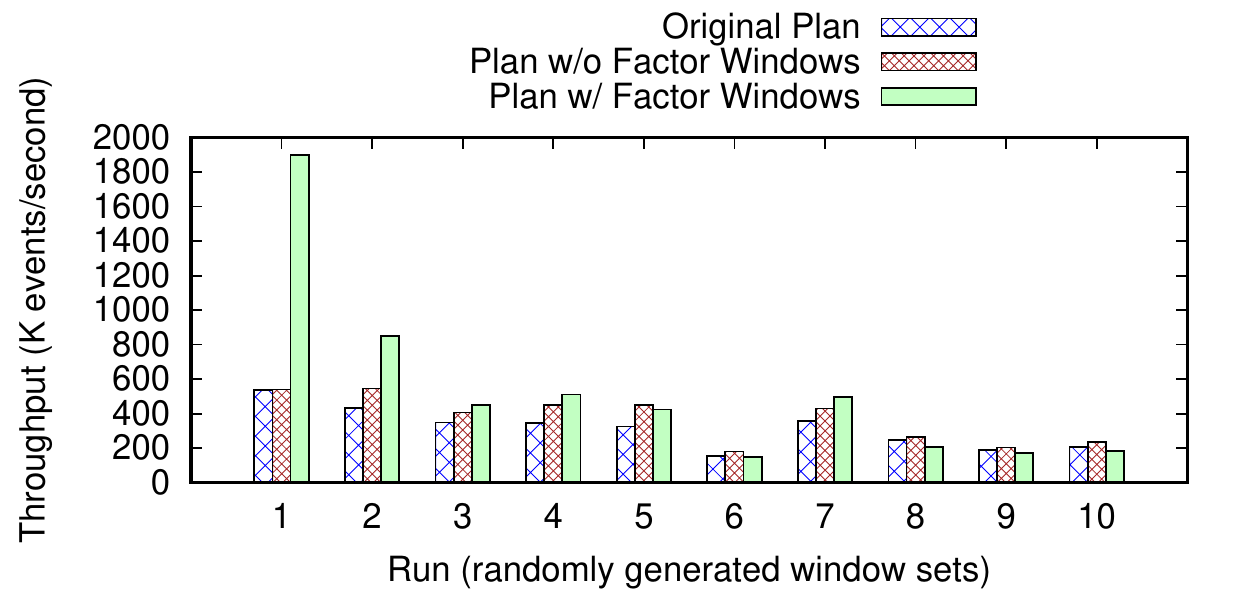}}
\subfigure[\textbf{SequentialGen}, ``partitioned by'']{ \label{fig:DEBS:throughput:seq:partitioned-by:W10}
    \includegraphics[width=0.45\textwidth]{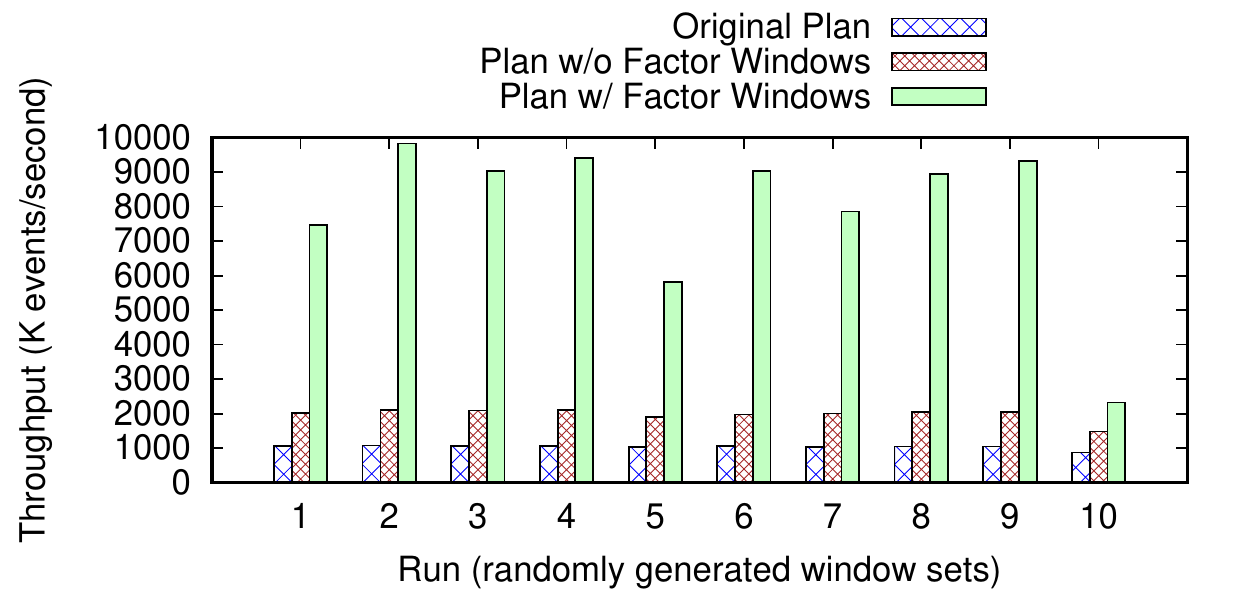}}
\subfigure[\textbf{SequentialGen}, ``covered by'']{ \label{fig:DEBS:throughput:seq:covered-by:W10}
    \includegraphics[width=0.45\textwidth]{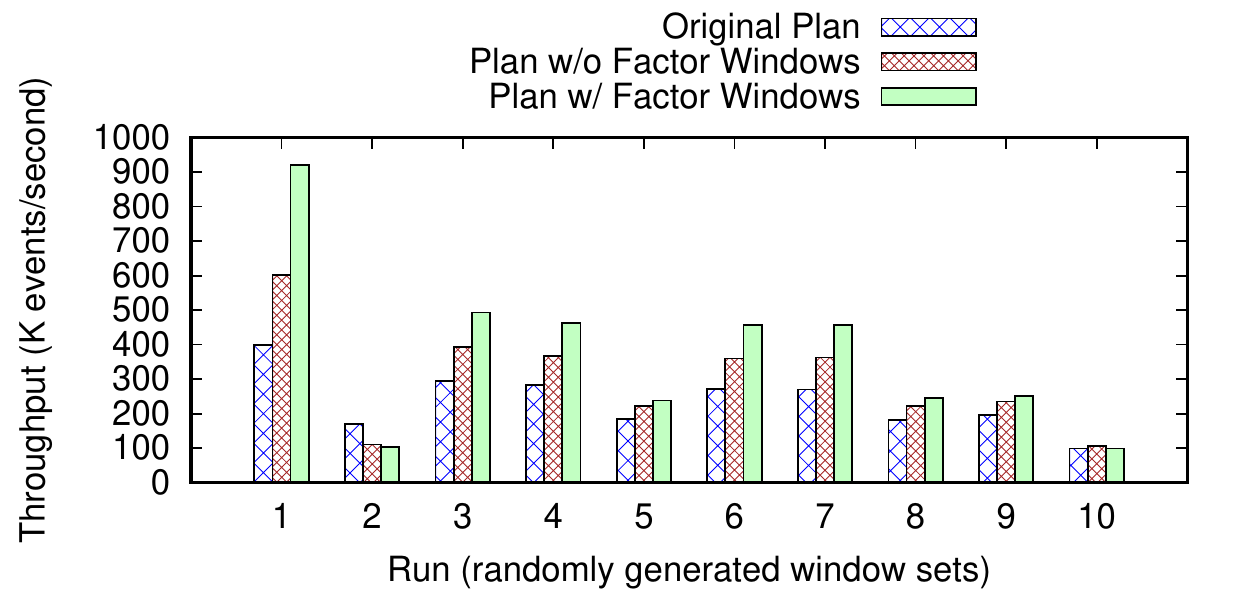}}
\vspace{-0.5em}
\caption{\blue{Throughput when processing 32 million input events from \textbf{Real-32M} with $|\mathcal{W}|=10$.}} 
\label{fig:DEBS:ws10}
\vspace{-1em}
\end{figure*}

\begin{figure*}[t]
\centering
\subfigure[\textbf{RandomGen}, ``partitioned by'' ($r=0.98$)]{ \label{fig:cc:random:partitioned-by}
    \includegraphics[width=0.4\textwidth]{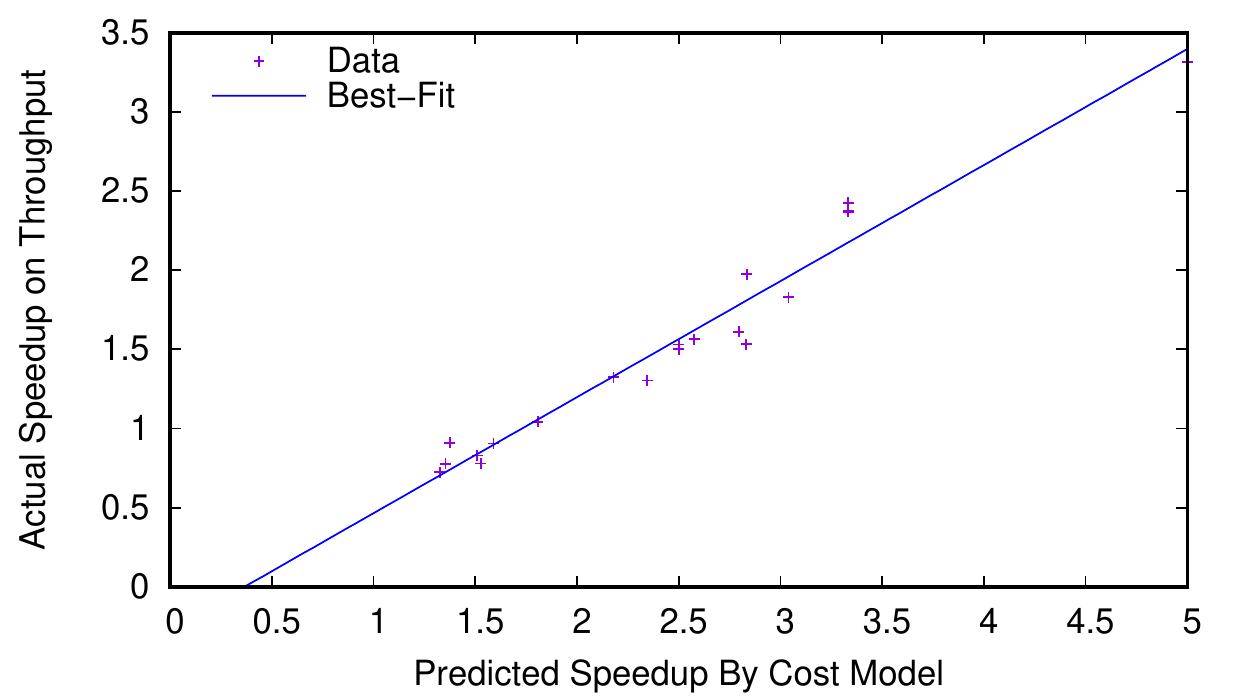}}
\subfigure[\textbf{RandomGen}, ``covered by'' ($r=0.95$)]{ \label{fig:cc:random:covered-by}
    \includegraphics[width=0.4\textwidth]{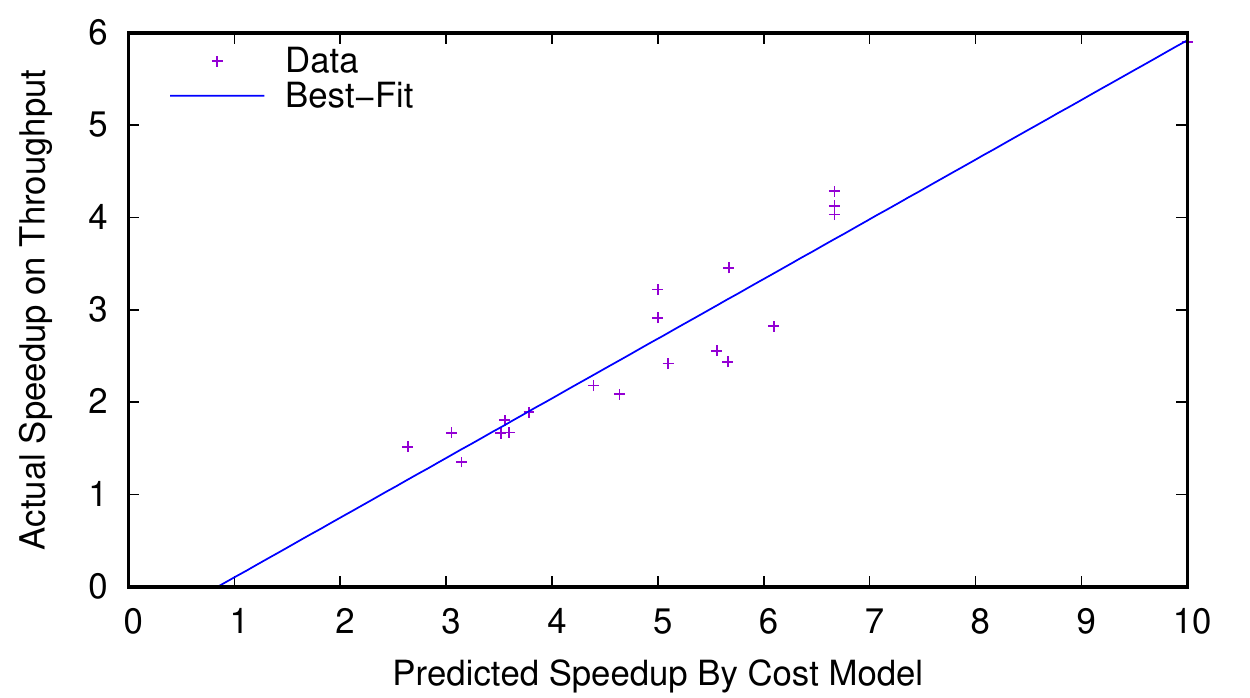}}
\subfigure[\textbf{SequentialGen}, ``partitioned by'' ($r=0.94$)]{ \label{fig:cc:sequential:partitioned-by}
    \includegraphics[width=0.4\textwidth]{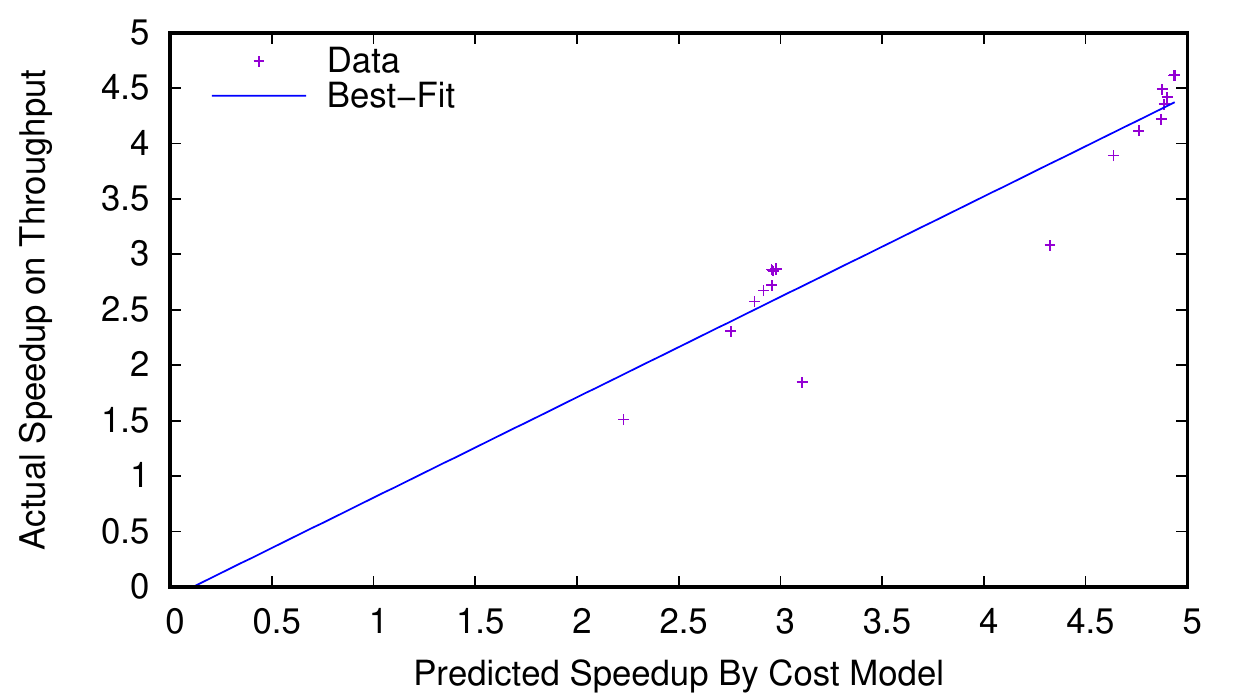}}
\subfigure[\textbf{SequentialGen}, ``covered by'' ($r=0.94$)]{ \label{fig:cc:sequential:covered-by}
    \includegraphics[width=0.4\textwidth]{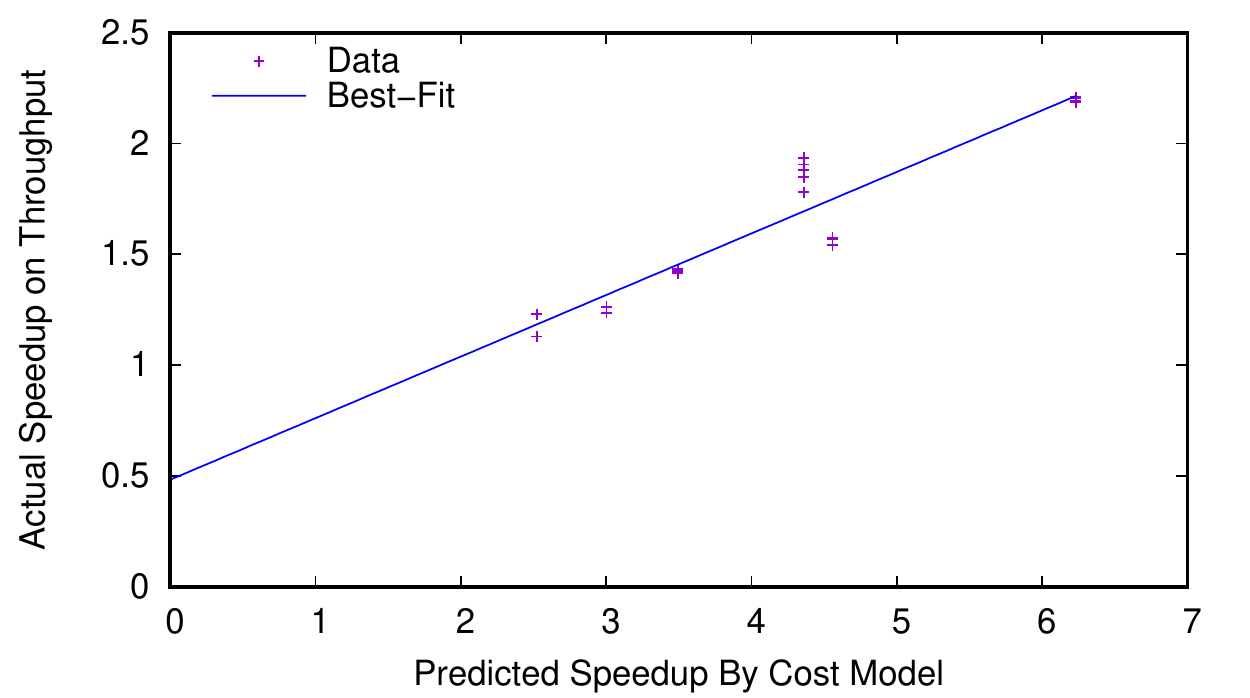}}
\vspace{-0.5em}
\caption{Correlation between predicted `speedup' by the cost model and observed `speedup' on throughput over \textbf{Synthetic-10M}.}
\label{fig:correlation}
\vspace{-1em}
\end{figure*}

\begin{figure*}[t]
\centering
\subfigure[\textbf{RandomGen}, ``partitioned by'']{ \label{fig:scalability:random:partitioned-by:W15}
    \includegraphics[width=0.45\textwidth]{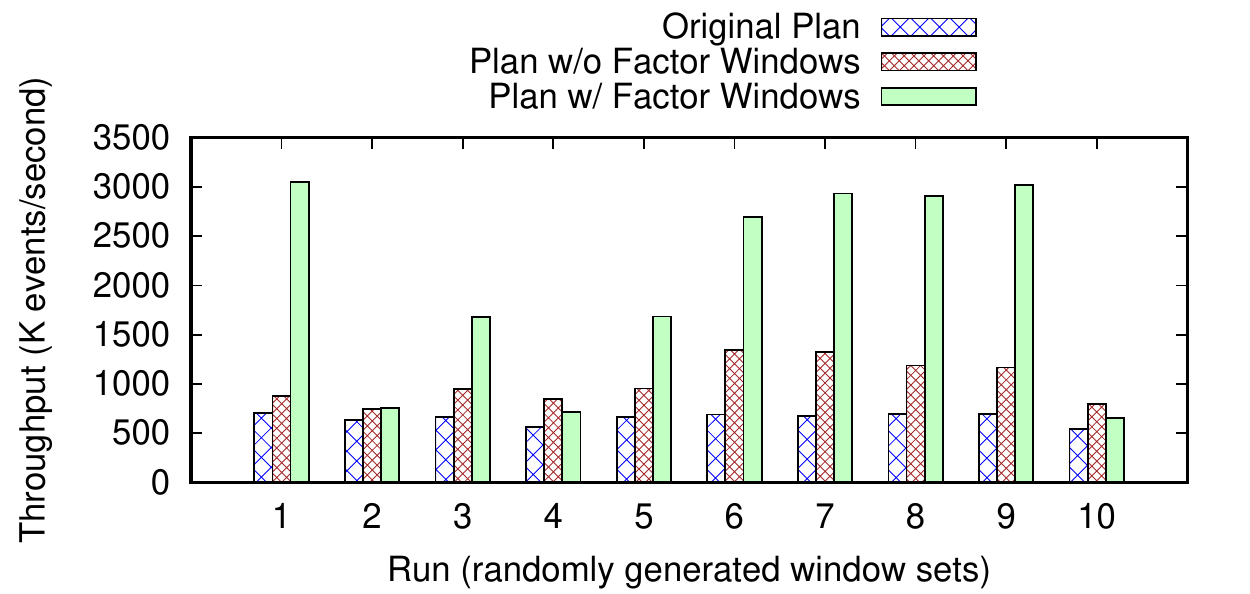}}
\subfigure[\textbf{RandomGen}, ``covered by'']{ \label{fig:scalability:random:covered-by:W15}
    \includegraphics[width=0.45\textwidth]{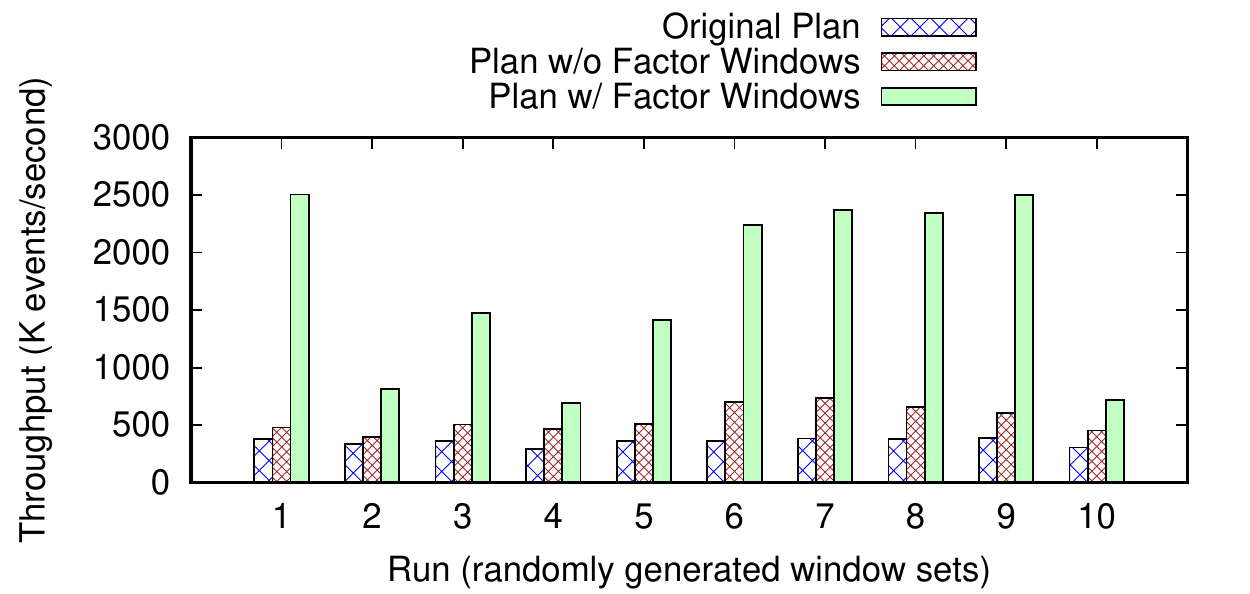}}
\subfigure[\textbf{SequentialGen}, ``partitioned by'']{ \label{fig:scalability:seq:partitioned-by:W15}
    \includegraphics[width=0.45\textwidth]{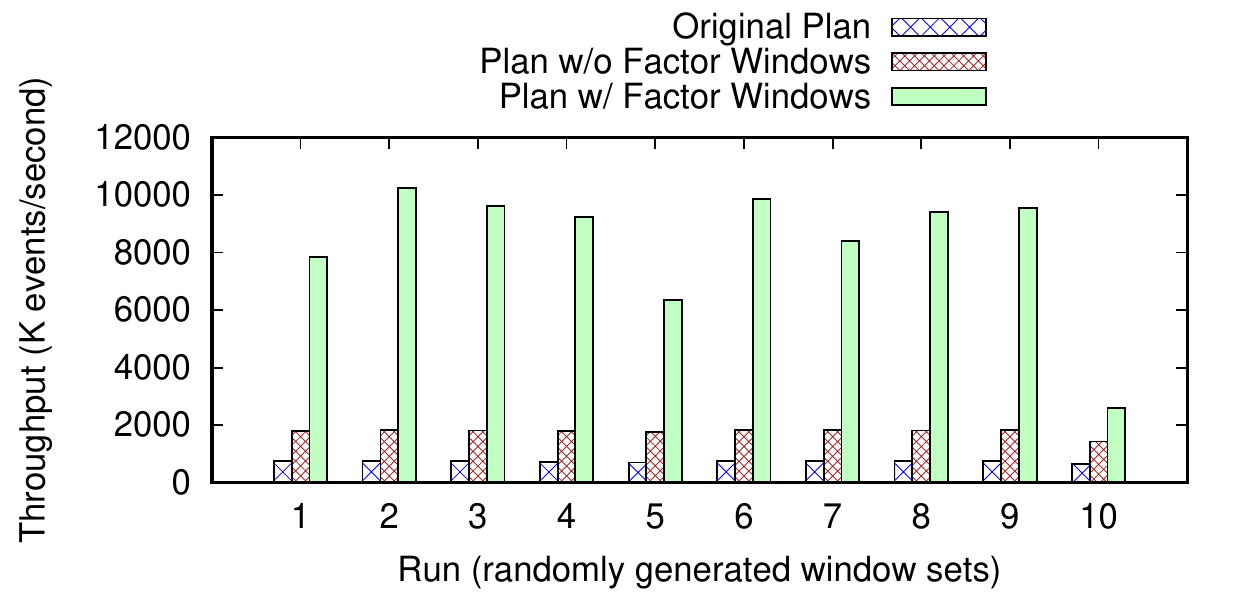}}
\subfigure[\textbf{SequentialGen}, ``covered by'']{ \label{fig:scalability:seq:covered-by:W15}
    \includegraphics[width=0.45\textwidth]{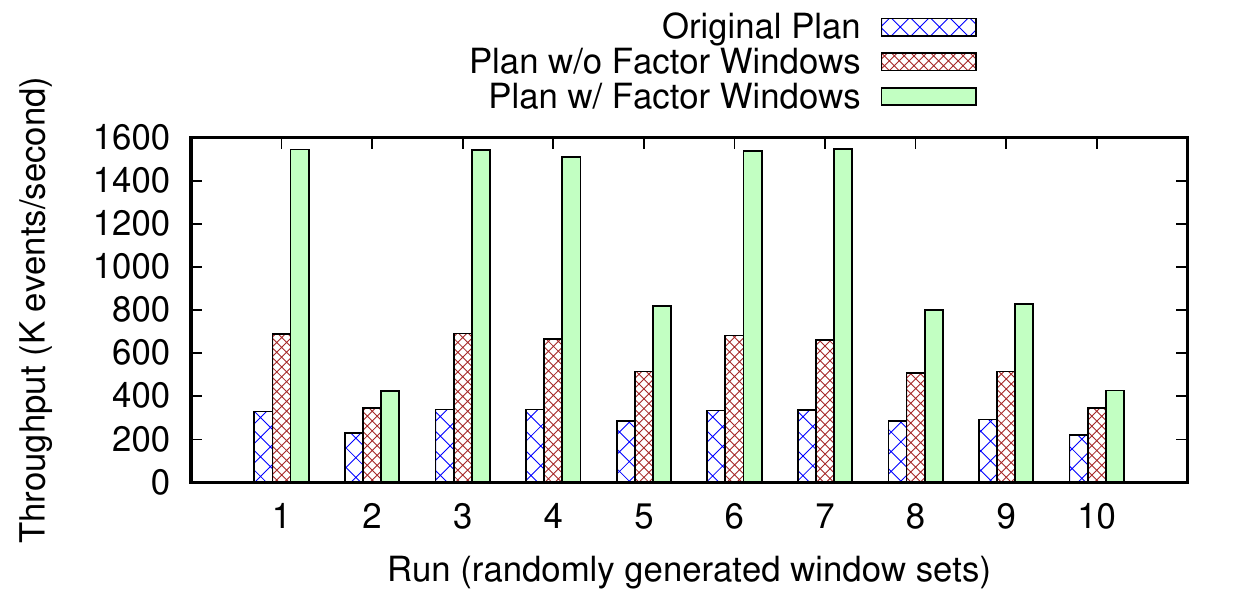}}
\vspace{-0.5em}
\caption{\blue{Throughput on window sets when processing 10 million input events from \textbf{Synthetic-10M} with $|\mathcal{W}|=15$.}}
\label{fig:scalability:ws15}
\vspace{-1em}
\end{figure*}

\begin{figure*}[!ht]
\centering
\subfigure[\textbf{RandomGen}, ``partitioned by'']{ \label{fig:scalability:random:partitioned-by:W20}
    \includegraphics[width=0.45\textwidth]{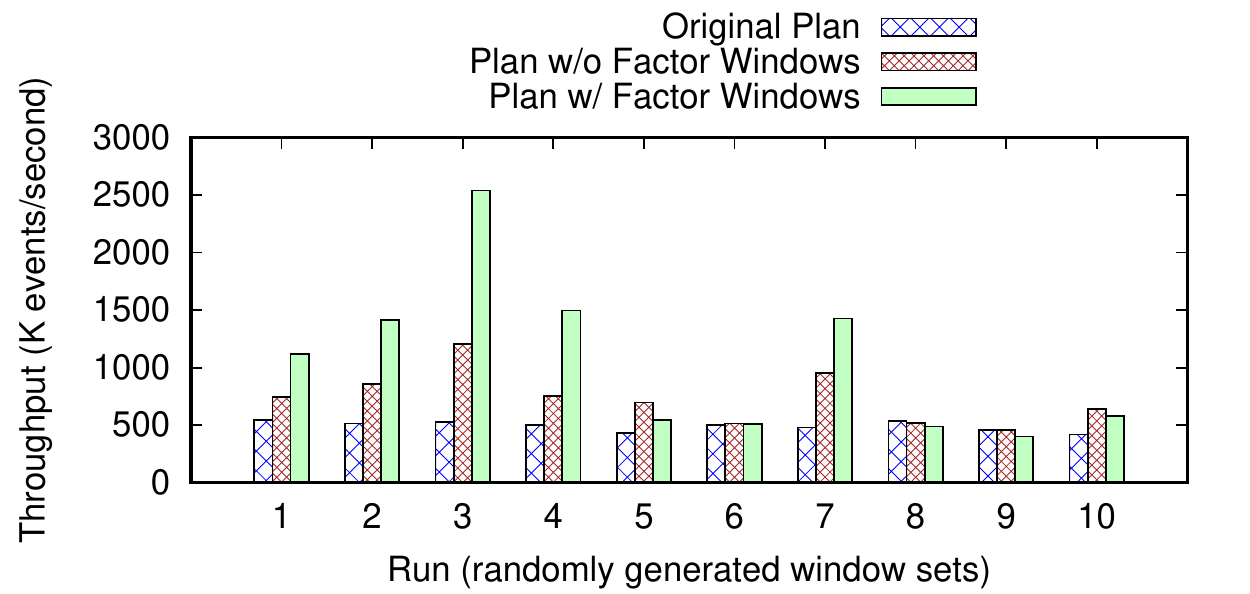}}
\subfigure[\textbf{RandomGen}, ``covered by'']{ \label{fig:scalability:random:covered-by:W20}
    \includegraphics[width=0.45\textwidth]{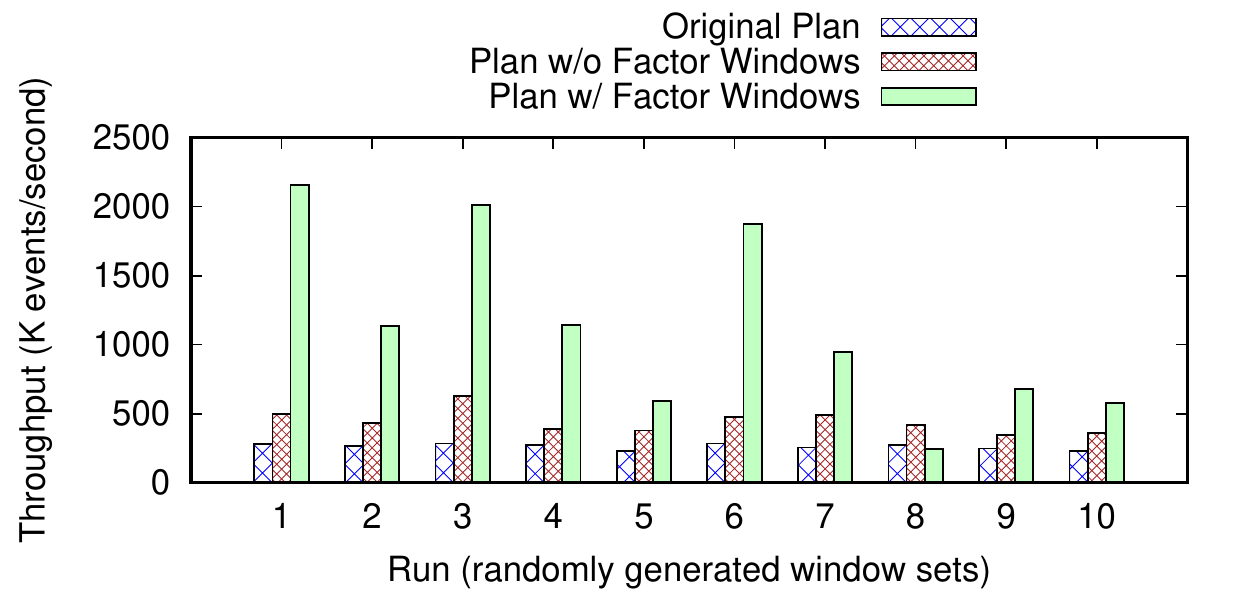}}
\subfigure[\textbf{SequentialGen}, ``partitioned by'']{ \label{fig:scalability:seq:partitioned-by:W20}
    \includegraphics[width=0.45\textwidth]{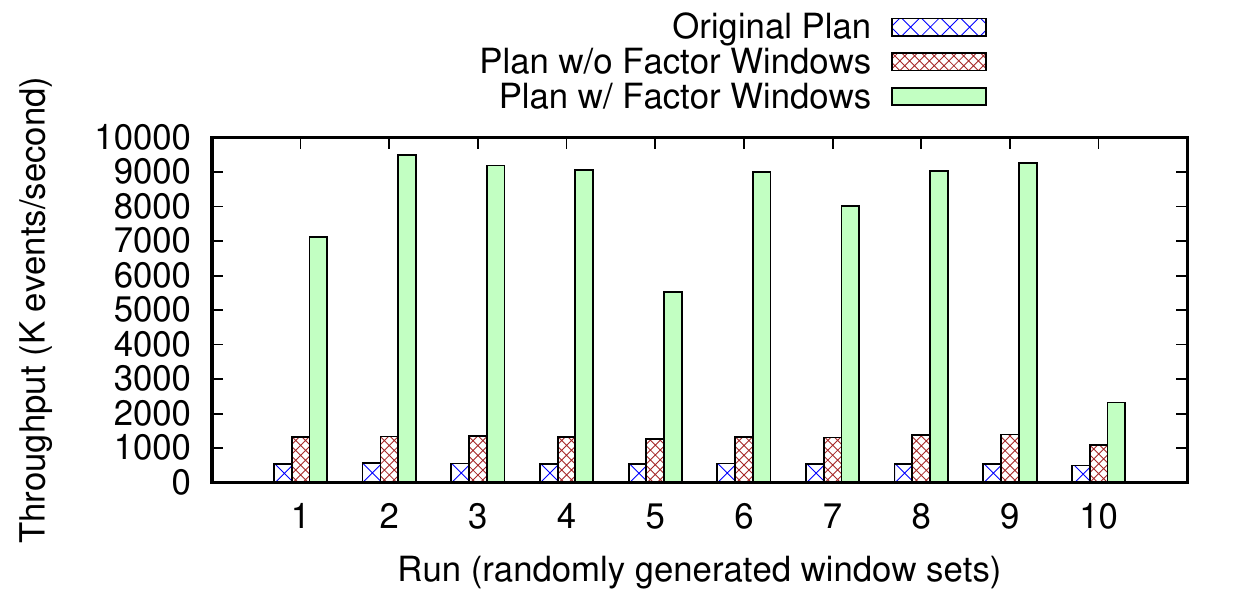}}
\subfigure[\textbf{SequentialGen}, ``covered by'']{ \label{fig:scalability:seq:covered-by:W20}
    \includegraphics[width=0.45\textwidth]{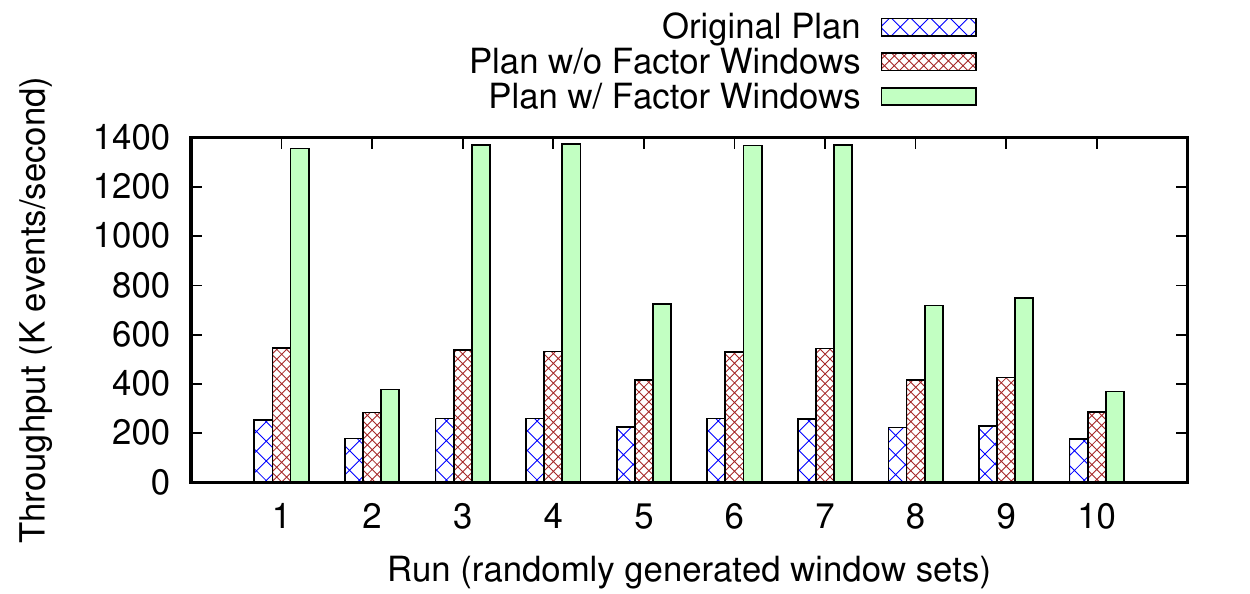}}
\vspace{-0.5em}
\caption{\blue{Throughput on window sets when processing 10 million input events from \textbf{Synthetic-10M} with $|\mathcal{W}|=20$.}} 
\label{fig:scalability:ws20}
\vspace{-1.5em}
\end{figure*}

\begin{figure*}[t]
\centering
\subfigure[\textbf{RandomGen}, ``partitioned by'']{ \label{fig:scotty-compare:random:partitioned-by:W5}
    \includegraphics[width=0.45\textwidth]{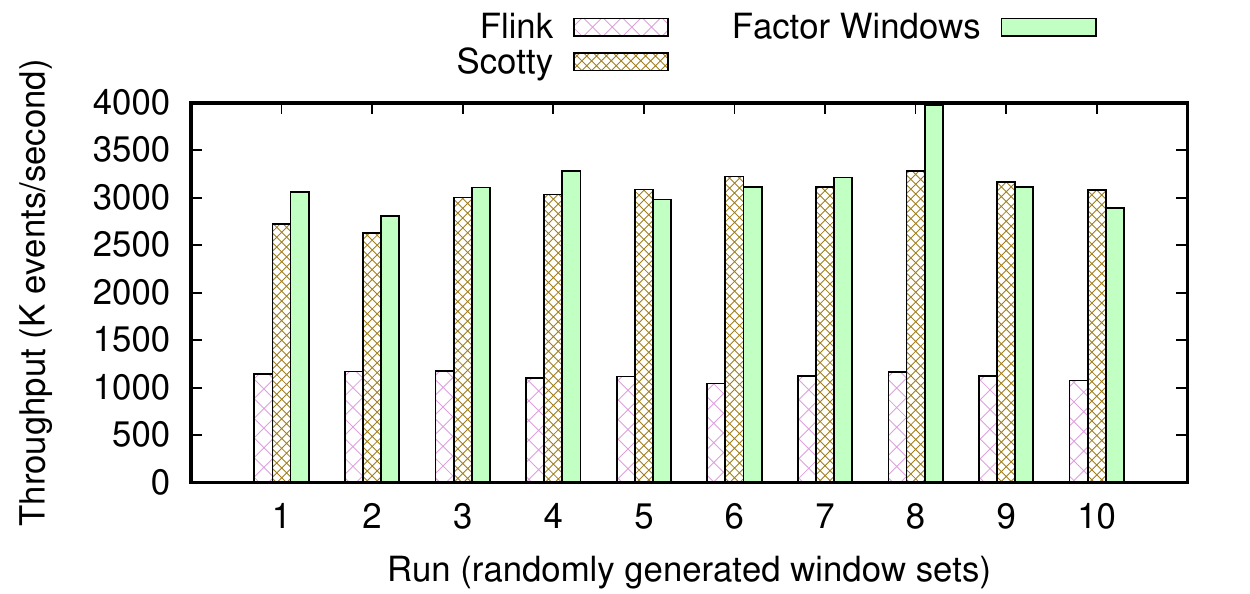}}
\subfigure[\textbf{RandomGen}, ``covered by'']{ \label{fig:scotty-compare:random:covered-by:W5}
    \includegraphics[width=0.45\textwidth]{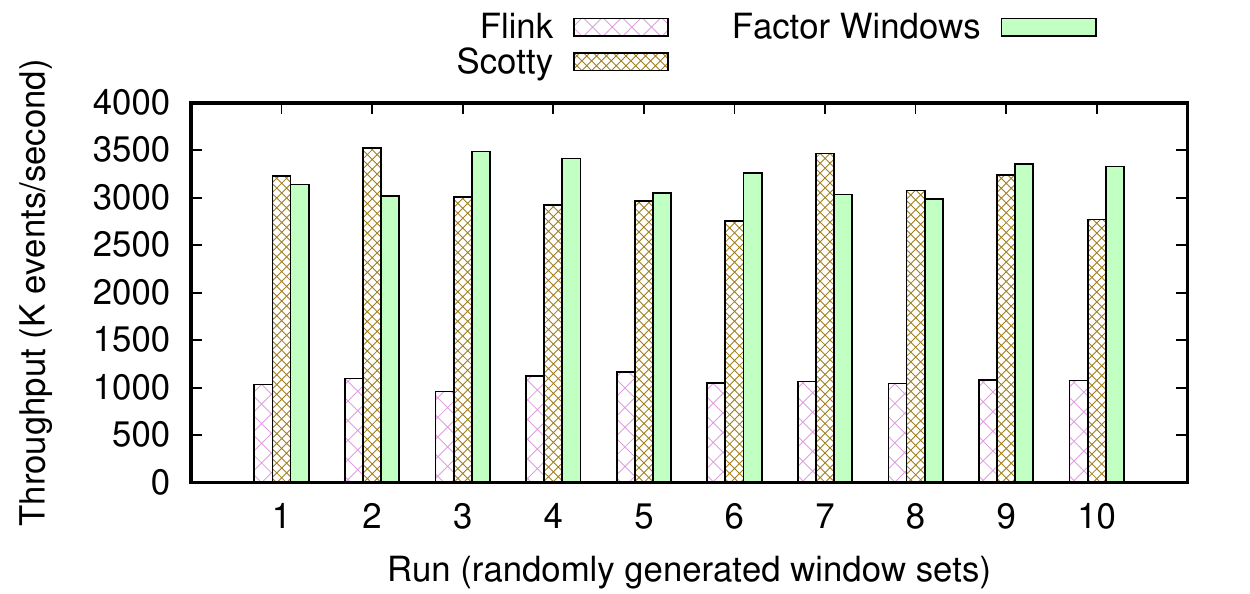}}
\subfigure[\textbf{SequentialGen}, ``partitioned by'']{ \label{fig:scotty-compare:seq:partitioned-by:W5}
    \includegraphics[width=0.45\textwidth]{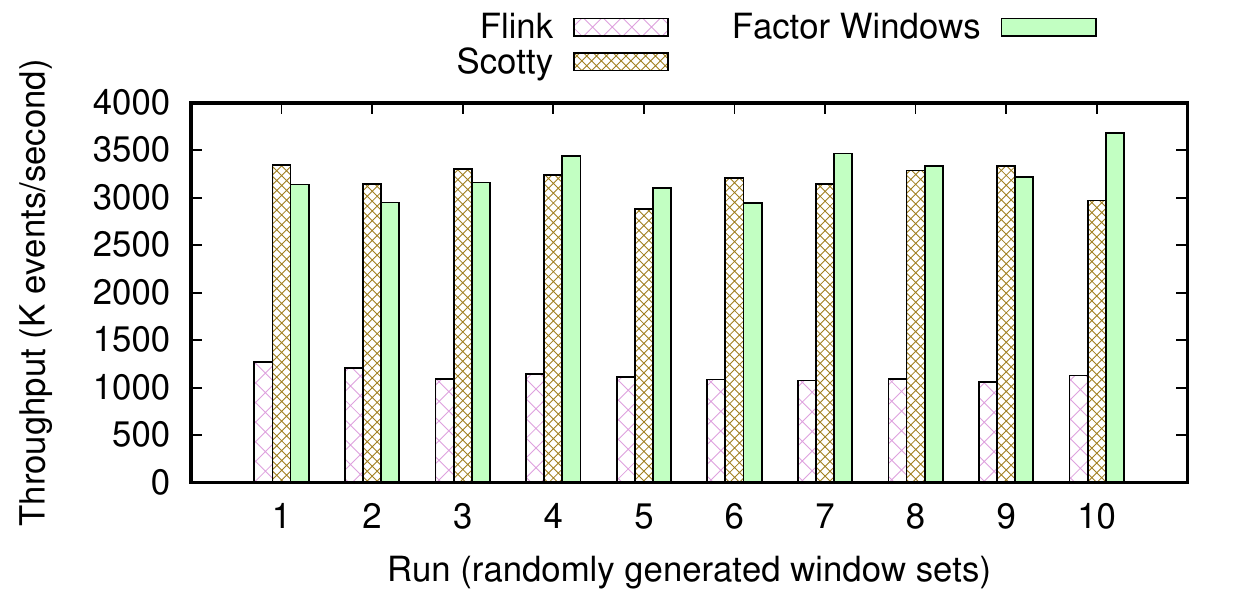}}
\subfigure[\textbf{SequentialGen}, ``covered by'']{ \label{fig:scotty-compare:seq:covered-by:W5}
    \includegraphics[width=0.45\textwidth]{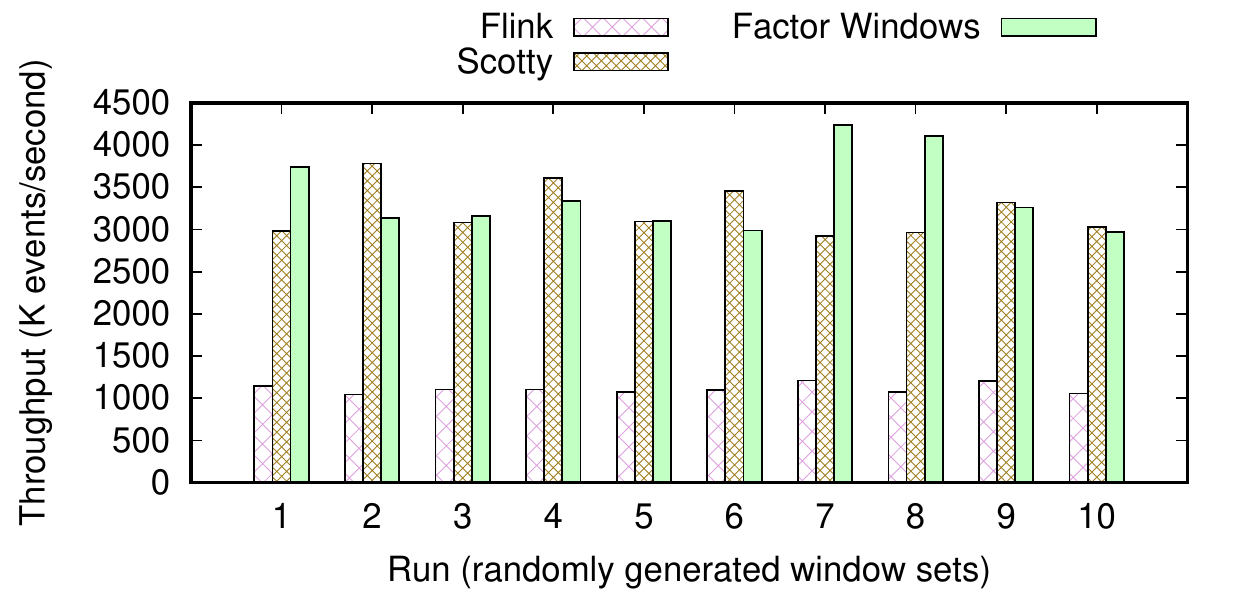}}
\vspace{-0.5em}
\caption{\blue{Comparison with Scotty~\cite{TraubGCBKRM21} in terms of throughput on window sets with $|\mathcal{W}|=5$.}}
\label{fig:scotty-compare:ws5}
\vspace{-1em}
\end{figure*}

\section{Proofs}

\subsection{Proof of Theorem~\ref{theorem:window-coverage}}\label{proof:theorem:window-coverage}

\begin{proof}
Consider an arbitrary interval $I=[a,b)\in W_1$. By the interval representation of $W_1$, we have $a=m_1\cdot s_1$ and $b=m_1\cdot s_1 + r_1$ for some integer $m_1\geq 0$.
\begin{enumerate}
    \item \textbf{The ``if'' part} $\Rightarrow$: Since $s_1$ is a multiple of $s_2$, we have $s_1=k\cdot s_2$ for some integer $k\geq 1$. As a result,
    $$m_1\cdot s_1 = m_1 \cdot k\cdot s_2 = (m_1\cdot k) \cdot s_2.$$
    Similarly, since $\delta_r=r_1-r_2$ is a multiple of $s_2$, $r_1-r_2=k'\cdot s_2$ for some integer $k'\geq 1$. As a result,
    \begin{eqnarray*}
    m_1\cdot s_1 + r_1 &=& (m_1\cdot k)\cdot s_2 + k'\cdot s_2 + r_2\\
    &=&(m_1\cdot k + k')\cdot s_2 + r_2.
    \end{eqnarray*}
    Set $m_2=m_1\cdot k$ and $m'_2=m_1\cdot k + k'$. Now consider two intervals $I_a=[a, x) = [m_2\cdot s_2, m_2\cdot s_2 + r_2)$ and $I_b=[y, b)=[m'_2\cdot s_2, m'_2\cdot s_2 + r_2)$ that belong to $W_2$.
    Clearly, we have
    $$m_2\cdot s_2=m_1\cdot s_1 = a$$
    and $$m'_2\cdot s_2 + r_2=m_1\cdot s_1 + r_1=b.$$
    Moreover, since $m'_2 > m_2$, we have
    $x=m_2\cdot s_2 + r_2 < b$ and $y=m'_2\cdot s_2 > a$.
    Therefore, $W_1$ is covered by $W_2$, by Definition~\ref{definition:window-coverage}.
    \item \textbf{The ``only if'' part} $\Leftarrow$: Since $W_1$ is covered by $W_2$, by Definition~\ref{definition:window-coverage} there exist two intervals $I_a=[a, x)$ and $I_b=[y, b)$ in $W_2$ such that $x<b$ and $y>a$.
    As a result, there is some $m_2\geq 0$ such that $m_2\cdot s_2 = a = m_1\cdot s_1$. That is,
    $$m_2 = m_1 \cdot (s_1 / s_2).$$
    Since both $m_1$ and $m_2$ are integers, $s_1 / s_2$ is also an integer. As a result, $s_1$ must be a multiple of $s_2$.

    On the other hand, similarly there is some $m'_2> m_2$ such that
    $$m'_2\cdot s_2+r_2=b=m_1\cdot s_1 + r_1.$$
    We then have
    $$m'_2\cdot s_2 + r_2 = m_2\cdot s_2 + r_1,$$ which yields
    $$m'_2=m_2 + (r_1-r_2) / s_2.$$
    Since both $m'_2$ and $m_2$ are integers, $(r_1-r_2)/s_2$ must be an integer.
    Hence, $\delta_r=r_1-r_2$ is a multiple of $s_2$.
\end{enumerate}
This completes the proof of the theorem.
\end{proof}

\subsection{Proof of Theorem~\ref{theorem:transitivity}}\label{proof:theorem:transitivity}

\begin{proof}
We prove the three properties one by one.
\begin{enumerate}
    \item \textbf{Reflexivity}: Clearly, by Definition~\ref{theorem:window-coverage} a window $W$ is covered by itself.
    \item \textbf{Antisymmetry}: Suppose that $W_1\leq W_2$ and $W_2\leq W_1$. Consider an arbitrary interval $[a, b)$ contained by $W_1$. Since $W_1\leq W_2$, there exist two intervals $I_x=[a, x)$ and $I_y=[y, b)$ in $W_2$. On the other hand, since $W_2\leq W_1$, for $I_x$ there exist intervals $I_{x'}=[a, x')$ and $I_{x''}=[x'', x)$ in $W_1$. Since no two intervals in a window start from the same time point but end at different time points, we conclude that
    $$x'=b.$$
    Since $x'\leq x\leq b$ by Definition~\ref{definition:window-coverage}, we have $$x=x'=x''=b.$$ Using similar arguments we can show that $y=y'=y''=a$. As a result, we have proved that $W_1=W_2$.
    \item \textbf{Transitivity}: Suppose that $W_1\leq W_2$ and $W_2\leq W_3$. Again, consider an arbitrary interval $[a, b)$ in $W_1$. Since $W_1\leq W_2$, there exist two intervals $I_x=[a, x)$ and $I_y=[y, b)$ in $W_2$. Moreover, since $W_2\leq W_3$, there exist two intervals $I_{x'}=[a, x')$ and $I_{x''}=[x'', x)$ in $W_3$, and there also exist two intervals $I_{y'}=[y, y')$ and $I_{y''}=[y'', b)$ in $W_3$. Now consider $I_{x'}$ and $I_{y''}$. By Definition~\ref{definition:window-coverage}, we have $x'\leq x\leq b$ and $y''\geq y\geq a$.
    Since $[a, b)$ is an arbitrary interval in $W_1$, it follows that $W_1\leq W_3$.
\end{enumerate}
This completes the proof of the theorem.
\end{proof}

\subsection{Proof of Theorem~\ref{theorem:covering-multiplier}}\label{proof:theorem:covering-multiplier}

\begin{proof}
If we take a union of the intervals in $\mathcal{I}_{a,b}$, it is easy to see $I=\cup_{J\in\mathcal{I}_{a,b}}J$.
By Definition~\ref{definition:window-coverage}, we can further enumerate the intervals in $\mathcal{I}_{a,b}$ as $J_1=[x_1, y_1)$, ..., $J_n=[x_n, y_n)$ such that $x_1=a$, $y_n=b$, and $x_1 < \cdots < x_n$, where $n=|\mathcal{I}_{a,b}|$.
Therefore,
$$I=J_1\cup (J_2-J_1)\cup\cdots\cup (J_n-J_{n-1}).$$
Since the intervals $J_1$, $J_2-J_1$, ..., $J_n-J_{n-1}$ are mutually exclusive, it follows that
$$|I|=|J_1|+ |J_2-J_1|+\cdots + |J_n-J_{n-1}|.$$
We have $|I|=r_1$, $|J_1|=r_2$, and $|J_k-J_{k-1}|=s_2$ for $2\leq k\leq n$.
As a result, $r_1 = r_2 + (n-1)\cdot s_2,$
which yields $$M(W_1, W_2)=n=1+(r_1-r_2)/s_2.$$
This completes the proof of the theorem.
\end{proof}

\subsection{Proof of Theorem~\ref{theorem:window-partitioning}}\label{proof:theorem:window-partitioning}

\begin{proof}
We prove each direction separately.
\begin{enumerate}
    \item[(a)] \textbf{The ``if'' part} $\Rightarrow$: Suppose that conditions (1) to (3) hold. By (2) and (3), we know that $r_1-r_2$ must be a multiple of $s_2$ either. Combining with (1), $W_1$ is covered by $W_2$ according to Theorem~\ref{theorem:window-coverage}. Now consider an arbitrary interval $I$ in $W_1$. Let the covering set of $I$ in $W_2$ be $\mathcal{I}$.
    We next show that $\mathcal{I}$ is disjoint.
    By (2) and (3) we know that $r_1$ is a multiple of $r_2$. As a result, $r_1=k\cdot r_2$ where $k$ is an integer. To show that $\mathcal{I}$ is disjoint we only need to show that $|\mathcal{I}|=k$ (recall Figure~\ref{fig:window:partition}). We have
    \begin{eqnarray*}
    |\mathcal{I}|&=&1+(r_1-r_2)/s_2,\qquad \text{[by Theorem~\ref{theorem:covering-multiplier}]} \\
    &=&1 + (k\cdot r_2 - r_2) / s_2,\qquad \text{[by Condition (2)]}\\
    &=&1 + (k - 1),\qquad \text{[by Condition (3)]}\\
    &=&k.
    \end{eqnarray*}
    \item[(b)] \textbf{The ``only if'' part} $\Leftarrow$: Suppose that $W_1$ is partitioned by $W_2$. By Theorem~\ref{theorem:window-coverage}, condition (1) holds.
    Again, consider an arbitrary interval $I$ in $W_1$ and let its covering set in $W_2$ be $\mathcal{I}$. We know that $\mathcal{I}$ is disjoint, which implies condition (3), i.e., $r_2=s_2$, as well as that $r_1$ must be a multiple of $r_2$.
    Therefore, $r_1$ must also be a multiple of $s_2$ and condition (2) holds.
\end{enumerate}
This completes the proof of the theorem.
\end{proof}

\subsection{Proof of Theorem~\ref{theorem:distributive:min-max}}\label{proof:theorem:distributive:min-max}

\begin{proof}
We only prove \verb|MIN| is distributive over overlapping partitions, as the proof for \verb|MAX| is very similar.
We set both $f$ and $g$ in the definition of distributive aggregate function as \verb|MIN|.
It is easy to see that, if two sets $S_1$ and $S_2$ satisfying $S_1\subseteq S_2$, then $\verb|MIN|(S_2)\leq\verb|MIN|(S_1)$.\footnote{We treat each element in $T$ differently, even if some of them may have the same data value.}
Moreover, for any set $S$, $\verb|MIN|(S)\in S$ and thus $\{\verb|MIN|(S)\}\subseteq S$.
Therefore,
$$S=\{\verb|MIN|(T_1), ..., \verb|MIN|(T_n)\}\subseteq T_1\cup\cdots\cup T_n,$$
since $\verb|MIN|(T_1)\subseteq T_1$, ..., $\verb|MIN|(T_n)\subseteq T_n$.
As a result,
$$\verb|MIN|(T)\leq\verb|MIN|(S)=\verb|MIN|(\{\verb|MIN|(T_1), ..., \verb|MIN|(T_n)\}).$$

We now prove that $\verb|MIN|(S)\leq\verb|MIN|(T)$.
To see this, let
\begin{eqnarray*}
S_1 &=& T_1,\\
S_2 &=& T_2-T_1, \\
S_3 &=& T_3-(S_1\cup S_2), \\
&...& \\
S_n &=& T_n-(S_1\cup\cdots\cup S_{n-1}).
\end{eqnarray*}
We have $T=S_1\cup\cdots\cup S_n$, and $S_i\cap S_j=\emptyset$ for all $1\leq i,j\leq n$.
Therefore, $\verb|MIN|(T)=\verb|MIN|(S_1\cup\cdots\cup S_n)$.
Moreover, there exists some $j$ such that $\verb|MIN|(S_j)=\verb|MIN|(T)$.
Since $S_j\subseteq T_j$, $\verb|MIN|(S_j)\geq\verb|MIN|(T_j)$.
As a result,
\begin{eqnarray*}
\verb|MIN|(T)&=&\verb|MIN|(\{\verb|MIN|(S_1), ...,\verb|MIN|(S_n)\})\\
&\geq &\verb|MIN|(\{\verb|MIN|(T_1), ...,\verb|MIN|(T_n)\})\\
&=&\verb|MIN|(S).
\end{eqnarray*}

Since we have proved both $\verb|MIN|(S)\leq\verb|MIN|(T)$ and $\verb|MIN|(T)\leq\verb|MIN|(S)$, it must hold that $\verb|MIN|(S)=\verb|MIN|(T)$.
\end{proof}

\subsection{Proof of Theorem~\ref{theorem:factor-window:partition-by}}

Since both $W_f$ and $W$ in Figure~\ref{fig:factor-window} are now tumbling windows, $k_f=k_W=1$. Equation~\ref{eq:general-condition} then yields
\begin{equation*}
\sum\nolimits_{j=1}^{K}\frac{n_j}{n_f}\Big(\frac{r_j}{s_f}-\frac{r_j}{s_W}\Big)+\frac{r_f}{s_W}\leq 0.
\end{equation*}
Since $r_f=s_f$ and $r_W=s_W$, it follows that
\begin{equation*}
\sum\nolimits_{j=1}^{K}\frac{n_j}{n_f}\Big(\rho_j-\frac{r_j}{r_W}\Big)+\frac{r_f}{r_W}\leq 0.
\end{equation*}
Since \HL{$r_f=\frac{r_j}{\rho_j}$} by definition, we have \HL{$r_j=\rho_j r_f$ and thus}
\begin{equation}\label{eq:final-condition}
\sum\nolimits_{j=1}^{K}\frac{n_j\rho_j}{n_f}\Big(1-\frac{r_f}{r_W}\Big)+\frac{r_f}{r_W}\leq 0.
\end{equation}
Moreover, by definition of $n_f$ (Equation~\ref{eq:recurrence-count}) we have
$$n_f=(m_f - 1)k_f + 1= m_f=\frac{R}{r_f}=\frac{R\rho_j}{r_j}=m_j\rho_j.$$
Substituting into Equation~\ref{eq:final-condition}, it follows that
\begin{equation}\label{eq:final-condition-simplified}
\Big(1-\frac{r_f}{r_W}\Big)\cdot \lambda +\frac{r_f}{r_W}\leq 0,
\end{equation}
where $\lambda$ has been defined in Equation~\ref{eq:lambda}.
As a result, we have
\begin{equation}\label{eq:condition-lambda}
\frac{r_f}{r_W} \geq \frac{\lambda}{\lambda-1}.
\end{equation}

Since $n_j=(m_j-1)k_j + 1\geq m_j$, by Equation~\ref{eq:lambda} we have $\lambda\geq K$.
We distinguish two cases: $K\geq 2$ and $K=1$.

\paragraph*{The Case of $K\geq 2$}
When $K\geq 2$ we have
$$\frac{\lambda}{\lambda - 1}\leq\frac{K}{K-1}\leq 2.$$
Since $\frac{r_f}{r_W}\geq 2$, Equation~\ref{eq:condition-lambda} holds, which implies $c\leq c'$.
Note that the equality $c=c'$
only holds when $r_f=2r_W$ \emph{and} $\lambda=K=2$, which implies $n_j=m_j$ for $j=1,2$.
In this case, both downstream windows of $W$ (and thus $W_f$) are tumbling, and $W_f$ exactly doubles the range of $W$, which is a very special case.

\paragraph*{The Case of $K=1$}
When $K = 1$, $\lambda=\frac{n_1}{m_1}$. We distinguish two situations:
\begin{itemize}
    \item If $k_1=1$, which means that the (unique) downstream window is tumbling, then $n_1=m_1$ and thus $\lambda=1$. Equation~\ref{eq:final-condition-simplified} then implies that $1\leq 0$, which is impossible. As a result, $c\leq c'$ does not hold.
    \item If $k_1>1$, then $\lambda>1$ and thus the RHS of Equation~\ref{eq:condition-lambda} is well-defined. Note that we must have $m_1 >1$, since if $m_1=1$ then $n_1=(m_1-1)k_1 + 1=1$ and thus $\lambda=1$, a contradiction. Substituting $\lambda=\frac{n_1}{m_1}$, we obtain
    \begin{eqnarray*}
    \frac{\lambda}{\lambda-1}&=&1+\frac{m_1}{n_1-m_1}\\
    &=&1+\frac{m_1}{(m_1-1)(k_1-1)}\\
    &=&1+\frac{1}{k_1-1}+\frac{1}{(m_1-1)(k_1-1)}.\\
    \end{eqnarray*}
    As a result, when $k_1 \geq 3$ and $m_1 \geq 3$,
    $$\frac{\lambda}{\lambda-1}\leq 1+\frac{1}{2}+\frac{1}{4}<2,$$
    and thus Equation~\ref{eq:condition-lambda} holds without equality as $r_f\geq 2r_W$, which implies $c<c'$. For the other two special cases where one of $k_1$ and $m_1$ is 2 and the other is 3, we have to compare the LHS and RHS to determine whether Equation~\ref{eq:condition-lambda} holds.
\end{itemize}

\subsection{Proof of Theorem~\ref{theorem:factor-window:cost-compare}}\label{proof:theorem:factor-window:cost-compare}
Let $d=c_f-c'_f$. It then follows that
\begin{eqnarray}\label{eq:general-condition:cost-compare}
d &=&\sum\nolimits_{j=1}^{K} n_j\Big(M(W_j, W_f) -M(W_j, W'_f)\Big) + \Delta\\\nonumber
&=&\sum\nolimits_{j=1}^{K} n_j \Big(\frac{r_j-r_f}{s_f} -\frac{r_j-r'_f}{s'_f}\Big) + \Delta\\\nonumber
&=&\sum\nolimits_{j=1}^{K} n_j \Big(\frac{r_j}{s_f} -k_f -\frac{r_j}{s'_f}+k'_f\Big) + \Delta,
\end{eqnarray}
where
\begin{eqnarray*}
\Delta &=& n_f\cdot M(W_f, W)-n'_f\cdot M(W'_f, W)\\
&=& n_f\Big(1+\frac{r_f-r_W}{s_W}\Big)-n'_f\Big(1+\frac{r'_f-r_W}{s_W}\Big)\\
&=& n_f\Big(1+\frac{r_f}{s_W}-k_W\Big)-n'_f\Big(1+\frac{r'_f}{s_W}-k_W\Big).
\end{eqnarray*}
Clearly, $W_f$ is more beneficial if $d< 0$.

\begin{proof}
Since $W_f$, $W'_f$, and $W$ are all tumbling windows, $k_f=k'_f=k_W=1$.
Substituting into Equation~\ref{eq:general-condition:cost-compare} and using the facts $r_f=s_f$, $r'_f=s'_f$, and $r_W=s_W$ yields
\begin{eqnarray*}
c_f-c'_f &=&\sum\nolimits_{j=1}^{K} n_j \Big(\frac{r_j}{r_f}-\frac{r_j}{r'_f}\Big) + n_f\cdot\frac{r_f}{r_W} -n'_f\cdot\frac{r'_f}{r_W}\\
&=&n_f\Big(\sum\nolimits_{j=1}^{K} \frac{n_j}{n_f} \Big(\frac{r_j}{r_f}-\frac{r_j}{r'_f}\Big) + \frac{r_f}{r_W} -\frac{n'_f}{n_f}\cdot\frac{r'_f}{r_W}\Big).
\end{eqnarray*}
Again we consider when $c_f\leq c'_f$ holds.
Or equivalently,
\begin{equation*}
\sum\nolimits_{j=1}^{K} \frac{n_j}{n_f} \Big(\frac{r_j}{r_f}-\frac{r_j}{r'_f}\Big) + \frac{r_f}{r_W} -\frac{n'_f}{n_f}\cdot\frac{r'_f}{r_W}\leq 0.
\end{equation*}
Similarly, define
$$\rho_j=\frac{r_j}{r_f}, \quad \rho'_j=\frac{r_j}{r'_f},\quad \forall 1\leq j\leq K.$$
Since $W_f$ is tumbling,
$$n_f=m_f=\frac{R}{r_f}=\frac{m_jr_j}{r_f}=m_j\rho_j.$$
It therefore follows that
\begin{equation*}
\sum\nolimits_{j=1}^{K} \frac{n_j}{m_j\rho_j} (\rho_j-\rho'_j) + \frac{r_f}{r_W} -\frac{n'_f}{n_f}\cdot\frac{r'_f}{r_W}\leq 0.
\end{equation*}
Noting that
$$\frac{\rho'_j}{\rho_j}=\frac{r_j/r'_f}{r_j/r_f}=\frac{r_f}{r'_f}$$
and making some rearrangement of the terms yields
\begin{equation*}
\Big(1-\frac{r_f}{r'_f}\Big)\sum\nolimits_{j=1}^{K} \frac{n_j}{m_j} +\frac{r'_f}{r_W}\Big(\frac{r_f}{r'_f} -\frac{n'_f}{n_f}\Big)\leq 0.
\end{equation*}
As before, define $\lambda=\sum\nolimits_{j=1}^{K} \frac{n_j}{m_j}$.
It then follows that
\begin{equation*}
\frac{r_f}{r'_f}\geq\frac{\lambda-\frac{r'_f}{r_W}\cdot\frac{n'_f}{n_f}}{\lambda-\frac{r'_f}{r_W}}.
\end{equation*}
Moreover, since both $W_f$ and $W'_f$ are tumbling windows, we have
$n_f=m_f$ and $n'_f=m'_f$. Therefore,
\begin{equation*}
\frac{r'_f}{r_W}\cdot\frac{n'_f}{n_f}=\frac{r'_f}{r_W}\cdot\frac{m'_f}{m_f}=\frac{R}{r_Wm_f}=\frac{r_f}{r_W},
\end{equation*}
which yields
\begin{equation*}
\frac{r_f}{r'_f}\geq\frac{\lambda-\frac{r_f}{r_W}}{\lambda-\frac{r'_f}{r_W}}.
\end{equation*}
This completes the proof of the theorem.
\end{proof}

\section{\blue{Query Rewriting for Trill}}

Formally, given $\mathcal{G}_{\min}$ that captures the optimal window coverage relationships, the query rewriting algorithm works as follows.
Suppose that the original plan is
$$\text{Input Stream}\Rightarrow\verb|MultiCast|\Rightarrow\mathcal{W}=\{W_1,...,W_n\}\Rightarrow\verb|Union|.$$
We first replace $\mathcal{W}$ by the min-cost WCG $\mathcal{G}_{\min}$:
$$\text{Input Stream}\Rightarrow\verb|MultiCast|\Rightarrow\mathcal{G}_{\min}\Rightarrow\verb|Union|.$$
We then perform the following steps:
\begin{itemize}
    \item For each window $w$ (in $\mathcal{G}_{\min}$) without an incoming edge, create a link from \verb|MultiCast| to $w$. Remove the \verb|MultiCast| operator if there is only one such $w$.
    \item For each (intermediate) window $v$ with outgoing edges, insert a \verb|MultiCast| operator $M_v$. Create a link from $v$ to $M_v$ and a link from $M_v$ to \verb|Union|. For each $(v, u)$ of $v$'s outgoing edges, create a link from $M_v$ to $u$.
    \item For each window $w$ without outgoing edges, create a link from $w$ to \verb|Union|.
\end{itemize}

\section{\blue{More Evaluation Results}}

\subsection{\blue{Results on Synthetic Data}}

\blue{Figure~\ref{fig:throughput:ws10} presents the details of the throughput results on window sets over the \textbf{Synthetic-10M} dataset when setting the window-set size $|\mathcal{W}|=10$. Figures~\ref{fig:throughput:ws5:1m} and~\ref{fig:throughput:ws10:1m} further present the details of the throughput results on window sets over the \textbf{Synthetic-1M} dataset, and we summarize the results in Table~\ref{tab:throughput-summary:1m}. Overall, we observe similar patterns on \textbf{Synthetic-10M} and \textbf{Synthetic-1M}.}

\begin{table}
\small
\centering
\begin{tabularx}{\columnwidth}{|l|X|X|X|X|}
\hline
\textbf{Setup} & \textbf{w/o FW (Mean)} & \textbf{w/o FW (Max)} & \textbf{w/ FW (Mean)} & \textbf{w/ FW (Max)}\\
\hline
\hline
R-5-tumbling &	1.21$\times$ & 	2.01$\times$ & 	1.85$\times$ & 	2.41$\times$\\
R-10-tumbling &	1.36$\times$ &	1.72$\times$ &	1.94$\times$ &	3.13$\times$\\
R-5-hopping &	1.19$\times$ &	1.76$\times$ &	2.90$\times$ &	3.78$\times$\\
R-10-hopping &	1.31$\times$ &	1.54$\times$ &	2.94$\times$ &	5.14$\times$\\
\hline
\hline
S-5-tumbling &	1.63$\times$ &	1.79$\times$ &	3.82$\times$ &	4.43$\times$\\
S-10-tumbling &	1.91$\times$ &	2.07$\times$ &	6.27$\times$ &	\textbf{7.27}$\times$\\
S-5-hopping & 1.33$\times$ & 1.51$\times$ &	2.10$\times$ &	2.73$\times$\\
S-10-hopping &	1.54$\times$ &	1.69$\times$ &	2.75$\times$ &	3.65$\times$\\
\hline
\end{tabularx}
\caption{\blue{Summary of throughput boosts on \textbf{Synthetic-1M}, where `R' stands for window sets generated by \textbf{RandomGen}, `S' stands for window sets generated by \textbf{SequentialGen}, and `5' and `10' are the sizes of the window sets generated.}}
\vspace{-1em}
\label{tab:throughput-summary:1m}
\end{table}

\balance


\subsection{{Results on Real Data}}
\blue{Figures~\ref{fig:DEBS:ws5} and~\ref{fig:DEBS:ws10} present the details of the throughput results on the real dataset \textbf{Real-32M} when setting the window-set size $|\mathcal{W}|$ to $|\mathcal{W}|=5$ and $|\mathcal{W}|=10$, respectively.}
Again, we observe that the throughput when using factor windows is often significantly higher than that when factor windows were not considered.
We also observe that, even just using the rewritten plans without including factor windows can often significantly outperform the original plans.

\subsection{\blue{Effectiveness of Cost Model}}

To test the effectiveness of our cost model developed in Section~\ref{section:cost-based-eval}, we measure the \emph{correlation}~\cite{GuSW12,WuWHN14,abs-2003-04410} between the observed throughput boost and the estimated cost reduction using the cost model on \textbf{Synthetic-10M}, given that the constant input event rate of \textbf{Synthetic-10M} matches the underlying assumption of our cost model.
Specifically, let the throughput of the rewritten query plans, without and with factor windows, be $T_{\text{w/o}}$ and $T_{\text{w/}}$, respectively, and we define the \emph{throughput speedup} as $\gamma_T = \frac{T_{\text{w/}}}{T_{\text{w/o}}}$.
Moreover, let the estimated costs of the rewritten query plans, without and with factor windows, be
$C_{\text{w/o}}$ and $C_{\text{w/}}$, respectively, and we define the \emph{estimated speedup} as $\gamma_C=\frac{C_{\text{w/o}}}{C_{\text{w/}}}$.
We are interested in the correlation between $\gamma_T$ and $\gamma_C$.
Figure~\ref{fig:correlation} presents the results of this correlation test.
In each chart, the $x$-axis represents $\gamma_C$ whereas the $y$-axis represents $\gamma_T$, and we have merged data points from both $|\mathcal{W}|=5$ and $|\mathcal{W}|=10$.
Overall, we observe very strong correlation between $\gamma_T$ and $\gamma_C$, regardless of the experiment setup (e.g., \textbf{RandomGen} vs. \textbf{SequentialGen}, ``partitioned by'' vs. ``covered by'').
We further calculated the Pearson correlation coefficient $r$ for each chart, and we observe that $r\geq 0.94$ in all cases that we tested.

\subsection{\blue{Scalability Tests}}

\blue{Figures~\ref{fig:scalability:ws15} and~\ref{fig:scalability:ws20} present the details of our scalability test on \textbf{Synthetic-10M} when setting the window-set size $|\mathcal{W}|\in\{15, 20\}$.
We observe similar patterns to that when setting $|\mathcal{W}|\in\{5, 10\}$, which suggests that the optimized query plans produced by our cost-based approach can scale up smoothly with increasing window-set size.}

\subsection{\blue{Comparison with Window Slicing}}

\blue{Figure~\ref{fig:scotty-compare:ws5} presents the results that compare the throughput of Apache Flink, Scotty, and our factor-window based optimization when setting the window-set size $|\mathcal{W}|=5$.
We observe that our approach and Scotty result in similar throughput, and both approaches significantly outperform the default query plan used by Apach Flink that evaluates each window aggregate independently.}

\end{document}